\documentclass[onecolumn,secnumarabic,amssymb, nobibnotes, aps, prd]{revtex4-2}
\usepackage{graphicx}   
\usepackage{lipsum}

\usepackage{amsmath}

\usepackage{tikz,xcolor,hyperref}

\usepackage{caption}
\usepackage{subcaption}
\usepackage{aas_macros}
\usepackage{slashed}
\usepackage{enumitem}
\usepackage{slashed}
\usepackage{float}
\usepackage{tabularx}
\usepackage{hyperref}
\usepackage{amsmath}

\begin{document}
	
%

\preprint{TIFR/TH/23-16}

\title{\boldmath EDGES of the dark forest: A new absorption window into the composite dark matter and large scale structure}

\author{Anoma Ganguly,}
\email{anoma@theory.tifr.res.in}

\author{Rishi Khatri,}
\email{khatri@theory.tifr.res.in}

\author{Tuhin S. Roy}
\email{tuhin@theory.tifr.res.in}
\affiliation{Department of Theoretical Physics, Tata Institute of Fundamental Research,\\Homi Bhabha Road,
	Mumbai 400005, India}
\date{\today}





\begin{abstract}
We propose a new method to hunt for dark matter using dark forest/absorption features across the whole electromagnetic spectrum from radio to gamma rays, especially in the bands where there is a desert i.e. regions where no strong lines from baryons are expected. Such novel signatures can arise for dark matter models with a composite nature and internal electromagnetic transitions. The photons from a background source can interact with the dark matter resulting in an absorption signal in the source spectrum. In case of a compact source, such as a quasar, such interactions in the dark matter halos can produce a series of closely spaced absorption lines, which we call the dark forest. We show that the dark forest feature is a sensitive probe of the dark matter self-interactions and the halo mass function, especially at the low mass end. There is a large volume of parameter space where dark forest is more sensitive compared to the best current and proposed direct detection experiments. Moreover, the absorption of CMB photons by dark matter gives rise to a global absorption signal in the CMB spectrum. For dark matter transition energies in the range $2.5\times 10^{-4} \text{\,eV} - 5\times 10^{3}$ eV, such absorption features result in spectral distortions of the CMB in the COBE/FIRAS band of 60-600 GHz. If the dark matter transition frequency is in the 100-200 GHz range, we show that the absorption of CMB photons by dark matter can provide an explanation for the anomalous absorption feature detected by the EDGES collaboration in 50-100 MHz range.
\end{abstract}
		\maketitle
		\tableofcontents
	\section{Introduction}
	\label{sec:intro}
	Dark matter, although making up a major fraction of the matter content in our Universe, continues to remain a mystery. Owing to its elusive nature, the different experiments searching for it \cite{2022arXiv220306380B,2022PTEP.2022h3C01W} have till now only succeeded in placing stringent constraints on its possible interactions beyond the usual gravitational force. The allowed parameter space for possible dark matter candidates is huge, with masses roughly ranging from $10^{-22}$ eV ultralight bosons to $10^{43}$ GeV compact objects. A naive estimate for the smallest dark matter mass can be obtained by restricting the de Broglie wavelength of dark matter to galactic scales $\sim $ kpc. For dark matter velocity dispersion $\sim 100$ km/s, the lower limit on dark matter mass comes out to be $\sim10^{-22}$ eV. More careful analyses using observations of ultra faint dwarf galaxies \cite{2019PhRvL.123e1103M,2022PhRvD.106f3517D} and satellite galaxies of Milky Way \cite{2021PhRvL.126i1101N} improve the lower limit on the dark matter mass to be in $\sim 10^{-21}-10^{-19}$ eV range.
	Therefore a more clear picture of the true nature of dark matter can only start emerging if we find new methods to look for dark matter in experiments.
	The current searches for dark matter broadly fall into three categories, direct detection: where detectors search for nuclear/electronic recoil due to dark matter, indirect detection: where experiments look for emission signals of different Standard Model particles produced from dark matter annihilation, decay, etc., or collider searches: where high energy accelerators try to produce dark matter by colliding Standard Model particles. The absence of a distinct signature of dark matter in any of these searches so far not only hints at its far more nuanced nature, but also calls for new detection strategies. In this work, we propose a novel method to look for dark matter in the absorption lines of a background source. The main advantage of absorption lines is their ability to probe very weak interactions between dark matter and photons, which is possible if the background source is sufficiently bright.
	
	Absorption lines are a generic feature of a class of models where dark matter is a composite particle with a discrete energy spectrum. The presence of a small electromagnetic coupling can allow the transitions between different dark matter energy states via emission/absorption of a photon. As a specific example, we will consider dark matter to be a composite particle made of two elementary particles of the dark sector. A strong dark attractive potential between the constituents makes dark matter stable on cosmological scales. As a whole dark matter stays electromagnetically neutral, while the constituents carry a millicharge. This simple model allows us to describe dark matter as a bound state with weak electromagnetic transitions similar to a hydrogen atom. The generic signatures of such models will include both absorption as well as emission lines in experiments. While a lot of work in the past has been on composite dark matter induced emission lines \cite{2010JCAP...05..021K,2012PhRvD..86k5013C,2014JCAP...05..033F,2014PhRvD..89l1302C} and electromagnetic signals in colliders \cite{2022JHEP...06..047B,2010PhRvD..82g5019F}, only a few touch upon the absorption signatures of dark matter \cite{2007PhRvD..75b3521P,2010PhRvD..81i5001K}. In addition to electromagnetic/radiative transitions, the transitions between different energy states can happen via inelastic scattering between dark matter particles. This has been studied in the context of small scale structure problems like the core-cusp problem and the missing satellite problem \cite{1994Natur.370..629M,2008ApJ...676..920K,2008AJ....136.2648D,2011ApJ...742...20W,1996ApJ...462..563N,1997ApJ...490..493N,1999MNRAS.310.1147M,2013PhRvD..87j3515C,2014PhRvD..89d3514C,2016PhRvD..94l3017B,1999ApJ...522...82K,1999ApJ...524L..19M,2015PhRvD..91b3512F,2016JCAP...07..013F,2015JCAP...01..021S,2017JCAP...03..048B,2018PhRvD..97b3002D}. Additionally, composite dark matter models have also been invoked to resolve the Hubble tension \cite{2022PhRvL.128t1301C, 2022PhRvD.105i5005B}. We would like to emphasize that even though our dark matter is a bound state of milli-charged constituents, as a whole it is electromagnetically neutral. Therefore the existing direct and indirect constraints on millicharged dark matter models do not directly apply to us. 
	However, higher order electromagnetic moments of dark matter give rise to signals at direct detection experiments specially sensitive to small recoil such as XENON 10 \cite{2011PhRvL.107e1301A}, XENON 100 \cite{2016PhRvD..94i2001A, 2017PhRvD..96d3017E}, Dark-Side \cite{2018PhRvL.121k1303A}, SENSEI (protoSENSEI@surface \cite{2018PhRvL.121f1803C} and protoSENSEI@MINOS \cite{2019PhRvL.122p1801A}), and CDMS-HVeV \cite{2018PhRvL.121e1301A}. We discuss these constraints in section \ref{sec:paramspace2}.

	In this work, we focus on the less studied and more promising signature of composite dark matter: the absorption of light. The absorption line from dark matter inside a single galaxy cluster at gamma ray frequencies was studied in \cite{2007PhRvD..75b3521P,2010PhRvD..81i5001K}. However, there is no apriori reason to be confined to the gamma-ray band. In particular, the detection of an absorption line unidentifiable with the known transitions in baryonic atoms or molecules in any part of the electromagnetic spectrum is a tell-tale signature of such dark matter models. In addition to a single absorption line, we can even have a collection of absorption lines or \emph{dark forest} similar to the Ly-$\alpha$ or 21 cm forest generated by the neutral hydrogen atoms. The dark forest arises due to absorption of light by dark matter halos along the line of sight (LoS) to a quasar. We show that the dark forest opens a new window to the large scale structure as it traces the evolution of  dark matter temperature and distribution inside the dark matter halos through the cosmic history.
	
	We make a detailed study of the evolution of dark forest from redshift of 7 to 0 for dark matter transitions at radiowave frequency of $156$ GHz. Interestingly, we find that the amount of absorption by a dark matter halo of a given mass is sensitive to the presence of dark matter self-interactions. Moreover, the density of absorption lines has a strong dependence on the smallest dark matter sub-structures present in the Universe. In general, the dark forest can appear in any part of the electromagnetic spectrum including radio, microwave, infrared, optical, X-ray, and gamma ray bands.  In particular, the detection of absorption lines in the spectrum of a bright quasar at  $z\sim6$ at frequencies $<$ 200 MHz, below the 21 cm forest of neutral hydrogen, will be a smoking gun signature for this model. This may be possible with the upgraded Giant Meterwave Telescope (uGMRT) \cite{1991CSci...60...95S,2017CSci..113..707G} and the Square Kilometer Array (SKA) \cite{2020PASA...37....2W} which have lowest frequency bands in 50-350 MHz and 125-250 MHz respectively.
	
	When the isotropic cosmic microwave background (CMB) acts as a background source, the absorption of CMB photons by inelastic composite dark matter gives rise to a global absorption feature in the CMB spectrum. The origin of the global absorption signal from transitions in dark matter is similar to the global absorption feature caused by the hyper-fine transitions in neutral hydrogen during the Dark Ages \cite{2004MNRAS.352..142B, 2004ApJ...608..622Z,2006PhR...433..181F,2012RPPh...75h6901P}. After dark matter decouples from the electron-baryon plasma, it cools as $(1+z)^2$, with the temperature soon becoming much lower than the CMB temperature. At very high redshifts, the strong inelastic collisions between dark matter particles bring the two dark matter energy levels in kinetic equilibrium with the dark matter temperature. The dark matter temperature being much lower than the CMB temperature implies that the dark matter particles in the ground state can absorb the CMB photons and generate an absorption signal in the CMB spectrum. As the Universe cools and the number density of dark matter particles gets diluted, the radiative transitions due to CMB photons take over the dark matter collisional transitions, bringing the level population in equilibrium with the CMB temperature and the signal vanishes.  An important difference from 21 cm cosmology is the role of bremsstrahlung, which is important before recombination when the Universe has ample number of electrons and protons. This process can erase spectral distortions in the low frequency tail of the CMB before recombination and establish an almost perfect black body spectrum in the Rayleigh-Jeans tail. Thus the high redshift (low frequency) edge of the absorption signal is entirely determined by bremsstrahlung, while the low redshift (high frequency) edge of the signal depends on the ratio of collisional to radiative coupling of dark matter.

	Tantalizingly, we find that the absorption of CMB photons by dark matter with a transition frequency in the 100-200 GHz range at redshifts $\sim 4000-1000$ can produce a global absorption feature that is consistent with the measurements of the Experiment to Detect the Global Epoch of reionization Signal (EDGES) collaboration \cite{2018Natur.555...67B,2021AJ....162...38M}. The EDGES collaboration reported a strong absorption feature which is almost double in amplitude compared to the maximum absorption expected from 21 cm absorption by hydrogen in the Standard Model of cosmology. Another experiment called The Shaped Antenna measurement of the background Radio Spectrum (SARAS 3) \cite{2022NatAs...6..607S} has disfavored the EDGES absorption profile from being cosmological in origin. A recent paper from the EDGES collaboration \cite{2022MNRAS.tmp.2422M} does a Bayesian analysis jointly constraining the receiver calibration, foregrounds, and the measured signal reaffirming the presence of an absorption feature. Several other groups such as Large aperture Experiment to detect the Dark Ages (LEDA) \cite{2015ApJ...799...90B,2016MNRAS.461.2847B,2018MNRAS.478.4193P}, Probing Radio Intensity at high Z from Marion (PRIZM) \cite{2019JAI.....850004P}, Radio Experiment for the Analysis of Cosmic Hydrogen (REACH) \cite{2022NatAs...6..984D}, Sonda Cosmol\'ogica de las Islas para la Detecci\'on de Hidr\'ogeno Neutro (SCI-HI) \cite{2014ApJ...782L...9V}, and Cosmic Twilight Polarimeter  (CTP) \cite{2019ApJ...883..126N} are working on validating these claims. Even if the exact profile measured by EDGES turns out to be due to systematics, an anomalously large amplitude of the absorption signal would still require beyond Standard Model physics, if confirmed by other experiments. 
	The proposed solutions to the EDGES anomaly broadly fall into two classes: decreasing the spin temperature of neutral hydrogen by cooling the baryons via elastic scattering with millicharged dark matter \cite{2018Natur.555...71B}, or increasing the radiation temperature at radiowave frequencies \cite{2018ApJ...858L..17F}. A large chunk of the parameter space of the vanilla millicharged dark matter has been ruled out by the early Universe observations of CMB and BBN \cite{2018PhRvD..98j3529K, 2018PhRvD..98j3005B, 2019PhRvD.100b3528C, 2018PhRvL.121a1102B}. Recent works in the literature have tried to reopen this parameter space by introducing additional features such as two-component \cite{2019PhRvD.100l3011L, 2022ChPhC..46d5102L} and composite millicharged dark matter \cite{2022PhRvD.105g5020M}. 

	We begin with a discussion of the theoretical framework of composite dark matter in the section \ref{sec:model}. We make a detailed study of the unique absorption signatures of such dark matter models in section \ref{sec:expt}, where we discuss the physics of dark forest in subsection \ref{sec:quasar}, and the global absorption feature in the CMB spectrum in subsection \ref{sec:edges}. 
	We then proceed towards the implications of different astrophysical experiments in section \ref{sec:paramspace} and direct detection experiments in section \ref{sec:paramspace2} on the allowed parameter space for our dark matter model.  
	We conclude our results in section \ref{sec:conclusion}.
	
	We use the Planck 2018 \cite{2020A&A...641A...6P} cosmological parameters (Hubble constant: $H_0 = 100\,h = 67.66$ km s$^{-1}$ Mpc$^{-1}$, $\Omega_{m} = 0.3111$, and $\Omega_{b} = 0.049$). We also use the publicly available codes Recfast++ \cite{2011MNRAS.412..748C,2010MNRAS.407..599C}, Colossus \cite{2018ApJS..239...35D}, and FeynCalc \cite{2020CoPhC.25607478S,2016CoPhC.207..432S, Mertig:1990an} in our analysis.
	
	\section{A theoretical framework for the dark sector}
	\label{sec:model}
	In this section, we set up a theoretical framework for the underlying physics of the dark sector whose principal modes of observation are absorption features in the spectrum of a background source. We begin by describing the minimal (simplified) model of the dark sector. In particular, we chalk out the essential parametrizations needed in order to describe the physics of the absorption signatures of the dark sector quantitatively. For the purpose of phenomenological studies, this discussion is sufficient and the reader may skip subsection \ref{subsec:2.2}. The purpose of subsection \ref{subsec:2.2}  is to put the simplified model in subsection \ref{subsec:2.1} on a firmer theoretical ground. We present a \emph{proof of principle} model where the simplified model emerges dynamically at low energies because of confinement in a strongly coupled dark sector.
	\subsection{A simple two-state dark matter model}\label{subsec:2.1}
\begin{figure}[t]
		\centering
		\includegraphics[width=0.3\textwidth]{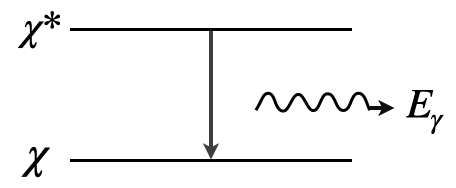}
		\caption{Electromagnetic transition in the simplified dark matter model characterized by two states $\chi$ and $\chi^*$ with an energy splitting $E_{\gamma}$.}
		\label{fig:scales}
	\end{figure}

	The main ingredients of the setup is a dark sector (shown in figure \ref{fig:scales}) characterized by two states, namely the ground state (state 0) and the excited state (state 1). Interchangeably we will refer to state 0 as particle $\chi$ (with mass $m_{\chi}$) and state 1 as particle $\chi^*$ (with mass $m_{\chi^*}$). As shown in the figure, the minimal model allows for transition $\chi^* \rightarrow \chi + \gamma$, which suggests that the total angular momentum ($J$) of $\chi$ and $\chi^*$ differ by 1. We will choose $\chi$ to be a scalar state ($J=0$) and $\chi^*$ to be a vector state ($J=1$).
	The existence of electromagnetic transitions also dictates that $\chi$ and $\chi^*$	must have identical quantum numbers for all other possible symmetry transformations. For example, if we invoke an additional symmetry that makes $\chi$ stable, then the particle $\chi^*$ must carry the same non-trivial charge as $\chi$, since the only distinction between $\chi$ and $\chi^*$ is the angular momentum quantum number. Therefore, without loss of generality, we can take the ratio of degeneracy of state 0 (say $g_0$) and state 1 (say $g_1$) to be $g_1/g_0 = 3$.

	We denote the rest energy of $\chi$ and $\chi^*$ by $E_0=m_\chi c^2$ and $E_1=m_{\chi^*}c^2$ respectively. In this work, we assume the energy splitting between the two states to be hierarchically smaller than the mass scales themselves. Mathematically, the energy of the emitted (absorbed) photon $E_{\gamma}$ satisfies,	
	\begin{equation}
		E_{\gamma} = h\nu_{0}\equiv k_BT_* = E_1-E_0 = (m_{\chi^*}-m_{\chi})c^2\ll m_{\chi}c^2<m_{\chi^*}c^2,\label{2.1}
	\end{equation}
	where $h$ is the Planck's constant and $k_{B}$ is the Boltzmann constant. We also took the opportunity to define the transition temperature ($T_{*}$) and frequency ($\nu_0$) in eq. \eqref{2.1}.

	Below, we discuss only the minimal set of parametrizations needed to describe the physics of absorption signatures of the dark sector:
	
	\begin{itemize}[leftmargin=10pt]
		
		\item The population of dark matter particles in the two states is decided by the collisional and radiative transition rates. The ratio of the number density of dark matter particles in the ground state ($n_{0}$) with respect to the excited state ($n_{1}$) is parameterized by the excitation temperature $T_{\text{ex}}$, 
		\begin{equation}
			\frac{n_{0}}{n_{1}} \equiv\frac{g_{0}}{g_{1}}\exp(T_*/T_{\text{ex}}).\label{29}
		\end{equation}
		In our simplified model, the occupation number of 0 and 1 states gives the total dark matter number density,
		\begin{equation}
			n_{\chi} = n_0 + n_1.\label{210}
		\end{equation}
		\item The transition between the two states can happen via emission/absorption of a photon which is parameterized in terms of Einstein A and B coefficients in the following way: 
		
		The number of radiative transitions per unit time per unit volume from level 0 to level 1 is proportional to the Einstein coefficient $B_{01}$,
		\begin{equation}
			\frac{dn_{0\rightarrow1}}{dt} = n_0B_{01}\bar{J},\label{3.3o}
		\end{equation}
		where $\bar{J}$ is the mean intensity of incident light.
		The number of radiative transitions per unit time per unit volume from level 1 to level 0 is a sum total of spontaneous emission which proportional to the Einstein coefficient $A_{10}$ and stimulated emission which is proportional to $B_{10}$,
		\begin{equation}
			\frac{dn_{1\rightarrow0}}{dt} = n_1(A_{10} + B_{10}\bar{J}).\label{3.4o}
		\end{equation}
		The Einstein coefficients $A_{10}$, $B_{01}$, and $B_{10}$ are related to each other via the Einstein relations which follow from the principle of detailed balance,
		\begin{align}
			A_{10} &= \frac{2h\nu_0^3}{c^2}B_{10}\nonumber,\\
			g_0B_{01} &= g_1B_{10}.
		\end{align}
		In this work, we parameterize the Einstein coefficient for hyper-fine transitions in the dark sector in terms of the Einstein coefficient for hyper-fine transitions in the hydrogen atom,
		\begin{equation}
			A_{10} = \alpha_A\, A_{10}^{\text{HI}}, \hspace{5pt} \text{where} \hspace{5pt} A_{10}^{\text{HI}} = 2.85 \times 10^{-15} \text{s}^{-1}. \label{a10}
		\end{equation}
	
		We will set $\alpha_A=0.35$ in our numerical computations (see section \ref{sec:paramspace} for justification) unless stated otherwise.
		\item The transition between the two states can also happen via inelastic collisions between dark matter particles parameterized in terms of the collisional excitation and de-excitation coefficients $C_{01}$ and $C_{10}$ respectively. The number of collisional transitions per unit volume from level $i$ to level $j$ is given by, 
		\begin{align}
			\frac{dn_{i\rightarrow j}}{dt} &= n_iC_{ij},\label{214}
		\end{align}
		where $i\neq j$ and $i, j$ run from 0 to 1. For a thermal velocity (Maxwell Boltzmann) distribution of dark matter particles at temperature $T_\chi$, the two collisional coefficients are related as,
		\begin{align}
			C_{ij}(T_\chi) &=\frac{g_{j}}{g_{i}}\exp(-T_*/T_{\chi})\,C_{ji}(T_\chi).
		\end{align}	
		In case of a quasar, we simply consider the implications for two extreme scenarios, one where inelastic collisions are completely absent (collisionless dark matter) and the other where the inelastic collisions are strong (collisional dark matter) (see eq.\eqref{tex} in subsection \ref{sec:quasar}). In case of CMB as a background source, we use intermediate collisional cross-section parameters as a function of dark matter temperature (see eq.\eqref{4.23} of subsection \ref{sec:edges} for details).
	\end{itemize}

	\subsection{The composite dark sector}\label{subsec:2.2}
	In this subsection, we provide a \emph{proof of principle} scenario when the simplified model described in the previous section emerges dynamically for a strongly coupled dark sector. As we show later in this subsection, our setup consists of an ultraviolet (UV) complete model with a non-abelian gauge theory (dark color) with specifically designed matter (dark quarks) which carry small electromagnetic charges. The low energy spectrum of this theory consists of color neutral bound states. Similarly, our candidates for dark matter ($\chi$ and $\chi^*$) are bound states with 0 electric charge. However, the higher electromagnetic moment operators involving $\chi$, $\chi^*$, and electromagnetic field tensor $F_{\mu\nu}$ allow for the radiative transitions $\chi^* \rightarrow \chi + A_{\mu}$.
	
	In the effective model we describe below, we will not explicitly specify the mechanism by which the dark quark acquires a millicharge.
	There exist different mechanisms in literature for generating dark quarks with a small electromagnetic coupling \cite{2001PhRvL..86.4757A,2006PhRvD..73d5016B, 2016PhRvD..93h5007K,2017JHEP...02..036G}. An important point to note is that in these models the millicharge of dark quarks is not due to the kinetic mixing between the photon and a dark photon.
	In cases where dark photon exists, we assume the dark photon to be massive enough that it is non-relativistic at the time of big bang nucleosynthesis (BBN) and does not contribute to the relativistic degrees of freedom (N$_{\text{eff}}$). The dark photon also does not play any direct role in the cosmological scenarios considered in this work. 
	
	Before we build such a model, we note the following set of considerations which guided us in the model-building exercise. Even though each of these conditions need not be fulfilled strictly, stating them is useful. This not only allows us to stay general but also provides a direction for building the dark sector model as described later in this subsection.
	\begin{itemize}[leftmargin=15pt]
		\item[A.] Consider first the simplified scenario described in the previous section. As discussed before, all states linked via electromagnetic transitions must have the same quantum numbers apart from the total angular momentum. In our model we introduce an abelian symmetry (say $U(1)_D$) which gives $\chi$ a non-zero charge (call it darkness number). The entire tower of states (call it dark tower) (which in principle can be connected to $\chi$ via one or multiple electromagnetic transitions) must therefore have the same darkness number.
		\item[B.] By construction, we have $\chi$ and the entire tower of transitioning states electrically neutral. This allows us to evade strong bounds from CMB \cite{2004JETPL..79....1D,2013PhRvD..88k7701D}, virialization in dark matter halos, elliptical shape of galaxies \cite{2011PhRvD..83f3509M}, bullet cluster \cite{2004ApJ...606..819M}, etc. In our model, the states in the dark tower are composites of constituents having small electric charge. Therefore the operators that give rise to $\chi^* \rightarrow \chi + \gamma$ transition are irrelevant.
		\item[C.] We demand that our UV model must yield the composite state with,
		\begin{equation}
			E_{\gamma} \ll E_{\text{binding}} \lesssim \text{mass of composite state}. \label{21}
		\end{equation}
		Composite states with binding energy comparable to the transition energy (as described in previous section) yield signals of ionization (typically much stronger) in the vicinity of emission/absorption lines. In the early Universe, ionized dark matter will be subject to radiation pressure similar to the baryons which can modify the CMB acoustic peaks, imprint acoustic oscillation features in the dark matter power spectrum, and erase structure on small scales. At late times, the Coulomb scattering between ionized dark matter particles inside a halo can give rise to cored central density profiles (see section \ref{sec:paramspace} for details). This suggests that scenarios where the transition energy and binding/ionization energy are similar ($E_{\gamma}\simeq E_{\text{binding}}$), the signals from ionization are a far better probe. The mass of dark matter also plays an important role in the strength of the absorption signal. A smaller dark matter mass yields a larger number density which in turn gives rise to a stronger signature in the spectrum of a source.
	\end{itemize}
\begin{table}[t]
	\centering
	\begin{tabular}{|c|c||cccc|}
		\cline{1-6}
		\centering  
		 & $SU(N)$& $SU(2)^{D}_{L}$ & $SU(2)^{D}_{R}$ & $U(1)_{D}$ & $U(1)_{\text{em}}$       \\	\cline{1-6} 
		
		$q_D$       & $N $      & $2$     & $1$     & $0$     & +$\bar{\epsilon}$ \\
		$q^c_D$  &$\bar{N} $   & $1$     & $\bar{2}$& $0$     & $-\bar{\epsilon}$ \\
		$Q_D$     &$N $      & $1$     & $1$     & $+1$    & +$\bar{\epsilon}$ \\
		$Q^c_D$   & $\bar{N} $  & $1$     & $1$     & $-1$     & $-\bar{\epsilon}$ 
		\\ \cline{1-6}
	\end{tabular}
	\caption{The dark quarks in the Weyl representation and their charges under gauge and global symmetries.}
	\label{tab:ppm}
\end{table}

	We now describe our proof-of-principle UV complete model which at low energy can yield the simplified setup described in the previous section. In our model, the dark matter $\chi$ is a bound state of dark quarks, where the dark gluons of a non-abelian $SU(N)$ gauge theory provides the attractive potential. The matter content of the model under the gauge group as well as charges under additional global symmetries (flavor) is listed in table \ref{tab:ppm}. The model is characterized in terms of three scales:
	\begin{itemize}
		\item[1.]$\Lambda_{N}$: The intrinsic scale of dark color (roughly related to the scale at which the gauge coupling of dark color becomes strong),
		\item[2.] $M_{Q}$: The scale of heavy quarks ($Q_D$ and $Q_D^c$) and,
		\item[3.] $m_q$: The scale of light quarks ($q_D$ and $q_D^c$).
 	\end{itemize}
	The mass terms for the dark quarks in Weyl representation is therefore, given by,
	\begin{align}
		\mathcal{L} \supset  M_{Q}\,Q_DQ^{c}_D + m_q\delta_{ab}\,q_{Da}q_{Db}^c + \text{h.c.},\label{24}
	\end{align}
	where $a$ and $b$ are the flavor indices. 
	
	The physics of the model at high energy is rather simple and straightforward. The infrared (IR) spectrum however requires a careful thinking. Below we summarize the essential features of the low energy behavior:
	\begin{itemize}[leftmargin=10pt]
		\item We utilize the mild hierarchy: $m_q<\Lambda_{N}<M_Q$. The mass term for dark quarks in eq.\eqref{24} ensures that the condensate $\langle q_D q_D^c \rangle$ is flavor diagonal signaling that the global $SU(2)_{L}^{D}\times SU(2)_{R}^{D}\times U(1)_L^{D} \times U(1)_R^{D}$ gets spontaneously broken into a diagonal $SU(2)_V^{D}\times U(1)_D$, with an additional axial $U(1)_A$ broken by the dark color itself. However, a non-zero $m_q$ itself explicitly breaks the axial $SU(2)_A^{D}$. Therefore, there are no massless Nambu Goldstone Bosons (NGBs), but there exist three light pseudo-NGBs or dark pions $\pi_D$ lying in the coset $SU(2)_L^{D}\times SU(2)_R^{D} \big/SU(2)_V^{D}$. One can use the usual exponential parametrization to express $\pi_D$ as,
		\begin{align}
			\Sigma_D\equiv \exp\bigg(\frac{2i\pi_D}{f_D}\bigg) \hspace{5pt}
			\text{where}\hspace{5pt} \pi_D\equiv \pi_D^a\,\tilde{t}^a , \hspace{3pt}\text{and}\hspace{5pt} \Sigma_D \xrightarrow{SU(2)_L^D\times SU(2)_R^D} L_D\Sigma_D R_D^\dagger \label{25}
		\end{align}
		The broken generators $\tilde{t}^a$ belong to the coset, $f_D$ is the energy scale of the dark condensate $\langle q_Dq_{D}^c\rangle$, and $L_D$ and $R_D$ are transformation operators for $SU(2)_L^D$ and $SU(2)_R^D$ respectively.
		\item We identify $U(1)_D$ as the global symmetry responsible for stability of dark matter, and charges under $U(1)_D$ as the darkness number. We take $\bar{Q}_{D}\,q_{D}$ bound state (darkness number -1) as the candidate for dark matter. Similar to the matter anti-matter asymmetry in the visible sector, the asymmetry in the number of dark (darkness number -1) versus anti-dark states (darkness number +1) yields the observed dark matter abundance. 
		\item The lowest lying bound state $\bar{Q}_{D}\,q_{D}$ contains both spin 0 (pseudo scalar) and spin 1 (vector) states. We take the spin 0 state as the candidate for $\chi$ and spin 1 state for $\chi^*$. A convenient way to represent these four physical states collectively is using a matrix field $\mathcal{X}_{v}$ which is an eigenstate of the velocity $v$ of the dark bound state,
		\begin{equation}
			\mathcal{X}_{v}\equiv 	P_{+}\big(\chi_{\mu}^*\gamma^\mu-\chi\gamma^5\big).\label{dm}
		\end{equation}
		The projection operator $P_{+}=\frac{1}{2}\big(1 + \slashed{v}\big)$ captures only the small fluctuations ($\ll M_{Q}$) for this Heavy Quark Effective Field theory (HQET) \cite{Isgur:1989vq, Georgi:1990um,1994NuPhB.412..181J,Mehen:2005hc}.
		
		\item $\chi$-$\chi^*$ exhibits a nearly degenerate system. The mass difference between $\chi$ and $\chi^*$, known in the literature as the hyper-fine splitting $E_{\gamma}$,  arises from operators shown in \eqref{C27} of Appendix \ref{app:C}. This splitting is suppressed by the heavy quark mass $M_Q$, but contains a number of unknown parameters and in the limit $M_Q \rightarrow \infty$, $\chi$ and $\chi^*$ become exactly degenerate.
		
		\item The spontaneous emission rate for $\chi^* \rightarrow \chi\,\gamma$ process is computable within the paradigm of the chiral Lagrangian for the heavy-light systems using the operator mentioned in eq. \eqref{C28} of Appendix A.
		
		\item In order to derive the $\chi^*\chi^*\pi_D$ or $\chi\chi^*\pi_D$ couplings, one needs to take into consideration the symmetry properties of the heavy-light bound states. It is conventional to write the interactions between the bound states and dark pions by defining the vectorial and axial currents which contain the dark pions. These couplings also contribute non-trivially to the hyper-fine splitting as well as the transition rates. In this work, we assume $E_{\gamma}<m_{\pi_D}$, which implies that as far as $\chi^* \leftrightarrow \chi$ transitions are concerned, we can disregard the dark pion transitions.
	\end{itemize}
	One cannot help but notice at this point the essential similarities of our dark sector model with the Standard Model of particle physics. In nature, the strong interactions of Quantum Chromodynamics (QCD) play a similar role in producing heavy-light bound states such as D-mesons or B-mesons. The spontaneous breaking of the approximate chiral symmetry associated with the light quarks of the visible sector gives rise to pions (pNGBs of the visible sector) with substantial interactions with the heavy light mesons. Following the formalism of the chiral Lagrangian for the heavy flavor \cite{Wise:1992hn,Yan:1992gz,1997PhR...281..145C}, one can write down the form of interaction between the dark states and the dark pions, and estimate the hyper-fine splitting, transition rates, etc.
	
	Even though the formalism of HQET seems to describe the different properties of the B and D meson states, such as, the pattern of their couplings and mass gaps extremely well, there is one crucial drawback. The physics of these states is described in terms of unknown constants. While in the context of states in the visible sector, the existence of data allows us to determine these constants (which in turn, makes it possible to predict several other observables rather precisely), such a procedure is not practical for our dark bound states. We also cannot just scale up QCD to predict these, since the number of colors, flavors, and pattern of masses are not the same. Of course, lattice QCD can make definitive statements about the size of splittings and photon transition rates, but such an exercise is beyond the scope of this work. 
	
	\section{Experimental signatures} \label{sec:expt}
	The composite dark matter particle can make an electromagnetic transition from the ground state to an excited state by absorbing a photon. Such transitions can give rise to unique experimental signatures in the form of absorption lines in the spectrum of a bright background source. In particular, the detection of a new absorption line, not identifiable with a known atomic or molecular transition in any part of the electromagnetic spectrum, would be a smoking gun signature for such dark matter models.
	
	The high dark matter density in structures like dark matter halos and dwarf galaxies makes these sites ideal targets that can generate such absorption signals. In particular, when one such object lies along the line of sight (LoS) to a compact source, the absorption of light by the composite dark matter particles inside these objects produces an absorption line in the source spectrum. The shape of the line is characterized by the density and velocity distribution profile of dark matter particles inside the absorber. In reality, we will have multiple such absorbers along the LoS to a distant quasar resulting in a series of absorption lines at different frequency locations in the observer's frame. The frequency location of the absorption lines is decided by the transition frequency and the redshift of the absorber. We study the absorption lines in the spectrum of a compact source for a single absorber by taking an example of a dwarf galaxy and a general dark matter halo in subsection \ref{sssec:halo}. We then proceed towards the case of multiple absorbers along the LoS to a quasar in subsection \ref{sssec:df}. 
	
	When the source is isotropic i.e. the CMB, the composite dark matter particles absorb the CMB radiation giving rise to a broad global absorption feature in the CMB spectrum. We study such a dark global absorption feature in subsection \ref{sec:edges}.

	\subsection{Absorption lines in the spectrum of a compact source}
	\label{sec:quasar}
	When the LoS to the compact source passes through an absorber, such as a dark matter halo, the composite dark matter particles inside these structures can absorb the incident light, imprinting an absorption feature in the source spectrum. Similar to absorption, we can also have emission lines from dark matter imposed on the average spectrum of a galaxy inside the dark matter halo.
	
	In the rest frame of a point-like absorber situated at redshift $z_0$, the absorption/emission happens at the transition frequency $\nu_0$. Due to the expansion of the Universe, the absorption/emission line is observed today at a frequency $\nu = \nu_0/(1+z_0)$. However, complication arises for absorbers of finite size and non-trivial density and velocity profiles in different possible astrophysical scenarios.
	\begin{figure}[t]
		\centering
		\includegraphics[width=0.4\textwidth]{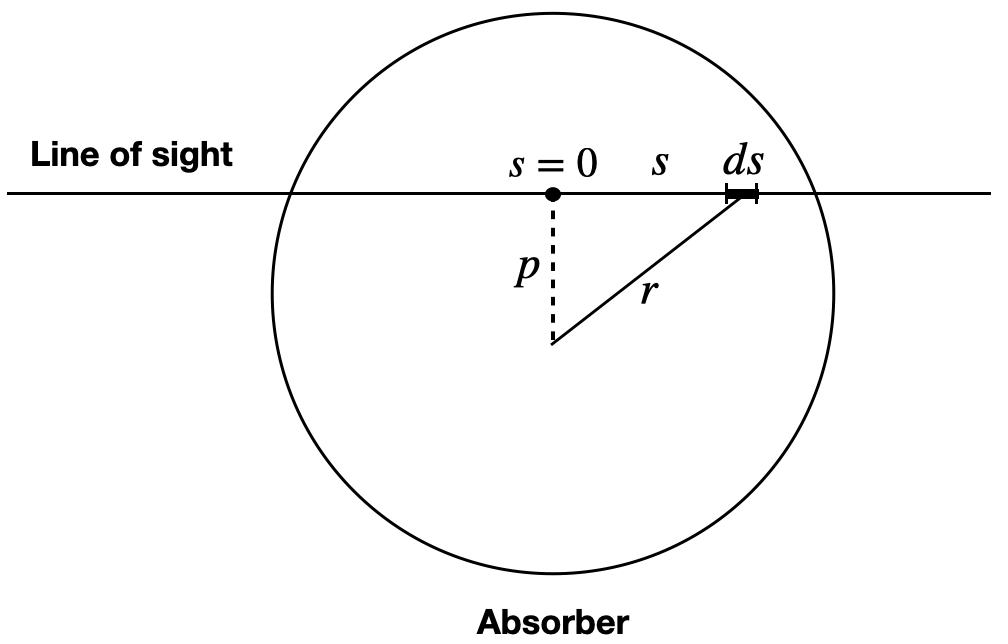}
		\caption{Schematic diagram of the line of sight intersecting an absorber at an impact parameter $p$.}
		\label{fig:halo}
	\end{figure}
	\begin{enumerate}[leftmargin=10pt]
		\item[$\bullet$] \textbf{Dark matter density profile:} For an extended absorber intersected by the LoS to the source at an impact parameter $p$ (as shown in figure \ref{fig:halo}), the cumulative net absorption (true absorption minus stimulated emission) gets contribution from all the particles present along the LoS, which is denoted by $s$. Lets consider a line element $ds$ (in figure \ref{fig:halo}) along the LoS. The true absorption (stimulated emission) is proportional to the number density of dark matter particles in the ground (excited) state, which in turn is proportional to the total number density of dark matter particles $n_{\chi}(r) = \rho(r)/m_{\chi}$, where $\rho(r)$ is the dark matter density at a distance $r$ from the center.
		
		\item[$\bullet$] \textbf{Excitation temperature:}
		The population of dark matter particles in the two states at a radius $r$  inside the dark matter halo is determined by the excitation temperature $T_{\text{ex}}$ defined in eq. \eqref{29}. The excitation temperature is determined by two processes, namely, the radiative transitions due to the CMB photons which try to bring the two levels in kinetic equilibrium with the CMB temperature ($T_{\gamma}(z_0)$), and the collisional transitions due to inelastic collisions between dark matter particles inside the halo, which try to bring the two levels in kinetic equilibrium with the temperature of the halo ($T_{h}(r)$). In this work we will study two extreme scenarios for dark matter (DM) inelastic self-interactions,
		\begin{equation}
			T_{\text{ex}}(r)=
			\begin{cases}
				T_{\gamma}\left(z_0\right) & \text{collisionless DM},\\
				T_{h}\left(r\right) & \text{collisional DM}.
			\end{cases}\label{tex}
		\end{equation}
		The general scenario would lie somewhere between these two extremes. We will indeed find that the absorption lines are sensitive to the collisional nature of dark matter.
		
		\item[$\bullet$] \textbf{Doppler broadening:} The non-trivial velocity profile of the dark matter particles along the LoS gives rise to the Doppler broadening of the absorption line around $\nu_0$ in the halo's rest frame. This broadening is characterized by the line profile,
		\begin{align}
			\phi\left(\nu_h, r, p\right) &= \frac{1}{\sqrt{\pi}\Delta\nu_D\left(r\right)}\exp\left(-\frac{\left(\nu_h-\nu_{0}\left(1+v_{\text{LoS}}(r, p)/c\right)\right)^2}{\Delta\nu_D\left(r\right)^2}\right),\nonumber \\
			\text{where} \hspace{7pt} \Delta\nu_D\left(r\right) &= \frac{\nu_{0}\left(1+v_{\text{LoS}}(r, p)/c\right)}{c}\sqrt{\frac{2k_BT_h\left(r\right)}{m_\chi}},\label{3.2o}
		\end{align}
		$\nu_h$ being the absorption frequency in the halo's rest frame, and $v_{\text{LoS}}$ being the peculiar velocity of the dark matter halo along the LoS. The effect of $v_{\text{LoS}}$ is simply to shift the frequency location of the line in observer's frame. In this work, we will not be calculating the two-point correlations but only the one-point statistics of the dark matter forest. So we will ignore the halo peculiar velocity and set $v_{\text{LoS}}=0$.
	\end{enumerate}
	In the presence of absorption, the flux density measured by the observer falls exponentially with the column density along the LoS. Conventionally, the observed absorption line is quantified by the optical depth $\tau_{\nu}$ which is defined as,
	\begin{equation}
		\tau_{\nu} = \log\left(\frac{F_\nu^0}{F_{\nu}}\right),
	\end{equation}
	where $F_{\nu}$ and $F_{\nu}^0$ are the flux densities of the source in the presence and absence of absorption respectively.
	
	For a halo intersected at an impact parameter $p$ (as shown in figure \ref{fig:halo}), the optical depth profile in the halo's rest frame is given by \cite{2002ApJ...579....1F}, 
	\begin{align}
		\tau(\nu_h, p)
		&=\int_{-d_{s}}^{\,d_{o}}\frac{g_1}{g_0}\frac{\alpha_A A_{10}^{\text{HI}}\,c^{2}}{8\pi\nu_{0}^{2}}\frac{\rho\left(r\right)}{m_{\chi}}\phi\left(\nu_h, r, p\right)\left(\frac{1-e^{-\frac{T_{*}}{T_{\text{ex}}\left(r\right)}}}{1 + \left(g_{1}/g_{0}\right)e^{-\frac{T_*}{T_{\text{ex}}\left(r\right)}}}\right)ds,\label{4.6o}
	\end{align}	
	where $r^{2} = p^{2} + s^{2}$. We set the origin $s=0$ to be the position where the impact parameter intersects the LoS. We integrate the LoS from the source to the observer. The distance between the source and the absorber is denoted by $d_s$ and the distance between the observer and the absorber is denoted by $d_o$. In the frame of the observer on Earth, the optical depth profile is obtained by mapping $\nu_h \rightarrow \nu_h/(1+z_0)$. 
	
	In the rest of the analysis we choose the following dark matter model parameters:\\
	$m_\chi$ = 1 MeV, $\nu_0$ = 156.2 GHz, and $\alpha_A = 0.35$ (see section \ref{sec:paramspace} for justification) for our study. Our main results are however quite general and we leave the full exploration of the parameter space for future work.
	
	\subsubsection{A dark line: absorption by a single dark matter halo}
	\label{sssec:halo}
	The dark matter halos are gravitationally bound structures which form the building block of the non-linear matter distribution. We want to study how the different properties of dark matter halos influence the absorption profile generated by them.
	
	A dark matter halo is characterized by its mass parameter ($M_h$), a length parameter (virial radius $r_{\text{vir}}$), a temperature ($T_h$), and a dark matter density profile $\rho(r)$. Some of these parameters are related (see Appendix \ref{app:CA} for exact expressions). We assume a Nevarro-Frenk-White (NFW) density profile \cite{1997ApJ...490..493N} for the dark matter halo. The Doppler line profile for the halo is decided by the halo temperature $T_h$ (see eq. \eqref{4.4oo} in Appendix \ref{app:CA} for definition).
	
\begin{figure}[t]
		\centering
		\begin{subfigure}[b]{7.73cm}
			\includegraphics[width=\textwidth]{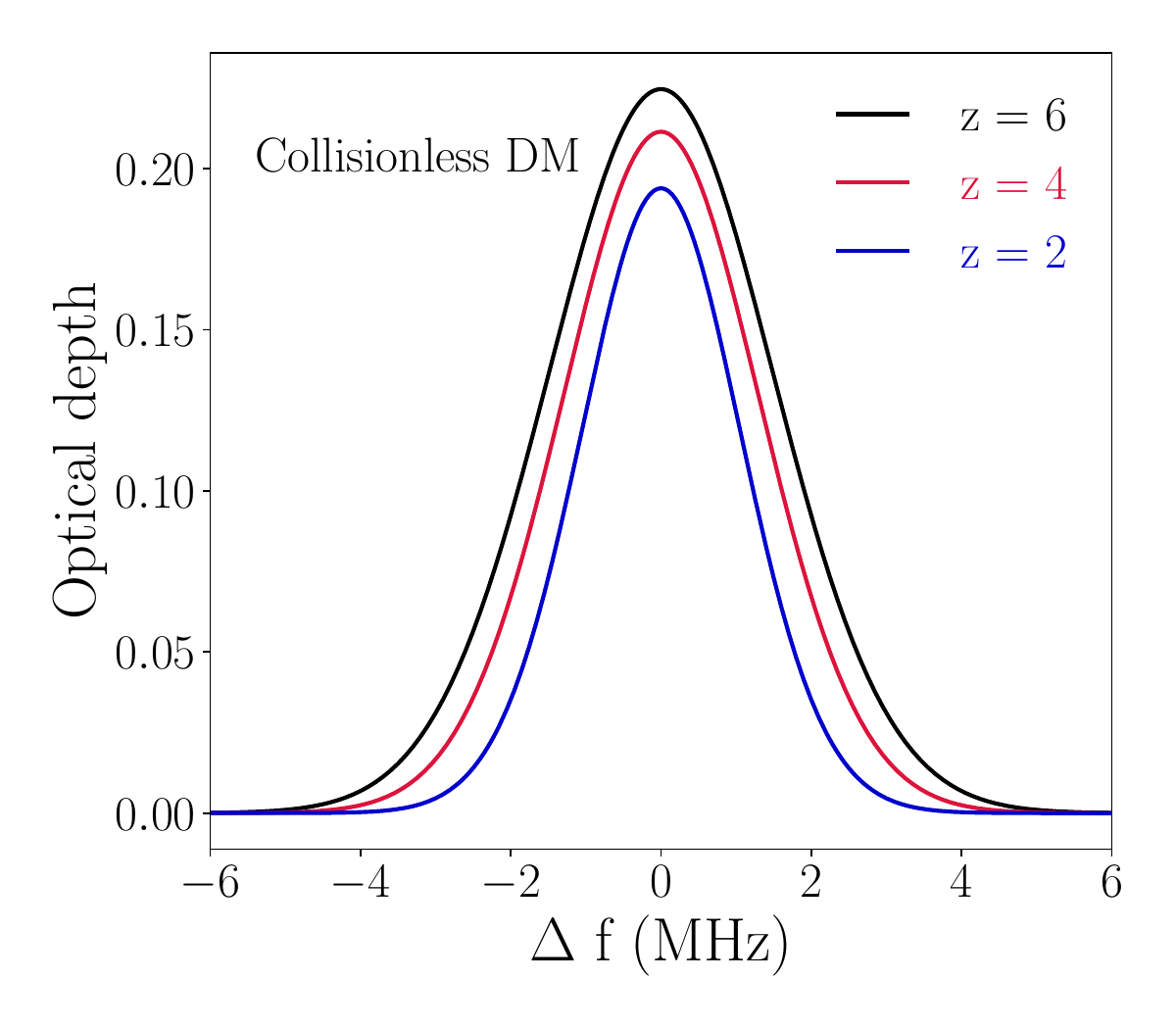}
		\end{subfigure}
		\begin{subfigure}[b]{7.73cm}
			\includegraphics[width=\textwidth]{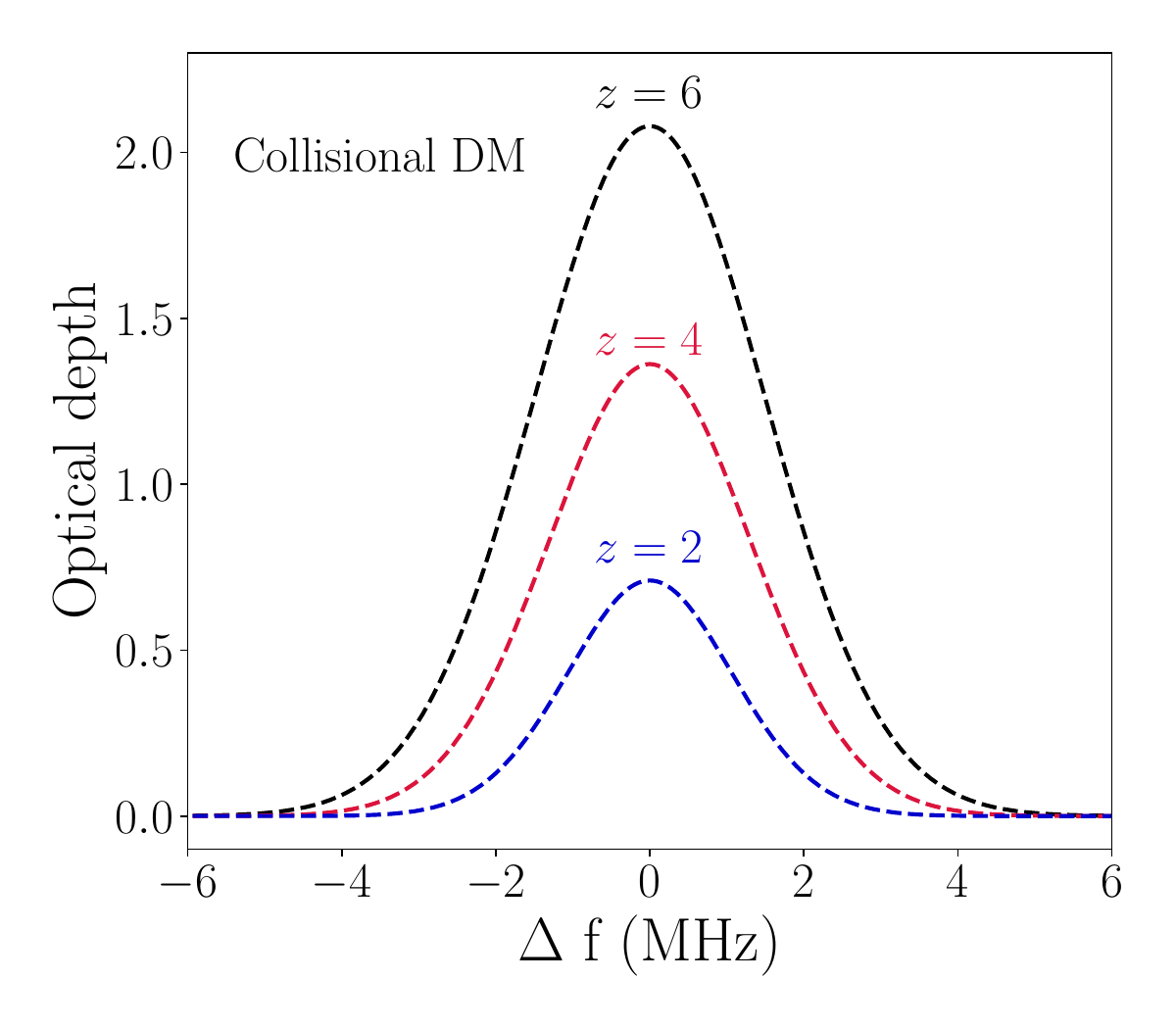}
		\end{subfigure}
		\begin{subfigure}[b]{7.73cm}
			\includegraphics[width=\textwidth]{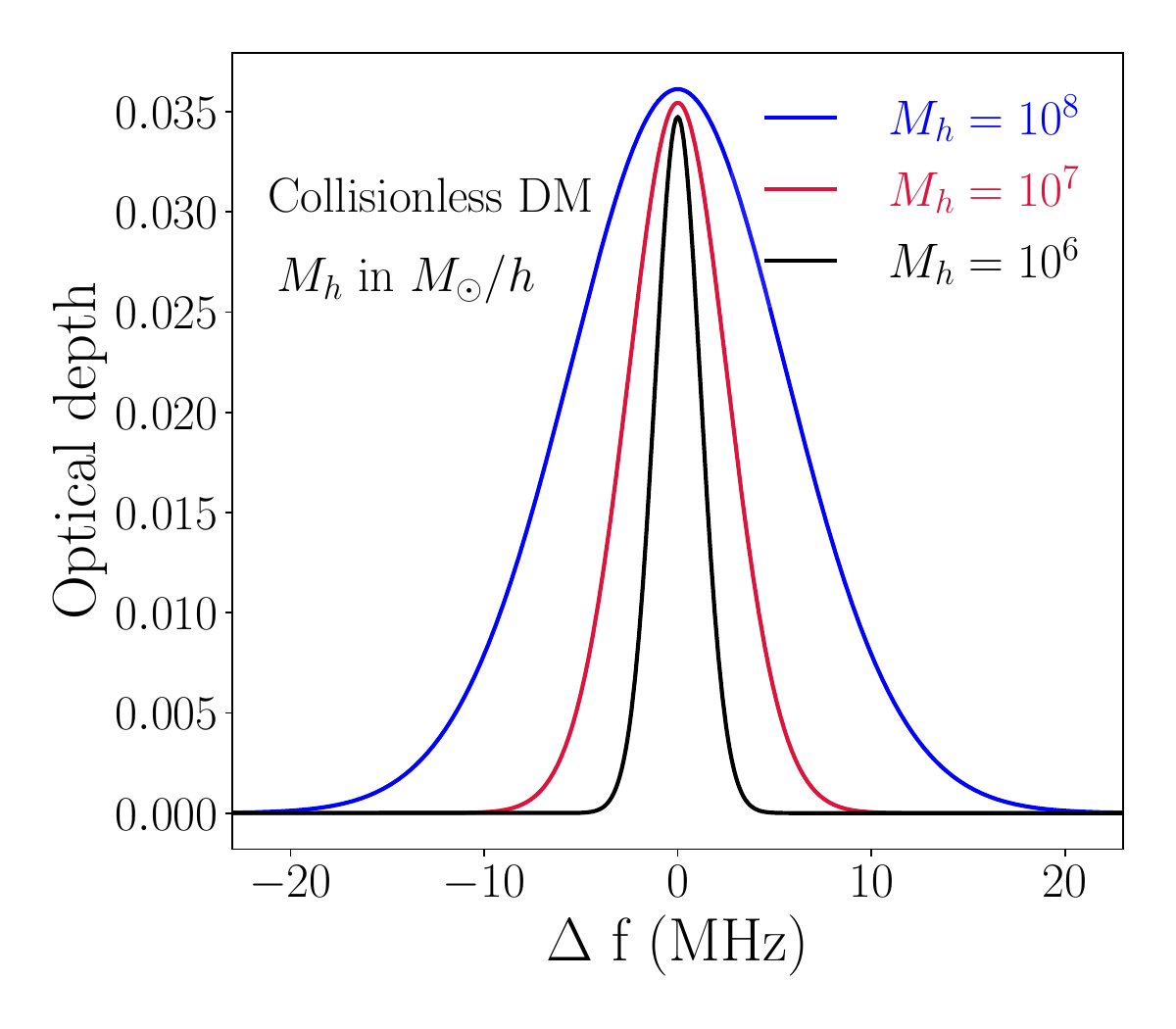}
		\end{subfigure}
		\begin{subfigure}[b]{7.73cm}
			\includegraphics[width=\textwidth]{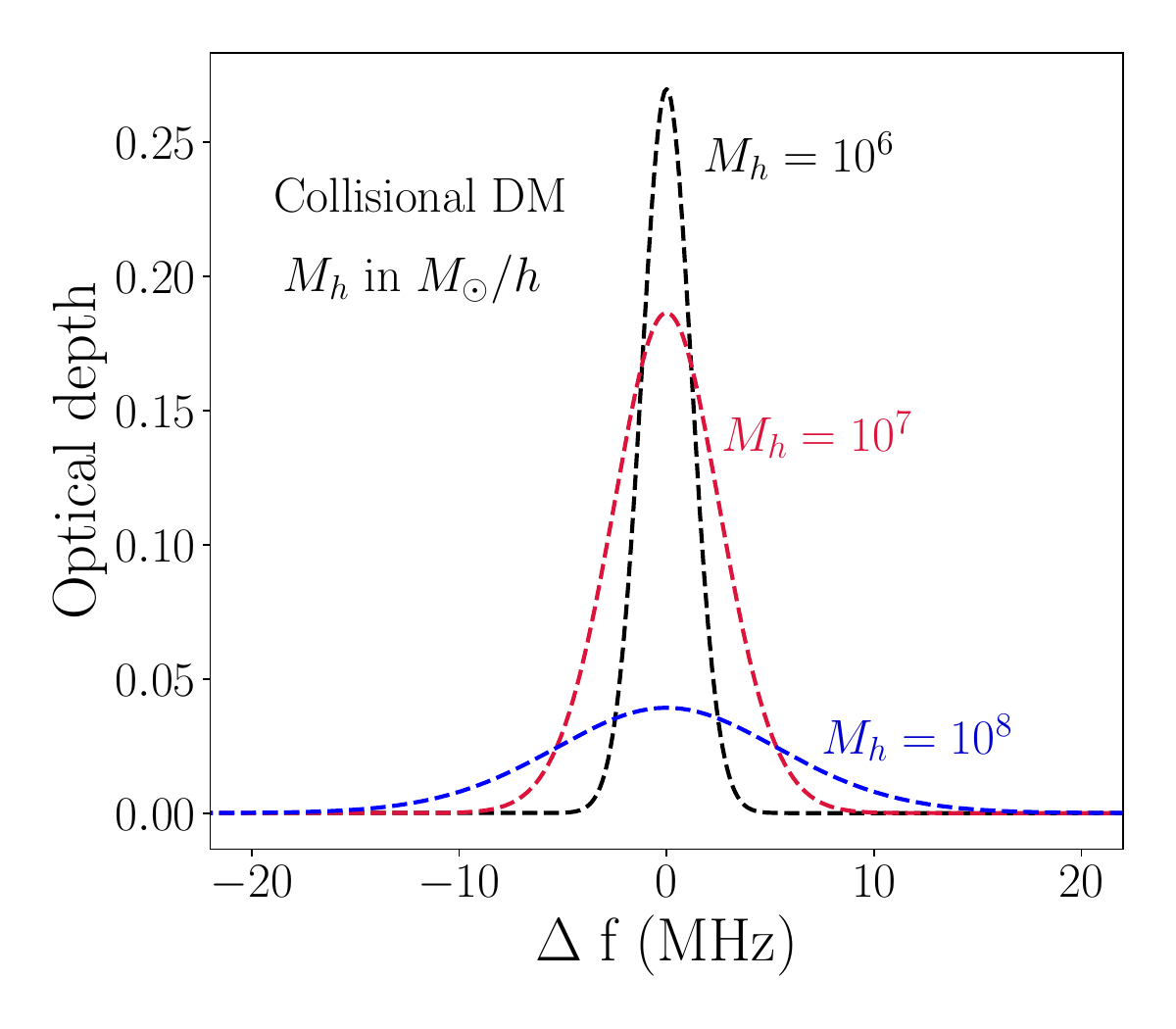}
		\end{subfigure}
		\caption{Top: The optical depth profiles for $10^6 M_{\odot}/h$ halo at impact parameters $0.1\,r_{\text{vir}}$ in the halo's rest frame. Bottom: The optical depth profiles for different halo masses intersected at $0.5\, r_{\text{vir}}$ at $z=5$ in the halo's rest frame. The solid lines refer to collisionless DM and dashed lines refer to collisional DM. }
		\label{fig:od_profile}
	\end{figure}

	We proceed towards calculating the optical depth or the absorption profile generated by a given halo mass $M_h$ using eq.\eqref{4.6o}. The optical depth depends crucially on the dark matter density profile, the impact parameter ($p$ expressed in $r_{\text{vir}}$ units), and the Doppler broadening due to random motion of the dark matter particles parameterized by the effective temperature ($T_h$) of the dark matter halo. We present our results in figure \ref{fig:od_profile}. We plot the optical depth profiles for a given halo mass ($10^6 M_{\odot}/h$) at different redshifts ($z=6$, $4$ and $2$) intersected at $p = 0.1\,r_{\text{vir}}$ in the top two panels. In the bottom two panels, we plot the optical depth profiles at a given redshift ($z=5$) for different halo masses (in $M_{\odot}/h$ units) intersected at $p = 0.5\,r_{\text{vir}}$.
	We make the following observations:
	\begin{enumerate}
		\item[(i)] There is stronger absorption in collisional DM compared to collisionless DM.
		\item[(ii)] As we go to higher redshifts, the total absorption by a halo, which is equal to the area under the optical depth profile, grows (top two panels of figure \ref{fig:od_profile}).
		\item[(iii)] The width of the absorption profile increases with halo mass and redshift.
		\item[(iv)] In collisionless DM case, the peak amplitude of the absorption profile increases with the halo mass (third panel of figure \ref{fig:od_profile}). 
		\item[(v)] In collisional DM case, the peak amplitude of the absorption profile decreases with the halo mass (fourth panel of figure \ref{fig:od_profile}). 
	\end{enumerate}
	We explain these findings using figure \ref{fig:m6_profile} where we compare the halo temperature at the virial radius $r_{\text{vir}}$ for different halo masses with the CMB temperature in the first panel. We also compare the dark matter number density profile and the halo temperature profile at different redshifts in the last two panels. Corresponding to the above observations, the explanations are as follows:
	\begin{figure}[t]
		\centering
			\includegraphics[width=0.329\textwidth]{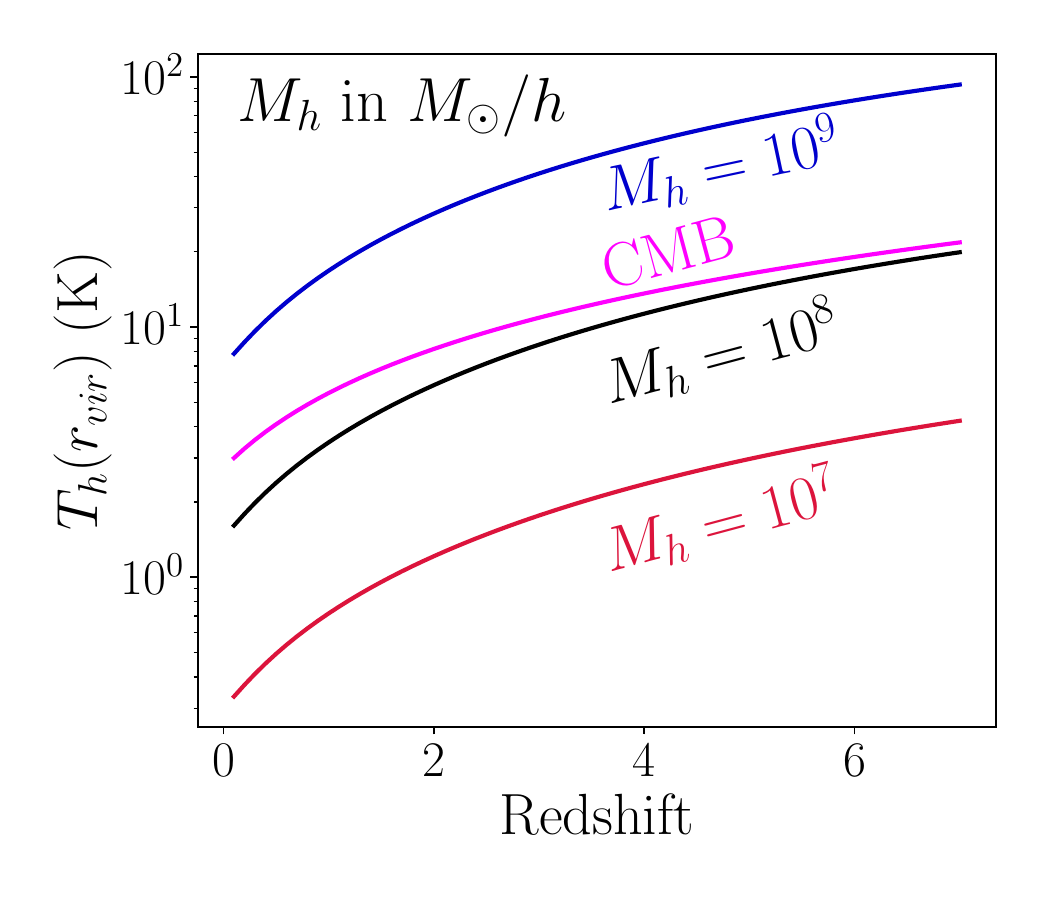}
			\includegraphics[width=0.328\textwidth]{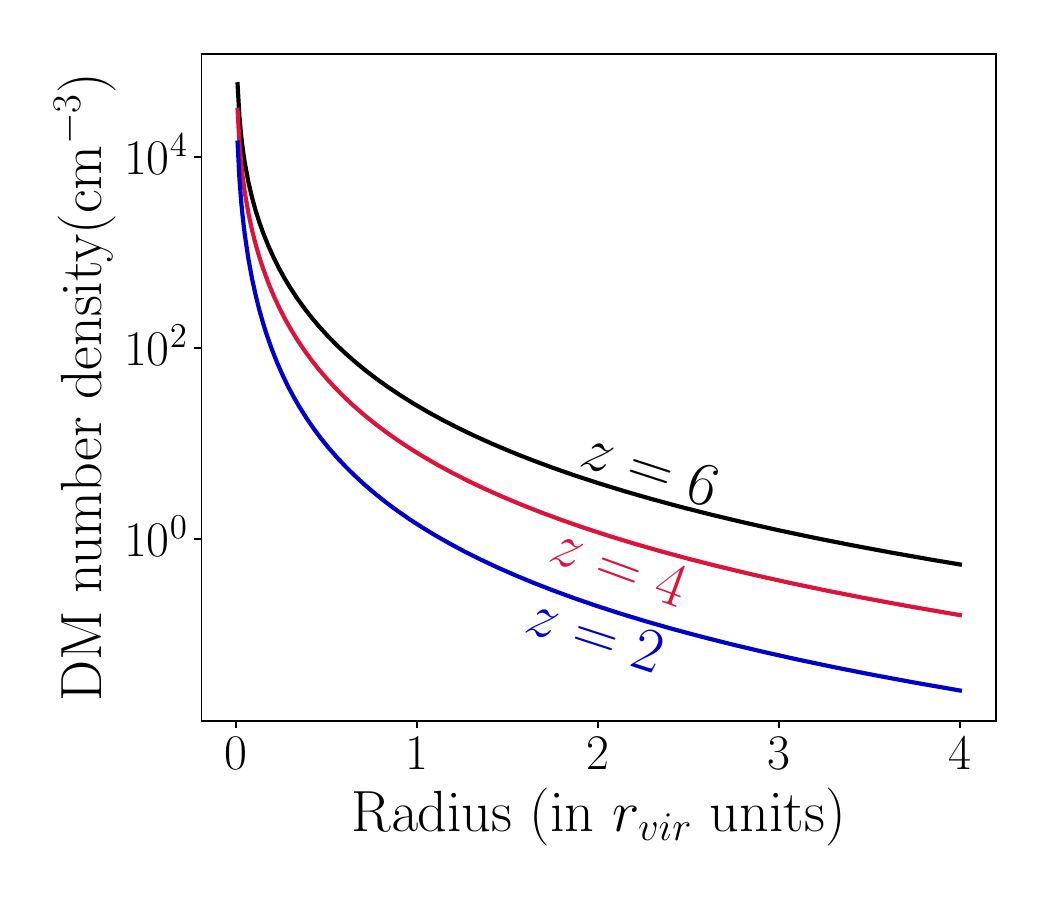}
			\includegraphics[width=0.328\textwidth]{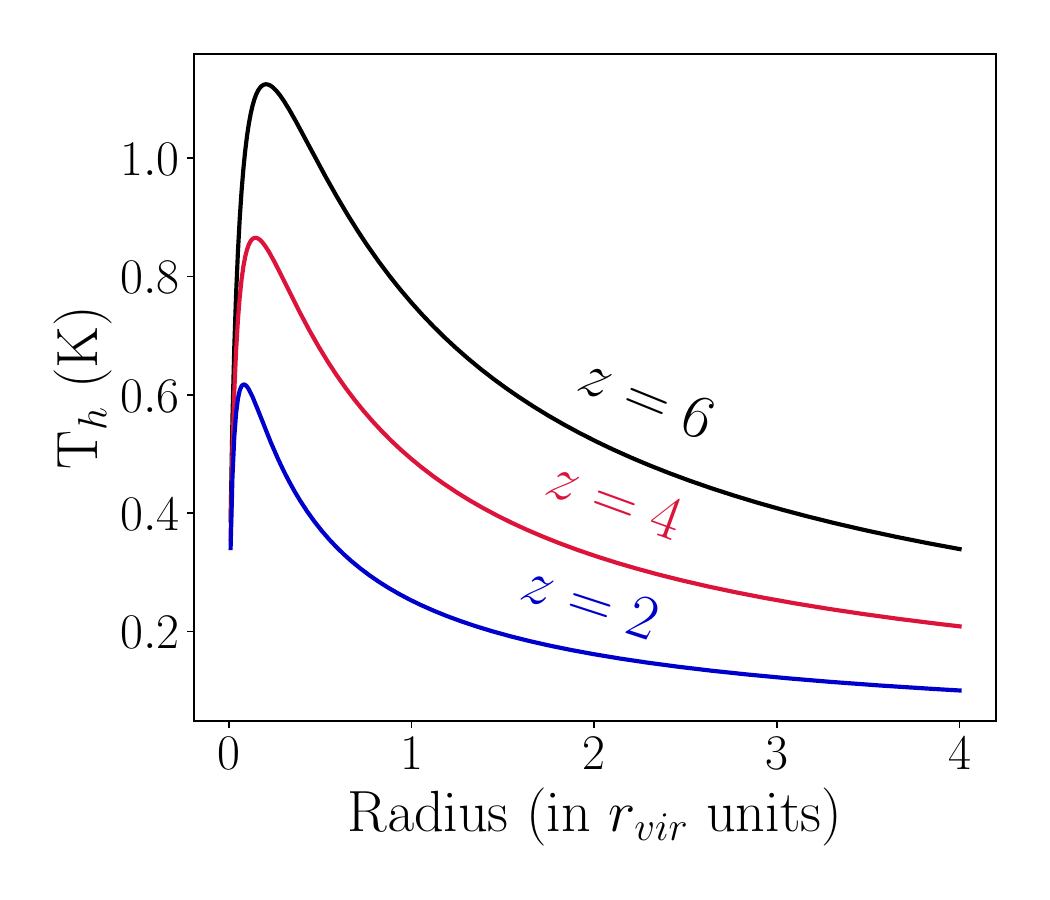}
		\caption{From left: The evolution of halo temperature at $r_{\text{vir}}$ as a function of redshift for halo masses $10^7$, $10^8$, and $10^9$ $M_{\odot}/h$, the dark matter number density profile and the halo temperature profile for $10^6\, M_{\odot}/h$ halo as a function of radius (in $r_{\text{vir}}$ units) at different redshifts $z=6, 4$ and $2$.}
		\label{fig:m6_profile}
	\end{figure}
	\begin{enumerate}
		\item[(i)] For halos of mass $\lesssim 10^8 M_{\odot}/h$, $T_h < T_{\gamma}$ (first panel of figure \ref{fig:m6_profile}). Therefore the excitation temperature (defined in eq.\eqref{tex}) for collisional DM is less than that for collisionless DM, resulting in stronger absorption in the case of collisional DM.
		\item[(ii)] As we go to higher redshifts, the number density of dark matter particles increases which results in stronger absorption (second panel of figure \ref{fig:m6_profile}).
		\item[(iii)] The width of the optical depth profile in the halo's rest frame $\propto \sqrt{T_h}$. The halo temperature, $T_h$ increases with both redshift and halo mass (third panel of figure \ref{fig:m6_profile}).
		\item[(iv)] In collisionless case, $T_{\text{ex}} = T_{\gamma}$ is independent of the halo mass. The number density of dark matter particles increases with the halo mass resulting in stronger absorption profiles  for collisionless DM.
		\item[(v)] In the collisional case, $T_{\text{ex}} = T_h$ increases with the halo mass. Even though the total dark matter number density increases with the halo mass, a higher value of $T_{\text{ex}}$ implies less dark matter particles in the ground state, which combined with broadening of the profile results in smaller peak amplitude of the  absorption profile for higher halo masses.
	\end{enumerate}
	Note that in this analysis we have assumed the dark matter density and velocity distribution profiles to be the same in both collisionless and collisional cases. The presence of dark matter self-interactions does not change the Maxwellian velocity profile as discussed in \cite{2022arXiv221006498R}. If the velocity profile deviates from the Maxwellian distribution, it would modify the line shape, in particular, the line width and the amplitude of the absorption lines. In principle, strong dark matter collisions may modify the density profile of dark matter halos which will affect the amplitude of the absorption line.
	\color{black}
	\begin{figure}[t]
		\centering
		\includegraphics[width=0.52\textwidth]{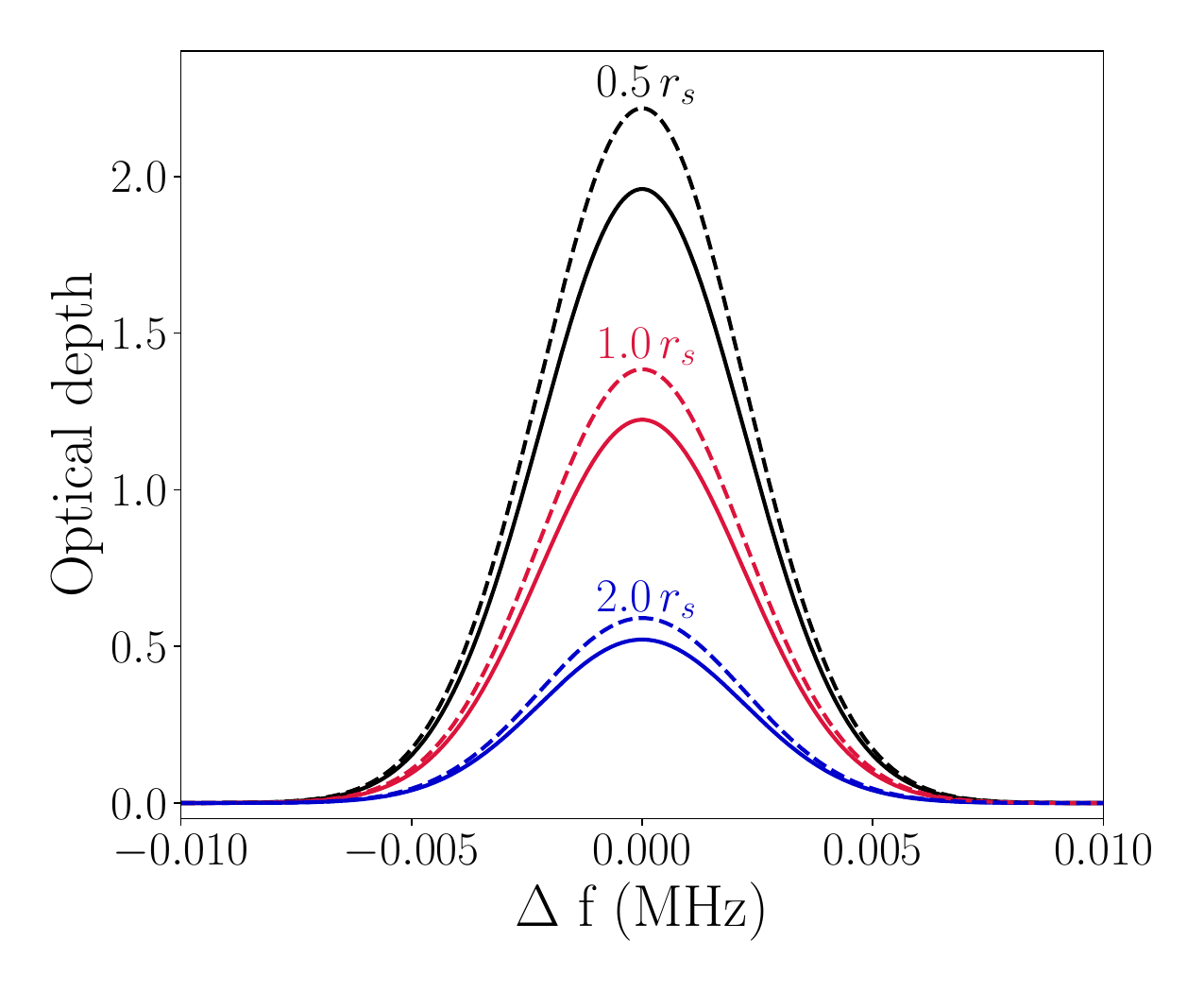}
		\caption{Optical depth profile for Leo T subhalo for impact parameters $p = 0.5\,r_s, 1.0\,r_s$ and $2.0\,r_s$ for collisionless DM represented by solid lines and collisional DM represented by dashed lines.} 
		\label{fig:od_profile_lt}
	\end{figure}
	
	As a detour, to showcase the possibility of hunting for dark matter absorption signatures in satellite galaxies of Milky Way, we take the example of absorption line generated by the dark matter subhalo that hosts the Leo-T dwarf galaxy.
	\paragraph{Leo-T dwarf galaxy:}	
	Low mass dwarf galaxies are excellent venues to study dark matter since they have low star formation activity and weak electromagnetic emission. Some of the best current constraints on emission signatures of dark matter come from the dwarf satellite galaxies of Milky Way. \cite{2007PhRvD..75b3513C, 2013MNRAS.431L..20Z,2017JCAP...07..025R,2018PhRvD..98h3024C,2021PhLB..81436075R,2021PhRvD.103l3028W}. Thus we also expect that the absorption of light from a background source by composite dark matter particles in dwarf galaxies would provide strong tests for such dark matter models. We consider one such MW satellite galaxy, namely, Leo-T.  We model its dark matter density profile using a Burkert profile from \cite{2019IAUS..344..483F}. We assume the velocity distribution of dark matter to be Maxwellian with a velocity dispersion ($\sqrt{\langle v^2\rangle}$) equal to that of hydrogen $\sim$ 6.9 km/s \cite{2008MNRAS.384..535R}. The temperature of the halo is defined by the relation, $k_BT_{h} = \frac{1}{3}m_{\chi}\langle v^{2} \rangle$. For $m_\chi=$ 1 MeV, we find $T_{h}\sim 2.1 $ K. In figure \ref{fig:od_profile_lt}, we show the absorption profiles of Leo-T intersected at impact parameters $0.5 \,r_{s}$, $1.0\, r_{s}$ and $2.0\, r_{s}$, where $r_s$ is the scale radius of the halo. We note that since $T_{h} \sim T_{\gamma}$ for Leo-T, the absorption profiles in collisional and collisionless cases are similar. However the small difference in $T_{\text{ex}}$ still gives a noticeably stronger absorption in case of collisional dark matter compared to collisionless dark matter. 
	
	\subsubsection{Dark forest: absorption by multiple dark matter halos}
	\label{sssec:df}
	If the LoS to the source passes through multiple halos (located at different redshifts $z_0$), each intersection gives rise to an absorption profile at $\nu = \nu_h/(1+z_0)$ to an observer on Earth. Collectively, a large number of absorption lines coming from the same transition in dark matter at different redshifts, and hence separated in frequency, are called \emph{forest} in spectroscopy. In this section we describe the procedure to simulate a \emph{dark forest} and discuss its qualitative and quantitative aspects.
	
	The simulation consists of discretized frequency bins in a given frequency range with the bin width adjusted such that each bin has an identical probability of net absorption. In a pseudo experiment, we then simulate the absorption line by generating a random number to select the bin where absorption occurs. We plot the observed dip in intensity in terms of the relative transmission $e^{-\tau_{\nu}}$. We summarize the algorithm for generating the dark forest spectra below \cite{2002ApJ...579....1F,2006MNRAS.370.1867F,2011MNRAS.410.2025X}:
	\begin{enumerate}[leftmargin=10pt]
		\item[$\bullet$] We begin by selecting the frequency range of simulation. For an instrument sensitive in $\nu_{\text{min}}$ to $\nu_{\text{max}}$ range, the absorption lines correspond to halos in $z_{\text{max}} = \nu_0/\nu_{\text{min}}-1$ to $z_{\text{min}} = \nu_0/\nu_{\text{max}}-1$ redshift range.
		\item[$\bullet$] We find the equiprobable bin width $\Delta \nu$ at a given $\nu$ by relating it to the probability of finding a halo in redshift bin $\Delta z$ centered at $z =  \nu_0/\nu-1$. This probability is equal to the fraction of the area on the sky covered by halos of all masses in $\Delta z$ redshift bin. Thus the probability of intersecting a halo in a frequency range $\nu$ to $\nu+\Delta\nu$ is given by,
		\begin{align}
			\Delta N_h = \Delta\nu \frac{dN_h}{d\nu} = \Delta z\frac{dN_h}{dz} &=  \Delta z\frac{c\left(1+z\right)^2}{H\left(z\right)}\int_{M_{\text{min}}}^{M_{\text{max}}}dM_h\,\frac{dn}{dM_h}(z)A\left(M_h, z\right),\nonumber\\
			\text{where}\hspace{10pt} A\left(M_h, z\right) &= \pi r_{\text{max}}\left(M_h, z\right)^2.\label{4.1}
		\end{align}
		The halo mass function $dn/dM_h$ in co-moving units is taken from \cite{2008ApJ...688..709T}, $M_{\text{min}}$ and $M_{\text{max}}$ denote the minimum and maximum halo mass at a given redshift respectively, and $r_{\text{max}}$ is the physical radius of the halo at which the dark matter number density is equal to the mean dark matter number density in the Universe.
		We choose the bin width $\Delta\nu$ at each $\nu$ such that the probability of absorption $\Delta N_h = 0.1$.
		\item[$\bullet$] We generate a random number from a uniform distribution in $[0, 1]$ in each frequency bin. The bin is selected for absorption if the random number is $\leq 0.1$.
		\item[$\bullet$] The absorption profile is characterized by the halo's redshift $z_0$, mass $M_h$, and impact parameter $p$. For the selected bin, we choose $M_h$ from the probability distribution function of the area fraction occupied by halos of mass $M_h$ at redshift $z_0$,
		\begin{equation}
			p\left(M_h, z_0\right) \propto \frac{dn}{dM_h}A\left(M_h, z_0\right). \label{3.4}
		\end{equation}
		We choose the impact parameter from a uniform distribution over the cross-sectional area of the halo $A(M_h, z)$. 
		\item[$\bullet$] We then generate the absorption profile in the halo's rest frame using eq.\eqref{4.6o} and map it to the observer's frame by transforming $\nu_h$ $\rightarrow$ $\nu_h$/(1+$z_0$). 
	\end{enumerate}
	We simulate the synthetic dark forest spectra for 100 different LoS in the redshift range 7 to 0 (see Appendix \ref{app:D} for one such sample spectra). We quantify the information in the dark forest by studying the distribution functions of the peaks ($\tau_{\text{peak}}$) and widths ($b^2$) of the absorption lines. The width is defined in terms of $b^2$ given by \cite{2015MNRAS.450.1465G},
	\begin{equation}
		b^2 = -2\left(\frac{c}{\nu_0}\right)^2\,\frac{\tau^2}{\tau\tau'' - \tau'^{2}} =  -2\left(\frac{c}{\nu_0}\right)^2\,\frac{\tau_{\text{peak}}}{\tau_{\text{peak}}''},\label{4.7o}
	\end{equation}
	where $\tau' = d\tau/d\nu$, $\tau'' = d^2\tau/d\nu^2$.   
	
\begin{figure}[t]
		\centering
		\includegraphics[width=0.45\textwidth]{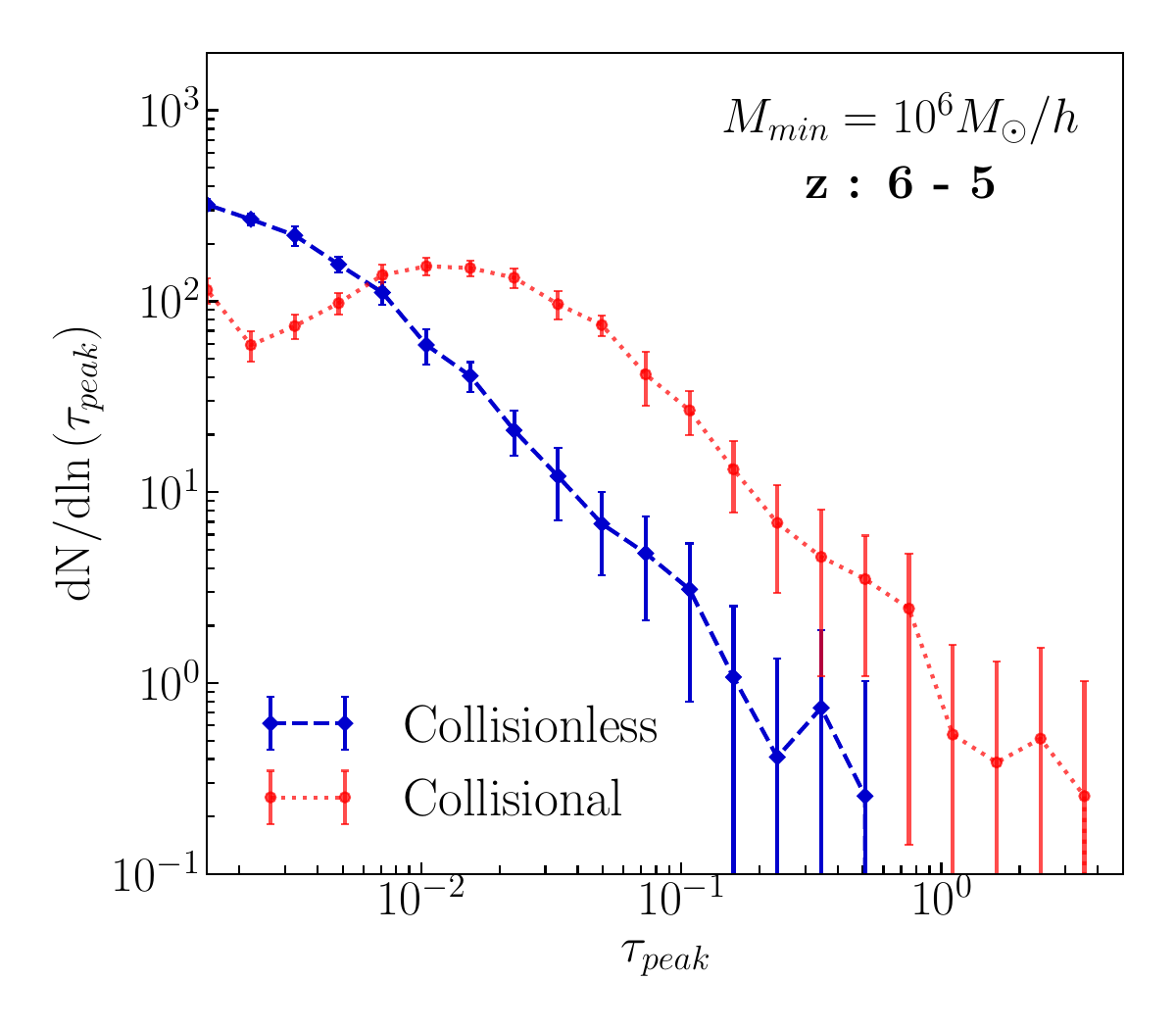}
		\includegraphics[width=0.45\textwidth]{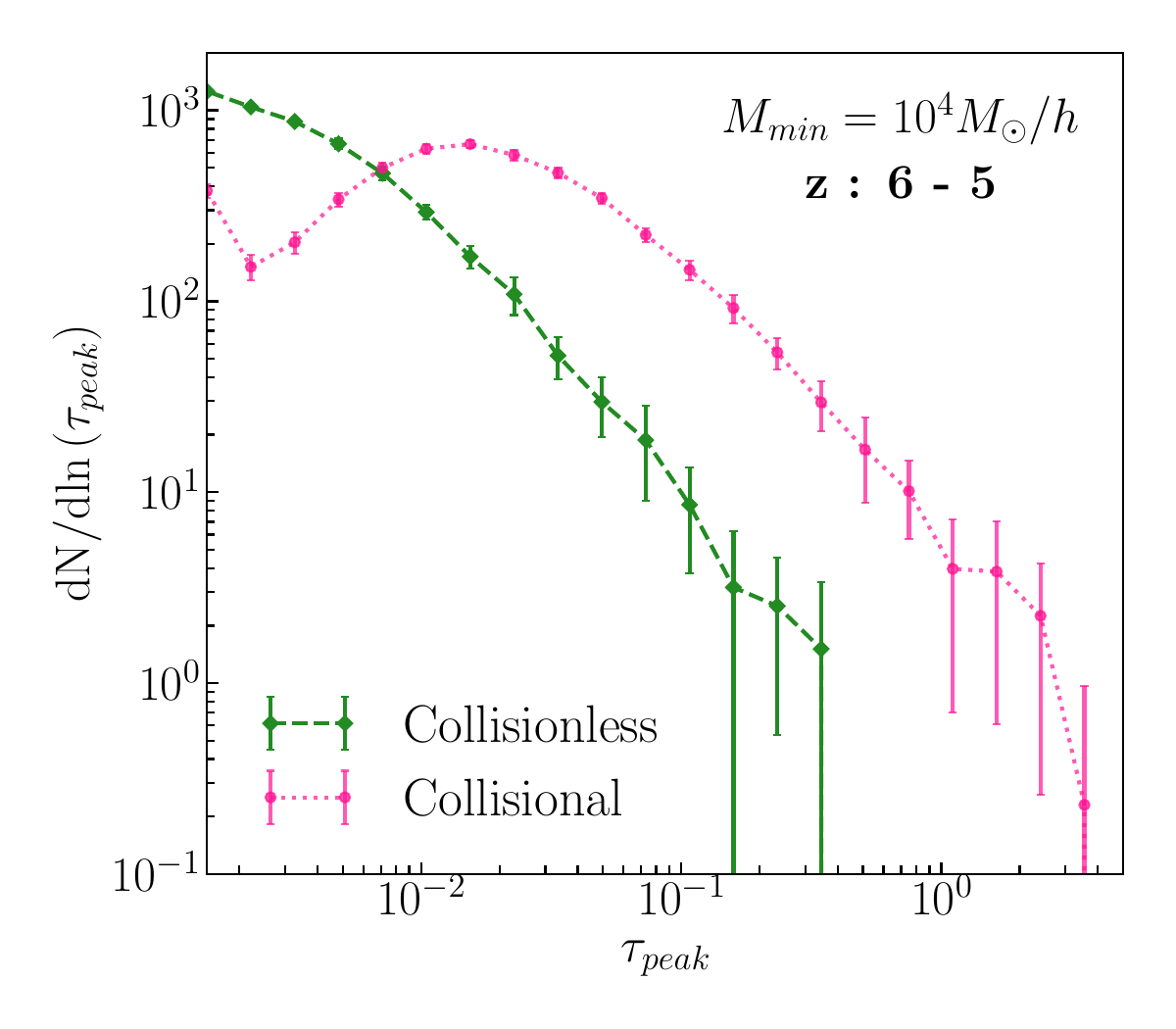}
		\includegraphics[width=0.45\textwidth]{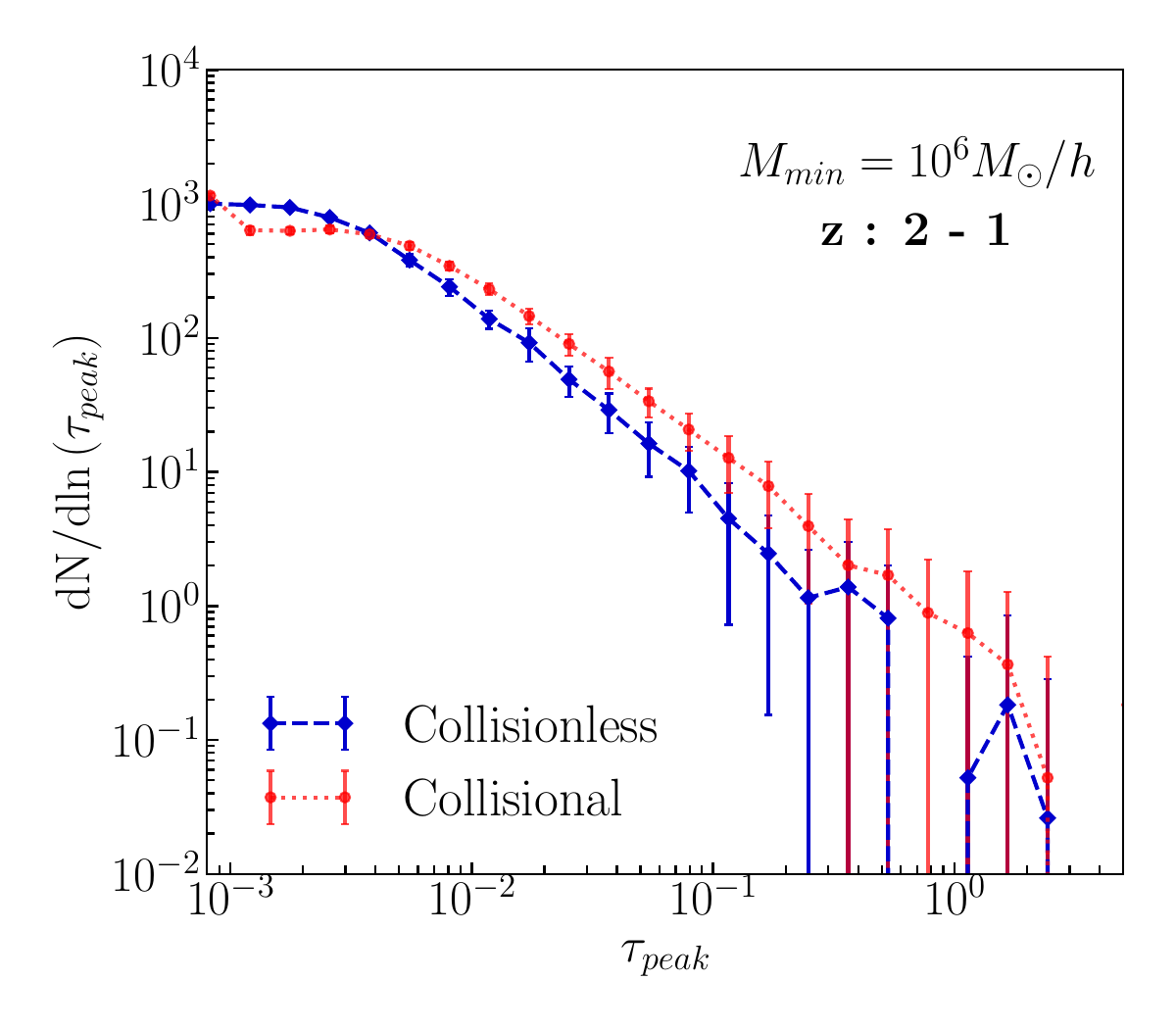}
		\includegraphics[width=0.45\textwidth]{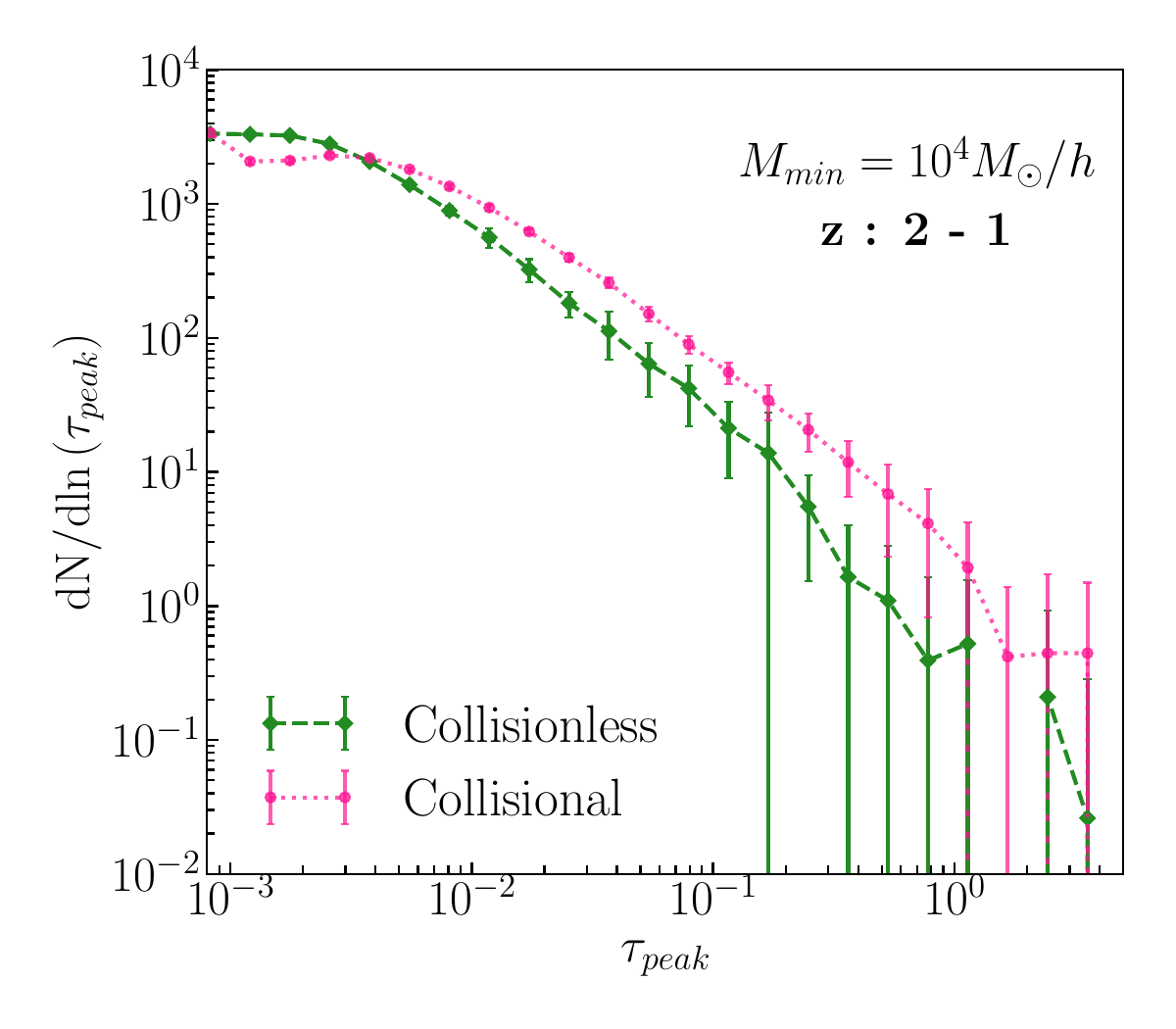}	
		\caption{Distribution function for the optical depth peaks for minimum halo mass $10^6\,M_{\odot}/h$ and $10^4\,M_{\odot}/h$ in redshift ranges 6-5 (top) and range 2-1 (bottom) respectively.}
		\label{fig:od_7621}
	\end{figure}

\begin{figure}[t]
		\centering
		\includegraphics[width=0.45\textwidth]{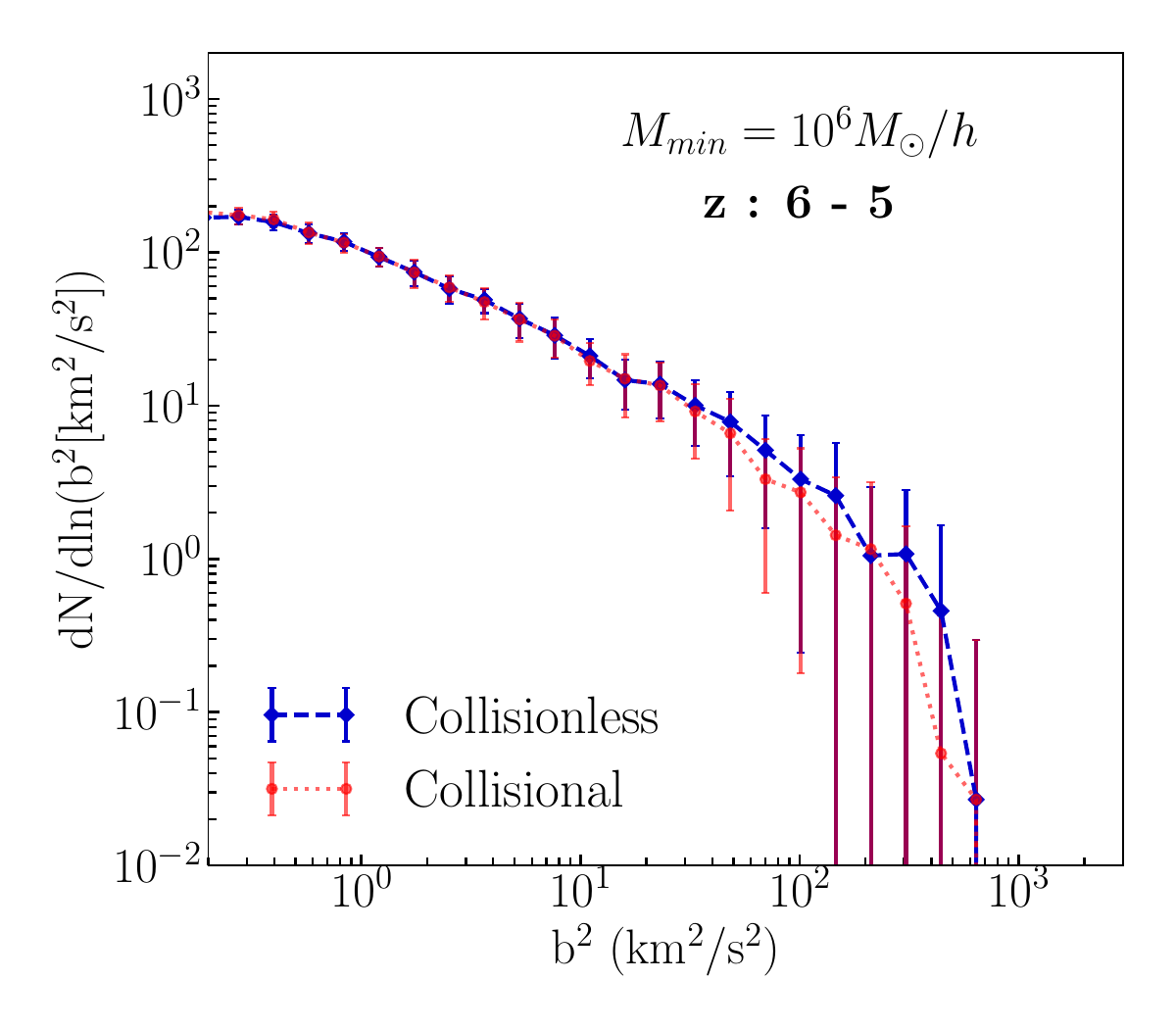}
		\includegraphics[width=0.45\textwidth]{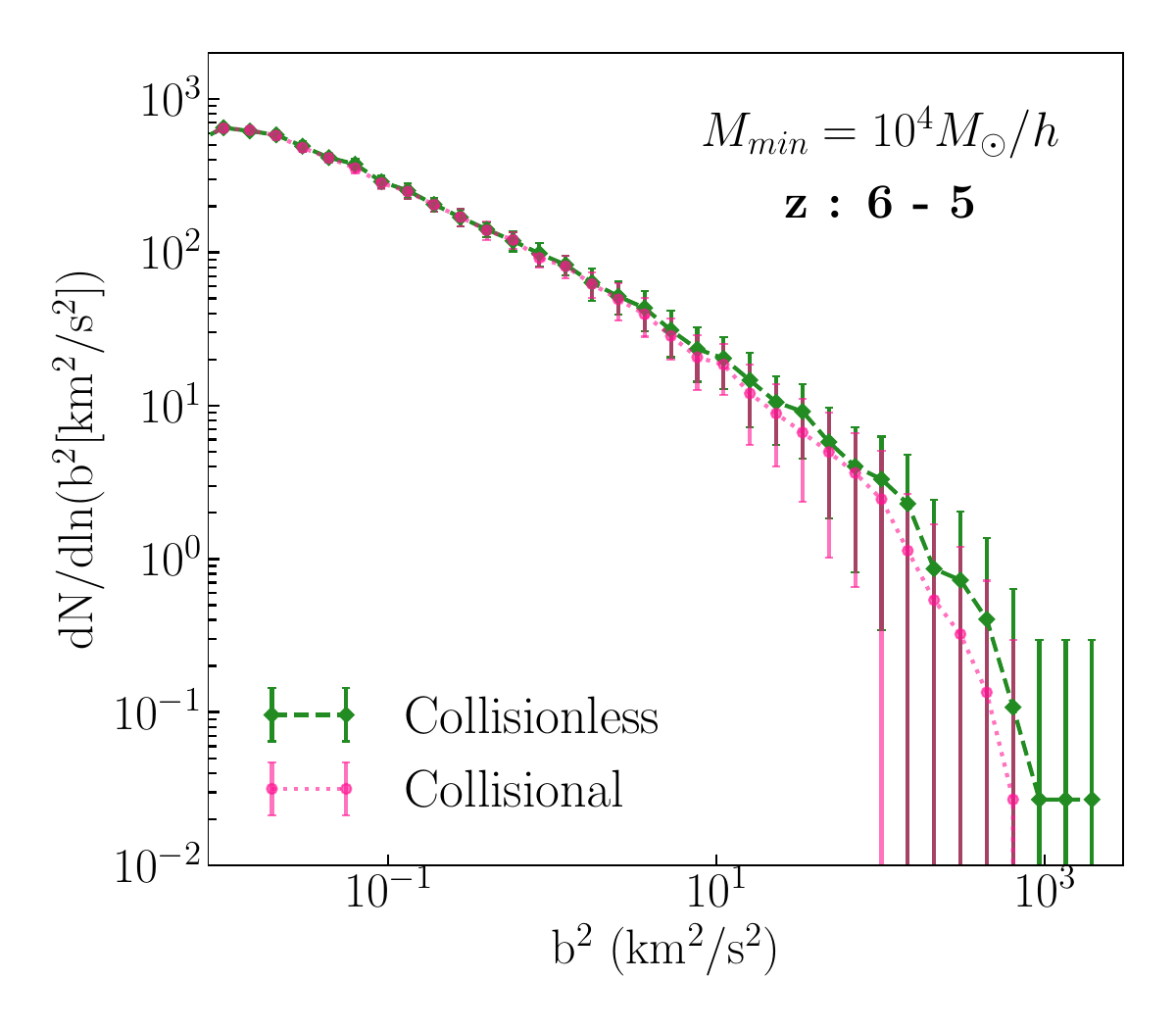}
		\includegraphics[width=0.45\textwidth]{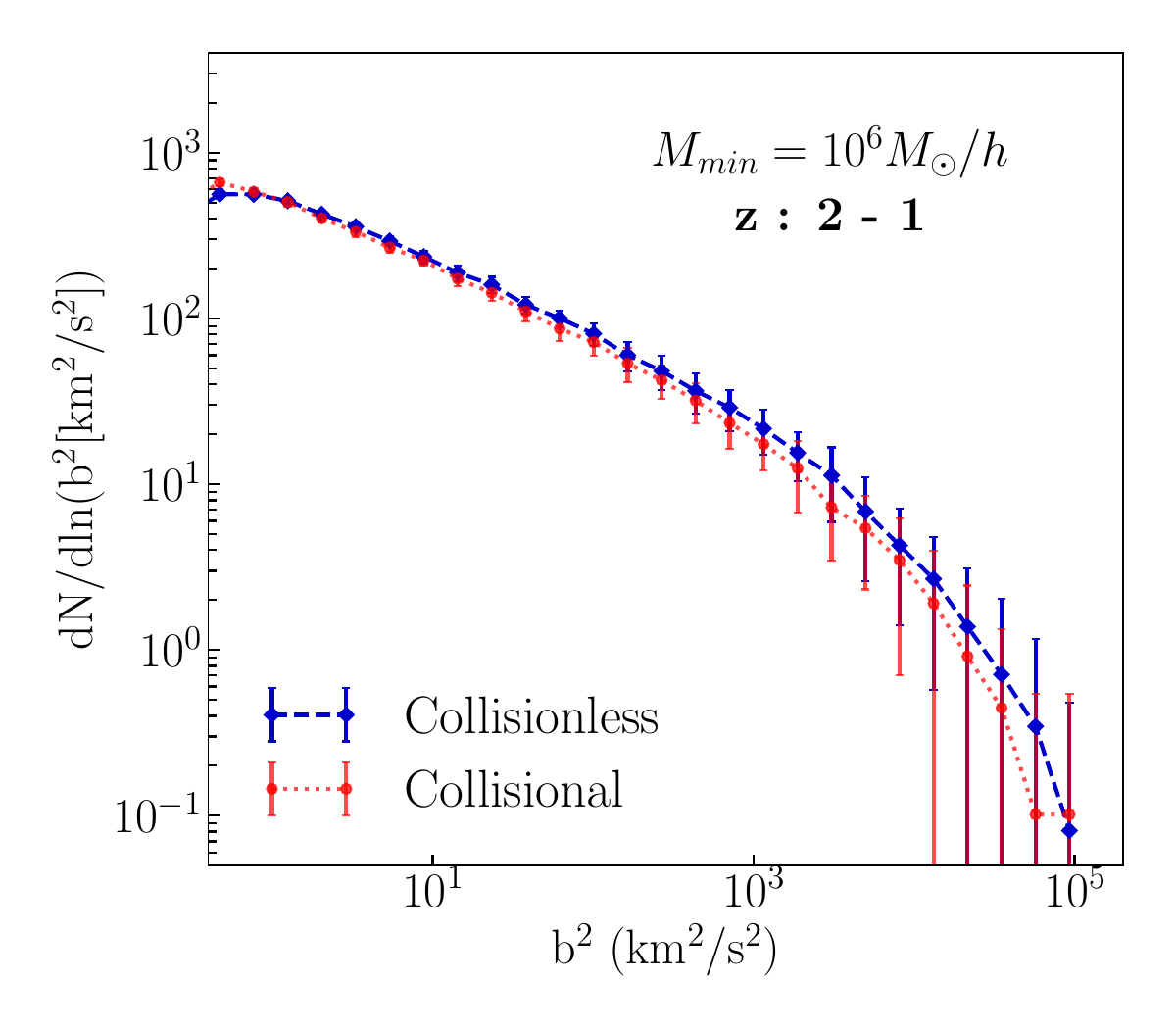}
		\includegraphics[width=0.45\textwidth]{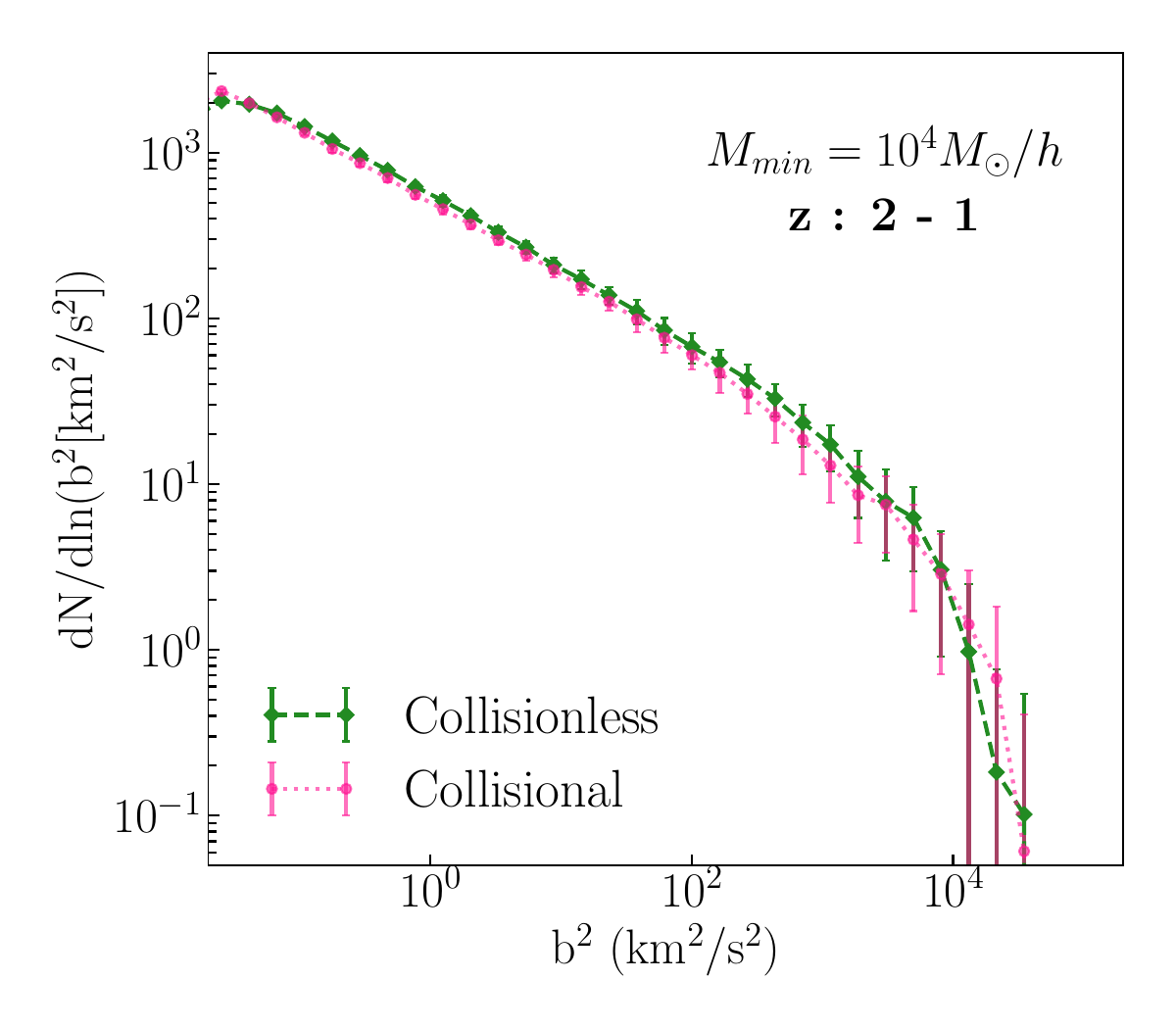}	
		\caption{Distribution function for line widths for minimum halo mass $10^6\,M_{\odot}/h$ and $10^4\,M_{\odot}/h$ in redshift ranges 6-5 (top) and 2-1 (bottom) respectively.}
		\label{fig:lw_7621}
	\end{figure}

	By combining the spectra for 100 different LoS, we calculate the mean count $N$ and standard deviation in $\sim 30$ bins to get the distribution functions for $\tau_{\text{peak}}$ and $ b^2$. Note that these distribution functions are unnormalized. The shape of the distribution functions at very low values is an artifact of the maximum impact parameter at which a halo is intersected ($\sim 4.5 r_{\text{vir}}$ in our case). So we place a cutoff on $\tau_{\text{peak}}$ and $ b^2$ at the lower end and plot the distribution functions only above this cutoff, where we are not affected by the choice of the maximum impact parameter. We present our results in figures \ref{fig:od_7621} and \ref{fig:lw_7621}.
	We observe:
	\begin{enumerate}
		\item[(i)]The distribution functions for both $\tau_{\text{peak}}$ and $b^2$ are higher for $M_{min}=10^4 M_{\odot}/h$ compared to $M_{\text{min}}=10^6 M_{\odot}/h$.
		\item[(ii)]  The tails of both $\tau_{\text{peak}}$ and $b^2$ distribution functions in both collisionless and collisional case extend to larger values at lower redshifts (redshift range 2-1 compared to redshift range 6-5).
		\item[(iii)] The $\tau_{\text{peak}}$ distribution function for collisional DM rises above the collisionless DM at large values of $\tau_{\text{peak}}$.
		\item[(iv)] In the collisionless DM case, the $\tau_{\text{peak}}$ distribution function rises monotonically at lower values of $\tau_{\text{peak}}$. For collisional DM, the curve rises at lower values of $\tau_{\text{peak}}$, reaches a peak, falls, and again rises at $\tau_{\text{peak}}\sim 10^{-3}$.
		\item[(v)] The $b^2$ distribution function for collisionless and collisional DM almost coincide.
		\item[(vi)] The tail of $b^2$ distribution function for $M_{\text{min}}=10^6 M_{\odot}/h$ extends to larger values compared to $M_{\text{min}}=10^4 M_{\odot}/h$.
	\end{enumerate}
	To explain the findings above, we make two plots in figure \ref{fig:mass_pdf}. In the first plot we compare the number of halos intersected per unit redshift for three different values of $M_{\text{min}}$ and in the second plot we compare the unnormalized probability of intersecting a halo of mass ($M_h$) at redshifts $1, 4$ and $7$. Corresponding to the above observations, the explanations are as follows:
	\begin{figure}[t]
		\centering
		\includegraphics[width=0.450\textwidth]{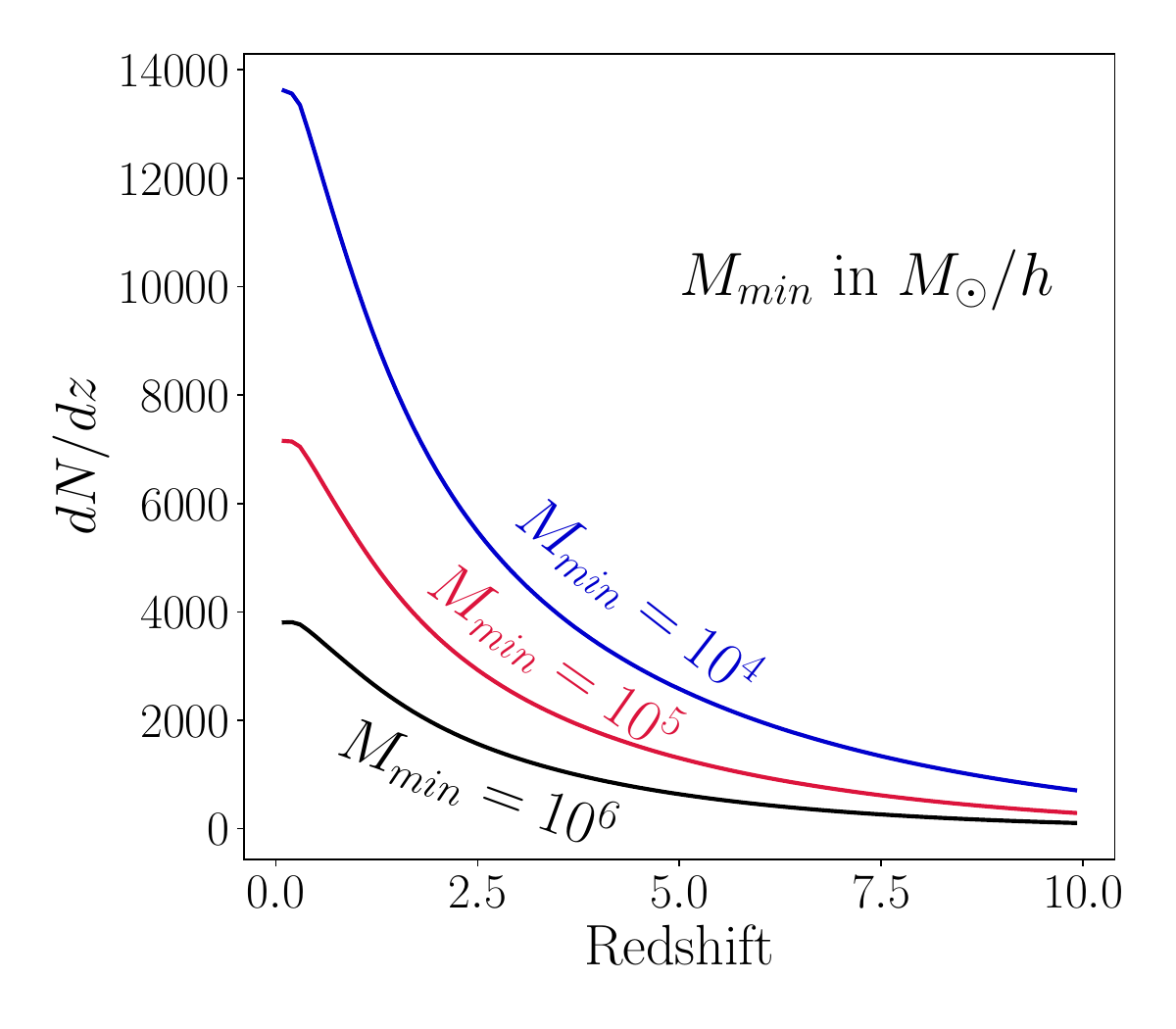}
		\includegraphics[width=0.450\textwidth]{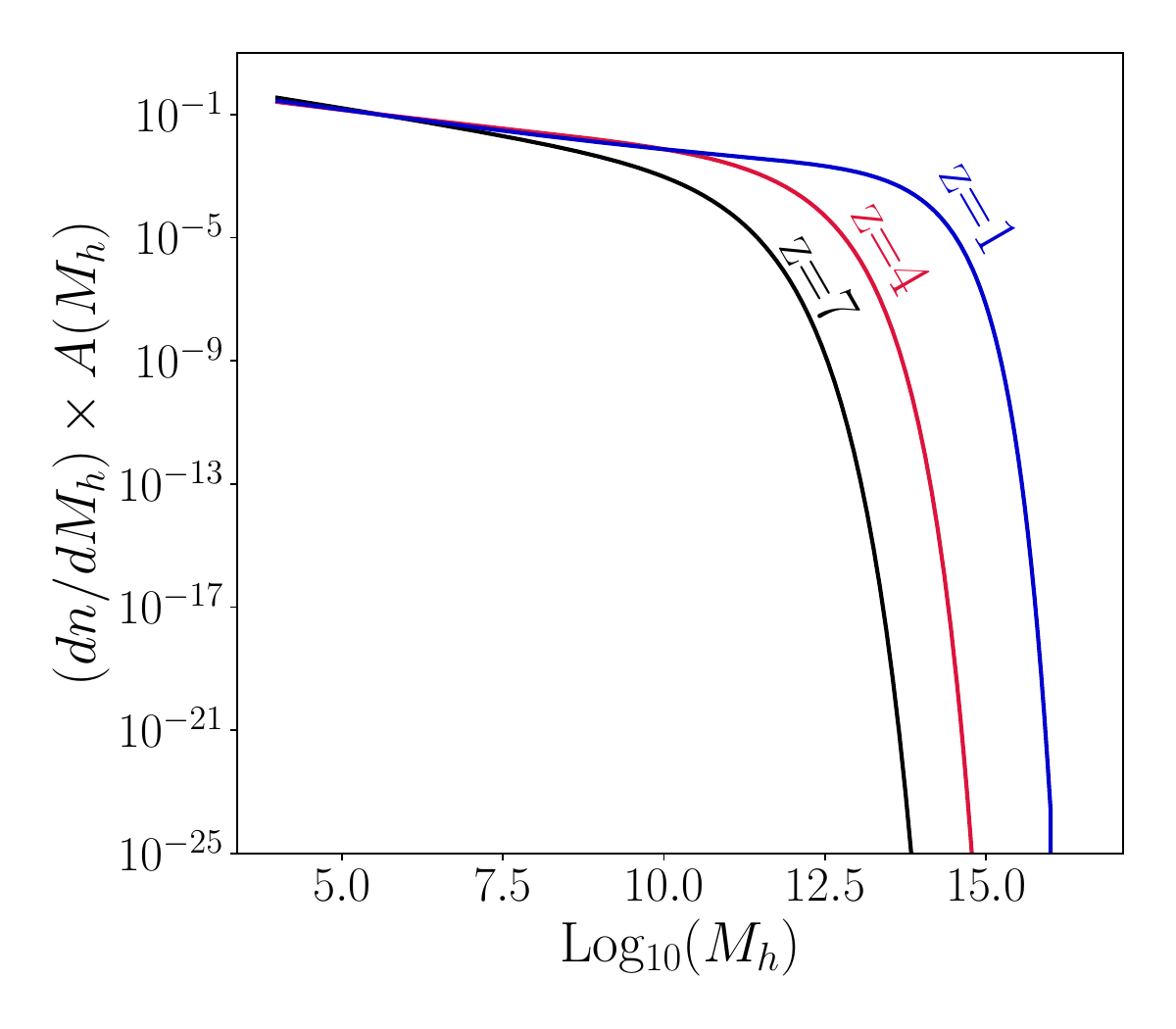}
		\caption{On the left we plot the number of halos intersected per unit redshift. On the right we plot the unnormalized probability distribution for intersecting a halo of mass $M_h$ along line of sight at redshifts $1, 4$ and $7$.}
		\label{fig:mass_pdf}
	\end{figure}
	\begin{figure}[t]
		\centering
		\includegraphics[width=0.7\textwidth]{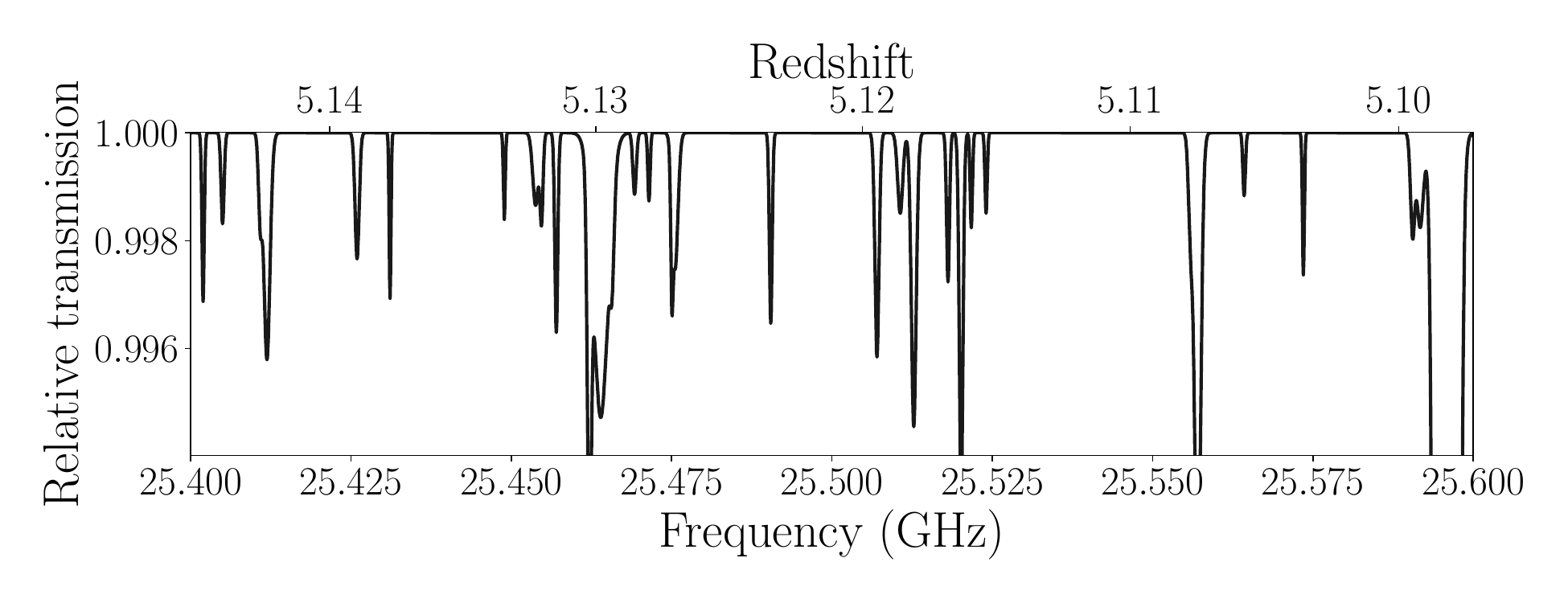}\\
		\includegraphics[width=0.7\textwidth]{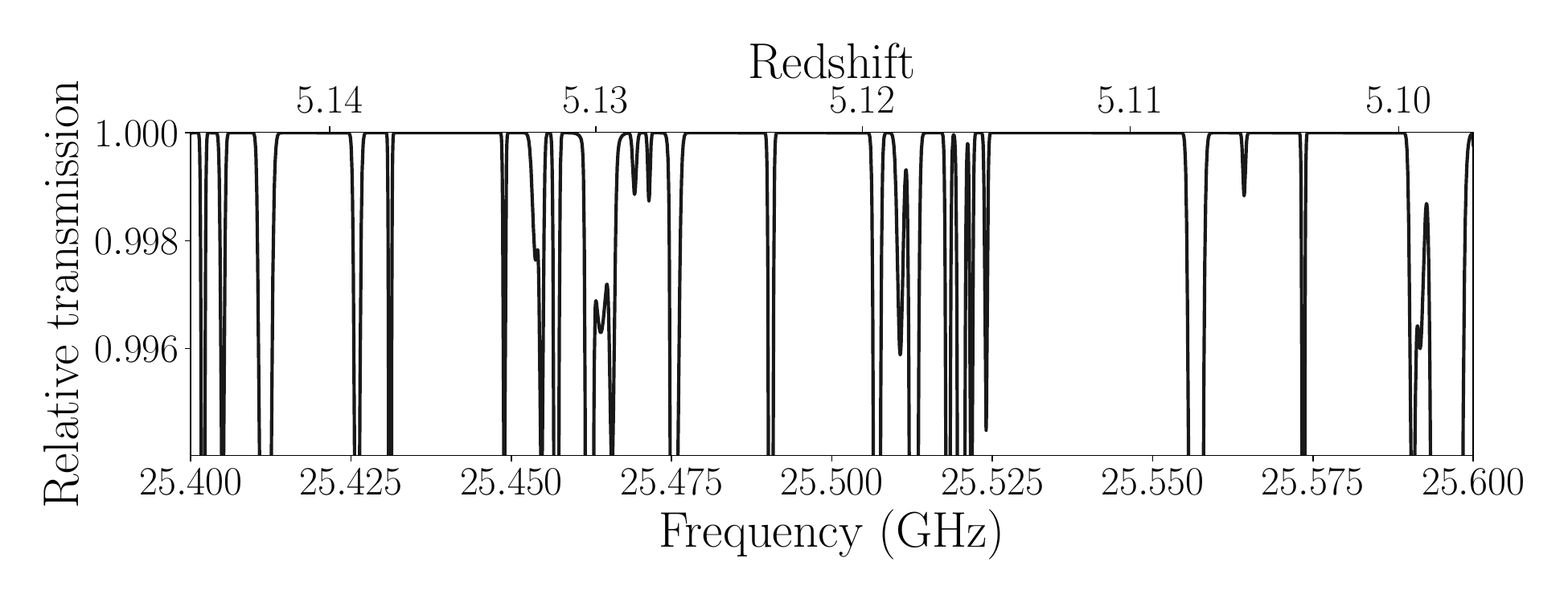}\\
		\caption{Dark forest spectrum for collisionless DM (above) and collisional DM (below) for $M_{\text{min}} = 10^6 M_{\odot}/h$. The absorption lines in the regions of overlap of absorption profiles of two halos are stronger in collisional case compared to collisionless case. Both the spectra are for identical halo intersections (i.e. same simulation), the only difference being the collisional property of dark matter.}
		\label{fig:bump}
	\end{figure}
	\begin{enumerate}
		\item[(i)] The lower limit of the halo mass function $M_{\text{min}}$ decides the contribution of the low mass end of the halo mass function to eq.\eqref{4.1}. Thus the probability of intersecting a halo is higher for $M_{\text{min}}=10^4 M_{\odot}/h$ compared to $M_{\text{min}}=10^6 M_{\odot}/h$ (first panel of figure \ref{fig:mass_pdf}).
		\item[(ii)] As the matter overdensities grow, the collapse fraction increases and the higher mass halos start contributing to the mass function at lower redshifts. In addition at lower redshifts, a given redshift interval $dz$ corresponds to a larger co-moving distance interval $d\eta = dz/H(z)$ resulting in more number of halo intersections (second panel of figure \ref{fig:mass_pdf}). This in turn gives rise to higher $\tau_{\text{peak}}$ and $b^2$ values at low redshifts as there is a greater chance of intersecting more massive halos as well as intersecting halos close to the center where dark matter density and halo temperature is high. Moreover, the line width for a halo also increases at lower redshifts due to smaller Doppler shift ($b^2 \propto 1/(1+z)^2$).
		\item[(iii)] At a given redshift, the probability of hitting low mass halos along the LoS is higher, since the halo mass function falls exponentially at larger masses (second panel of figure \ref{fig:mass_pdf}). As explained before, the absorption is stronger for collisional DM compared to collisionless DM in halos of masses $\lesssim 10^8 M_{\odot}/h$ (first panel of figure \ref{fig:m6_profile}).
		\item[(iv)] A new absorption peak is generated when the tails of two or more absorption profiles overlap. This can be seen in figure \ref{fig:bump} where we compare the dark forest spectrum for collisional and collisionless DM. Due to stronger absorption in collisional DM case, the new lines give rise to an extra feature in the $\tau_{\text{peak}}$ distribution function for collisional case compared to collisionless case at the low $\tau_{\text{peak}}$ end. 
		\item[(v)] We consider the same velocity distribution profiles for DM inside the halo for both collisionless and collisional DM. The small differences mostly arise when the absorption profiles of two or more halos overlap and give rise to new lines which can have different line widths in collisionless versus collisional case. 
		\item[(vi)] More massive halos have higher halo temperatures resulting in a larger line width ($b^2\propto T_h$).
	\end{enumerate}
	We check the convergence of the distribution functions by increasing the line of sight directions from 100 to 1000 in Appendix \ref{app:G}. 
	\subsubsection{Detectability of dark forest}
	The dark forest is a collection of absorption lines, where each line is characterized by a frequency, width and a peak amplitude. For our choice of $\nu_0 = 156.2$ GHz, the absorption lines were generated at radiowave frequencies with a typical width $\Delta\nu/\nu \approx 10^{-3}$. A different $\nu_0$ would give rise to the dark forest in a different part of the electromagnetic spectrum. The existence of a large number of spectroscopic experiments spanning different frequency ranges already make the detection of new dark absorption lines an exciting possibility.
	\begin{figure}[t]
		\centering
		\includegraphics[width=0.45\textwidth]{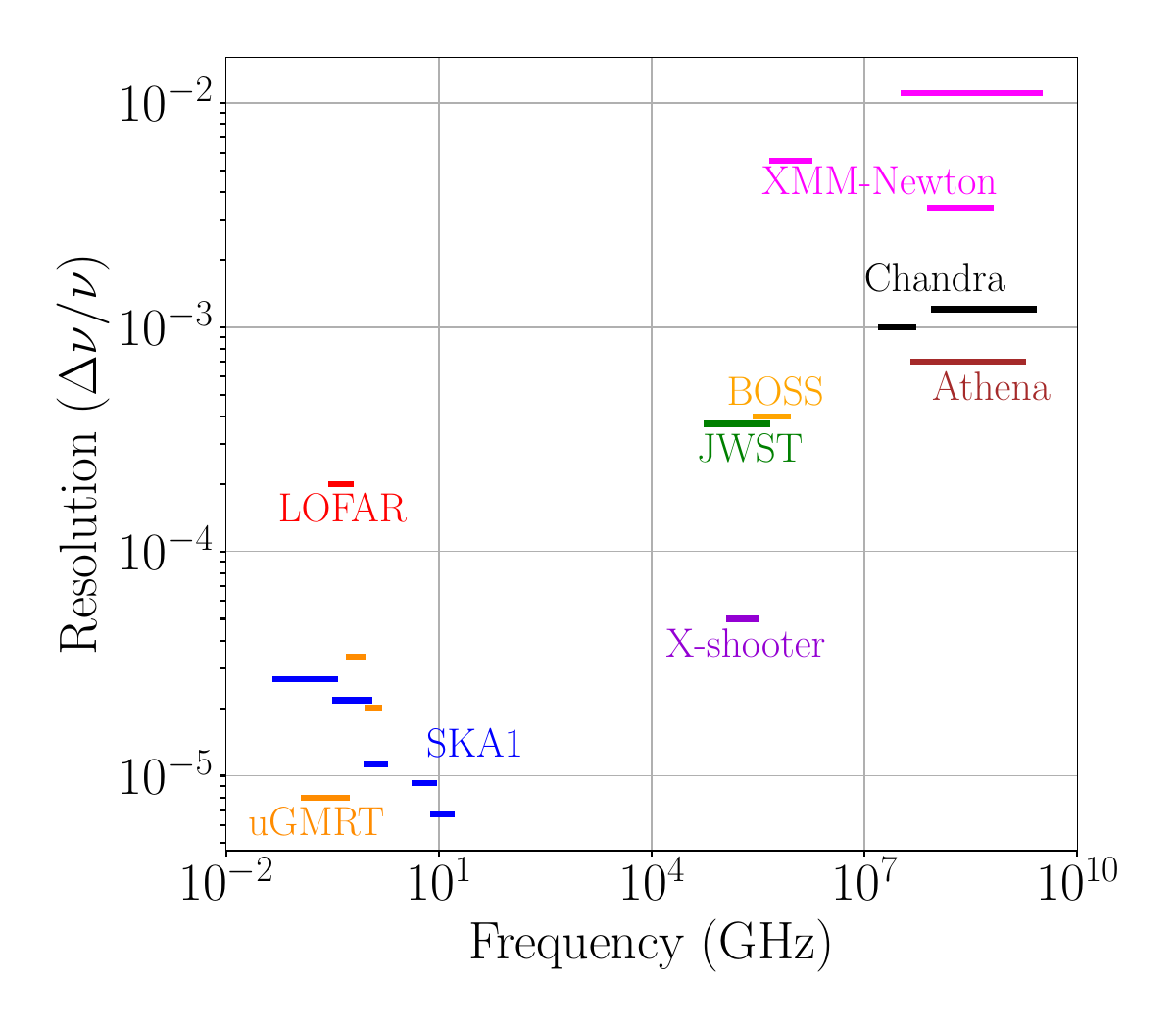}
		\includegraphics[width=0.45\textwidth]{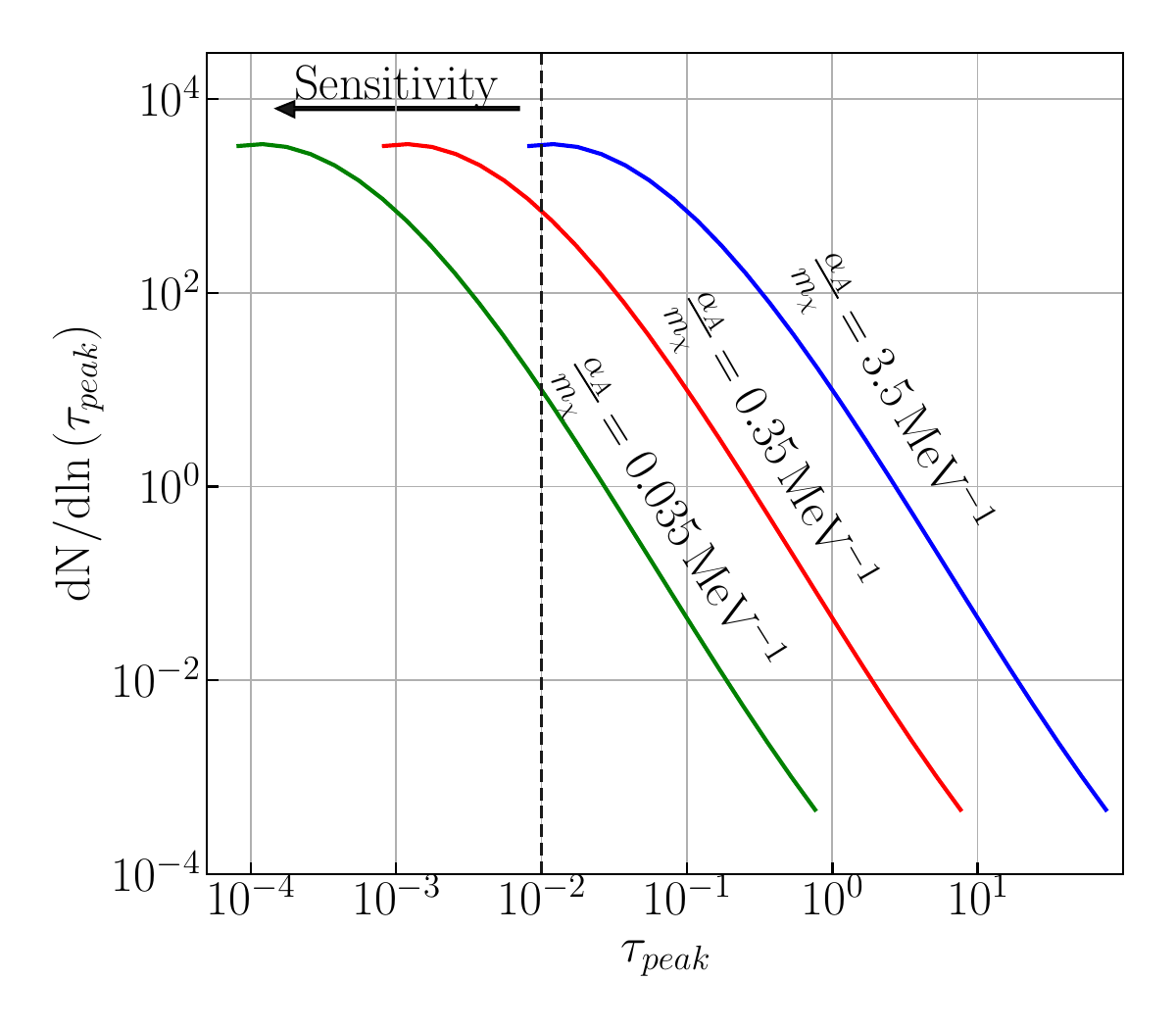}
		\caption{On the left we plot the spectral resolution of different spectroscopic experiments over the electromagnetic spectrum. On the right we show the scaling of the $\tau_{\text{peak}}$ distribution function (for collisionless DM in redshift range  2 to 1 and $M_{min} = 10^4 M_{\odot}/h$) with $\alpha_{A}/m_\chi$ (see eq.\eqref{a10} for the definition of $\alpha_{A}$). The dashed black line shows the minimum sensitivity ($\tau_{\text{peak}}>0.01$) of an experiment.}
		\label{fig:tau_min}
	\end{figure}
	
	To name a few, experiments like the Square Kilometer Array (SKA1) \cite{2020PASA...37....2W}, Low-Frequency Array (LOFAR) \cite{2019A&A...621A..56P}, and the upgraded Giant Meterwave Radio Telescope (uGMRT) \cite{1991CSci...60...95S,2017CSci..113..707G} operate at radiowave frequencies, the James Webb Space Telescope (JWST) \cite{2006SSRv..123..485G} covers the infra-red, Baryon Oscillation Spectroscopic Survey (BOSS) \cite{2002AJ....123..485S}  and X-shooter \cite{2011A&A...536A.105V} look for quasars at redshifts $\sim 2$ to $5$ at the optical frequencies, while Chandra \cite{2000SPIE.4012....2W} and X-ray Multi Mirror Mission (XMM-Newton) \cite{2001A&A...365L...1J} operate at UV and X-ray frequencies. The frequency coverage and spectral resolution for these experiments are shown in the first panel of figure \ref{fig:tau_min}. We can see that existing experiments operating in radio and optical frequencies already have the required spectral resolution ($\Delta\nu/\nu \gtrsim 10^{-3}$) to detect the new dark absorption lines in the spectrum of bright quasars/blazars. A high signal to noise ratio can be achieved by increasing the duration of observation or the integration time which would allow the detection of the weak absorption lines. Note that the peak amplitude of an absorption line scales as $\alpha_A/m_{\chi}$ as shown in eq. \eqref{4.6o}. Our choice of $\alpha_A$ is at the boundary of being disallowed for dark matter of mass 1 MeV by the CMB constraints (shown in the third panel of figure \ref{fig:c} in section \ref{sec:paramspace}). However, note that our constraints are very conservative and a more careful analysis would considerably weaken them. Also for a different value of $\nu_0$ outside the observable band of CMB, higher values of the radiative coupling $\alpha_{A}$ would be allowed. Thus, all our results and plots can be readily scaled for a different values of $\alpha_{A}$. 
	
	We show the scaling of the $\tau_{\text{peak}}$ distribution function  with $\alpha_A/m_\chi$ obtained by averaging over 100 different LoS directions for $\alpha_A/m_\chi = 0.035, 0.35$, and $3.5$ $\mathrm{MeV}^{-1}$ in the second panel of figure \ref{fig:tau_min}. We find that if an instrument can detect absorption lines with $\tau_{\text{peak}} \geq 0.01$, the full distribution function for $\alpha_A/m_\chi = 3.5\, \mathrm{MeV}^{-1}$ can be probed with the spectra of $\sim$ 100 quasars. The sensitivity decreases for weaker absorption lines and a sufficiently large quasar sample is required to probe the tail of the distribution function. For instance, in $ \alpha_A/m_\chi = 0.035 \, \mathrm{MeV}^{-1}$ case, only $\sim$ 1 in 100 quasar spectra contributes to $\tau_{\text{peak}}>0.01$ in the tail of the distribution function.
	
	\subsection{Global absorption signal in the CMB spectrum}
	\label{sec:edges}
	The absorption of photons of a particular frequency also leaves a tell-tale signature in the sky-averaged spectrum. Such features are called global signals. The much studied 21 cm global signal \cite{2004MNRAS.352..142B, 2004ApJ...608..622Z,2006PhR...433..181F,2012RPPh...75h6901P} in the CMB spectrum due to neutral hydrogen, for example, carries within important information about the growth of structure and first stars. Not surprisingly, we expect a similar global absorption in CMB in case of transition among the dark sector states.
	The underlying physics of the global absorption feature is more or less similar to the absorption along the LoS to a bright source. However, there are some crucial differences:
	\begin{enumerate}[leftmargin=10pt]
		\item[$\bullet$] In case of absorption along the LoS to a compact source, the observed signal is equal to absorption minus stimulated emission. The effect of spontaneous emission is negligible as it gets distributed along all directions in the $4\pi$ solid angle. However, spontaneous emission is important in case of CMB because it is an isotropic source and the observed signal along a given LoS gets contribution from spontaneous emission.
		\item[$\bullet$] We assume that dark matter is in kinetic equilibrium with the baryonic plasma and CMB till redshift of $z_{*} \approx 10^5$ (see eq. \eqref{cmb} in section \ref{sec:paramspace} for details). As long as dark matter is kinematically coupled to the CMB, its temperature is equal to the CMB temperature $\propto (1+z)$. Since the dark matter is non-relativistic at decoupling, it cools faster than the CMB with temperature evolving with redshift $\propto (1+z)^2$.  
		\begin{equation}
			T_{\chi}(z)=
			\begin{cases}
				T_{\gamma}(z)  & z\geq z_{*},\\
				T_{\gamma}(z_{*})\left(\frac{1+z}{1+z_{*}}\right)^{2} & z<z_{*} .
			\end{cases}\label{4.21}
		\end{equation}
		\item[$\bullet$] First consider the redshift at which dark matter starts absorbing CMB photons. Let $z_0$ be the redshift at which dark matter absorbs a photon of frequency $\nu_0$ as before. The optical depth per unit redshift is given by,
		\begin{align}
			\frac{d\tau_{\chi}}{dz}
			=-\frac{g_{1}/g_{0}}{1+\left(g_{1}/g_{0}\right)e^{-\frac{T_{*}}{T_{\text{ex}}(z_{0})}}}\left(1-e^{-\frac{T_{*}}{T_{\text{ex}}\left(z_{0}\right)}}\right)\frac{\alpha_A A_{10}^{\text{HI}}\,c^{4}\,n_\chi\left(z_{0}\right)}{8\pi \nu_{0}^{3}H\left(z\right)}\left(\frac{1+z_0}{1+z}\right)\delta\left(z-z_0\right).\label{4.17o}
		\end{align}
		The optical depth $\tau_\chi$ in eq.\eqref{4.17o} is obtained by integrating the line profile along the LoS in an expanding Universe using Sobolev approximation \cite{1957SvA.....1..678S,2004MNRAS.352..142B, 2004ApJ...608..622Z,2006PhR...433..181F,2012RPPh...75h6901P,2010MNRAS.407..599C,2011MNRAS.412..748C}. The Sobolev approximation is valid as long as the Doppler line width is negligible compared to the width of the global absorption feature.	
		\item[$\bullet$] In case of a single source we analyzed two extreme limits: collisionless DM and highly collisional DM. The effect of inelastic collisions is however essential to have a global absorption signal. The physics of dark matter inelastic collisions is described in detail in eq. \eqref{214} of section \ref{sec:model}.
		The exact functional form of the collisional coefficients depends on the details of the dark matter model. Even for a simple system of a hydrogen atom, collision cross-sections have a complicated temperature dependence \cite{2005ApJ...622.1356Z}. For simplicity, we will assume the dark matter collision cross-sections qualitatively similar to the inelastic cross-sections of hyper-fine transitions in hydrogen. We parameterize $C_{10}$ as a power law in $T_{\chi}$ at low redshifts:
		\begin{align}
			C_{10} &= n_{\chi}\langle \sigma v \rangle,\nonumber\\
			\text{where}\hspace{5pt}	\langle \sigma v \rangle &= 
			\begin{cases}
				\text{a}_1 \langle \sigma v\rangle _{\text{BC}}  \left(\frac{T_{\chi}\left(z\right)}{T_\chi\left(z_\text{sat}\right)}\right)^{\beta} & z< z_\text{sat};\\
				\text{a}_1 \langle \sigma v\rangle_{\text{BC}}  & z\geq z_{\text{sat}},\label{4.23}
			\end{cases}
		\end{align}
		where a$_1$ is used to parameterize the saturated collision cross-section in terms of the Bullet cluster bound \cite{2004ApJ...606..819M}  at redshifts $z\geq z_{\text{sat}}$,
		\begin{equation}
			\langle \sigma v\rangle_{\text{BC}} = 2 \,\text{cm}^2\text{gm}^{-1}m_\chi(\text{gm})\,v_{\text{BC}}, 
		\end{equation}
		where $v_{\text{BC}}$ is the relative velocity between two clusters in the Bullet cluster system. We take $v_{\text{BC}}=4700$ km/s \cite{2004ApJ...606..819M}. At redshifts $z< z_{\text{sat}}$, the collision cross-section is parameterized as a power law in dark matter temperature with $\beta$ as the power law index which is taken to be positive.
		\item[$\bullet$] In the presence of both radiative transitions as well as inelastic collisions, the change in the population of dark matter particles in the ground state ($n_{0}$) can be calculated from eq.\eqref{3.3o}, eq.\eqref{3.4o}, and eq.\eqref{214},
		\begin{align}
			\frac{dn_0}{dz}-\frac{3n_0}{1+z}
			&=-\frac{1}{H(z)(1+z)}\left(n_1C_{10}-n_0C_{01}+n_1A_{10}+(n_1B_{10}-n_0B_{01})\bar{J}\right),\label{4.2}
		\end{align}
		where $\bar{J}$ is the CMB intensity at $\nu_{0}$.
		\item[$\bullet$] From eq. \eqref{4.2}, we find that the level population of dark matter particles which is parameterized in terms of the excitation temperature $T_{\text{ex}}$ (see eq. \eqref{29}) is determined by the competition between the collisional rate and the radiative transition rate. By differentiating eq.\eqref{29} with respect to redshift and substituting eq.\eqref{4.2} into it, the evolution equation for excitation temperature $T_{\text{ex}}$ becomes,
		\begin{eqnarray}
			\frac{dT_{\text{ex}}}{dz}=\frac{T_{\text{ex}}^{2}}{T_{*}}\left\{\frac{1}{n_1}\frac{dn_1}{dz}-\frac{1}{n_0}\frac{dn_0}{dz}\right\}.\label{4.4}
		\end{eqnarray}
		We note that eq.\eqref{4.4} is quite general and applies to any two level system, not just the spin flip transitions. In particular, we have not made any assumptions about the smallness of $T_{*}$ with respect to other temperatures in the problem.
		\item[$\bullet$] It is customary to express the specific intensity at frequency $\nu$ i.e. $I_\nu$ in terms of the brightness temperature as,
		\begin{equation}
			T_{b} = \frac{c^2}{2\nu^2k_B}I_\nu.
		\end{equation}
		\item[$\bullet$] Prior to recombination, the collisions between the free electrons and ions create and destroy photons by the bremsstrahlung process. Bremsstrahlung plays a vital role in preserving the blackbody spectrum of CMB by erasing any distortion that may have originated in the past. Even when it becomes unimportant in maintaining the CMB blackbody spectrum over most of the frequency range at $z \leq 10^6$, it is still important in the low energy Rayleigh-Jeans tail of the CMB spectrum. The bremsstrahlung process tries to bring the brightness temperature in equilibrium with the gas/baryon temperature $T_g$, which is equal to the CMB temperature ($T_b \rightarrow T_g = T_{\gamma}$) until $z\approx 500$ \cite{1968ApJ...153....1P, 1969JETP...28..146Z}. 
		Consequently, the quantity
		\begin{equation}
			x \equiv \frac{h\nu\left(z\right)}{k_BT_{g}\left(z\right)}=\frac{h\nu_0}{k_BT_{\gamma}\left(z_0\right)}\label{3.14}
		\end{equation}
		remains invariant till $z\approx 500$. The optical depth due to bremsstrahlung ($\tau_{\text{br}}$) per unit redshift is given by \cite{1986rpa..book.....R,2011hea..book.....L},
		\begin{align}
			\frac{d\tau_{\text{\text{br}}}}{dz}(x) 
			= -\frac{c\sigma_{T}\alpha n_{e} n_{B}}{\left(24\pi^{3}\right)^{1/2}H\left(1+z\right)}g_{\text{br}}(x, z)\left(\frac{k_{B}T_{g}}{m_{e}c^{2}}\right)^{-7/2}\left(\frac{h}{m_{e}c}\right)^{3} \left(\frac{1-e^{-x}}{x^{3}}\right),\label{4.17}
		\end{align}
		where $\alpha$ is the fine structure constant, $\sigma_{T}$ is the Thomson scattering cross-section, $m_{e}$ is the mass of the electron, $n_{B}(z)$ and $n_{e}(z)$ are the number densities of baryons and electrons respectively, $g_{\text{br}}(x) \equiv Z_{i}^{2}n_{i}\langle g_{\text{ff}}(x)\rangle/n_{B}$, where $Z_{i}$ is the charge of the  $i^{\text{th}}$ ion having number density $n_{i}$, and $\langle g_{\text{ff}}(x)\rangle$ is the thermally averaged Gaunt factor which has been taken from \cite{2014MNRAS.444..420V}.
		\item[$\bullet$] At high redshifts, the number density of dark matter particles is high which results in stronger collisional transitions between the two dark matter states, compared to radiative transitions due to CMB photons. Thus, initially $T_{\text{ex}}$ is in kinetic equilibrium with the dark matter temperature which is much lower than the CMB temperature ($T_{\text{ex}} \rightarrow T_{\chi}\ll T_{\gamma}$) at $z<z_{\text{dec}}$. The dark matter particles absorb the CMB photons and a net flow of energy takes place from CMB to dark matter resulting in an absorption feature in the CMB spectrum. The absorption of CMB photons by dark matter at redshift $z_{0}$ generates an absorption line at $x$ (defined in eq.\eqref{3.14}).  If this line lies in the low frequency Rayleigh Jeans tail ($x\ll1$) of the CMB spectrum, it gets erased by the bremsstrahlung emission at subsequent times ($z<z_0$). As the number density and the temperature of dark matter falls due to expansion of the Universe, the collision rate falls and the radiative transitions involving the CMB photons begin to dominate over the collisions. This brings $T_{\text{ex}}$ in kinetic equilibrium with the CMB temperature ($T_{\text{ex}} \rightarrow T_{\gamma}$). When this happens there is no net emission or absorption of the CMB photons by dark matter and the absorption signal vanishes.  Thus we expect to see a broad absorption feature in the CMB spectrum, starting from the time dark matter decouples until a later time when the radiative transitions take over. 
		\item[$\bullet$]The evolution of brightness temperature at $\nu(z) = \nu_0(1+z)/(1+z_0)$ incorporating the effect of absorption by dark matter at $z=z_0$ (from eq.\eqref{4.17o}) and bremsstrahlung (from eq.\eqref{4.17}) is given by,
		\begin{align}
			\frac{d T_{b}(\nu)}{d z} - \frac{T_{b}(\nu)}{1+z} = \frac{d\tau_{\chi}}{dz}\left(-T_{b}(\nu)+\frac{h\nu}{k_B}\frac{1}{(e^{h\nu/\left(k_BT_{\text{ex}}\left(z\right)\right)}-1)}\right) \nonumber\\ +\frac{d\tau_{\text{br}}(x)}{dz}\left(-T_{b}(\nu)+\frac{h\nu}{k_B}\frac{1}{(e^{h\nu/\left(k_BT_{g}\left(z\right)\right)}-1)}\right).\label{4.12}
		\end{align}
		
	The differential brightness temperature $\delta T_{b}$ observed at frequency $\nu_{\mathrm{obs}} = \nu_{0}/(1+z_{0})$ is defined as the brightness temperature (obtained by solving eq. \eqref{4.12}) minus the brightness temperature of CMB today i.e. in the observer's frame,
	\begin{align}
			\delta T_{b}\left(\nu_{\mathrm{obs}}, z=0\right)_{\text{observer's frame}}
			&\equiv T_{b}\left(\nu_{\mathrm{obs}}, z=0\right) -\frac{h\nu_{\mathrm{obs}}}{k_B}\frac{1}{e^{h\nu_{\mathrm{obs}}/\left(k_BT_\gamma(z=0)\right)}-1}.\label{4.16}
		\end{align}	
	\end{enumerate}
	
	\subsubsection{The EDGES anomaly}
	The EDGES collaboration \cite{2018Natur.555...67B} reported a strong absorption feature in the CMB spectrum centered around 78 MHz, having an amplitude of $0.5^{+0.5}_{-0.2}$ K, and a full-width half maximum (FWHM) of $19^{+4}_{-2}$ MHz (where the bounds provide the 99\% confidence intervals). This feature is almost twice in amplitude compared to the maximum possible signal expected from the 21 cm transitions in neutral hydrogen during the cosmic dawn. In this work, we propose that this anomalous signal is caused by the absorption of CMB photons by composite dark matter. In particular, we show that we can get an absorption feature which has the frequency position of the dip at $78^{+10}_{-10}$ MHz and the maximum amplitude and FWHM within 99\% confidence intervals of the EDGES best fit signal. We note that the EDGES collaboration \cite{2018Natur.555...67B} does not have an independent measure of noise and error bars on their data points. They use the reduction in residuals after fitting the foreground + 21 cm signal as an evidence for the presence of a flattened Gaussian absorption profile in the data. We follow a similar procedure as them \cite{2018Natur.555...67B}. We first randomly sample the dark matter parameters over a considerably wide range and calculate the expected dark matter signal for the selected point in the parameter space. We take into account the Bullet cluster constraint \cite{2004ApJ...606..819M} on the dark matter collisional cross-section parameterized in eq. \eqref{4.23} by placing an upper cut-off on the collision parameter $a_1\leq 1$. 
The sampled parameter space of the dark matter model is given in table \ref{tab:edges}.
	\begin{table}[t]
	\centering
		\begin{tabular}{|l|l|l|}
			\hline
			\textbf{Parameter}                                            & \textbf{Sampling Range}                                        \\ \hline                                                                             
			Dark matter mass $\left(m_{\chi}\left(\text{MeV}\right)\right)$                                                & $10^{-1}-10^3$                                                                           \\
			Transition temperature $\left(\mathrm{T}_{*}\left(\text{K}\right)\right)$                  & $4.0-10.0$                                                                                 \\
			Radiative coupling $\left(\alpha_{\mathrm{A}}\right)$                                & $10^{-2}-10^2$      \\
			Collisional parameter $ \left(\text{a}_1 \right)$  & $10^{-5}-1$    \\
			Saturation redshift $ \left(z_{\text{sat}}\right)$  &  $10^3-10^4$     \\ 
			Power law index $ \left(\beta\right)$  &  $0-5$     \\\hline
		\end{tabular}
		\caption{The sampled parameter space of the dark matter model.}
		\label{tab:edges}
	\end{table}
	\begin{figure}[t]
	\centering
	\includegraphics[width=1.0\textwidth]{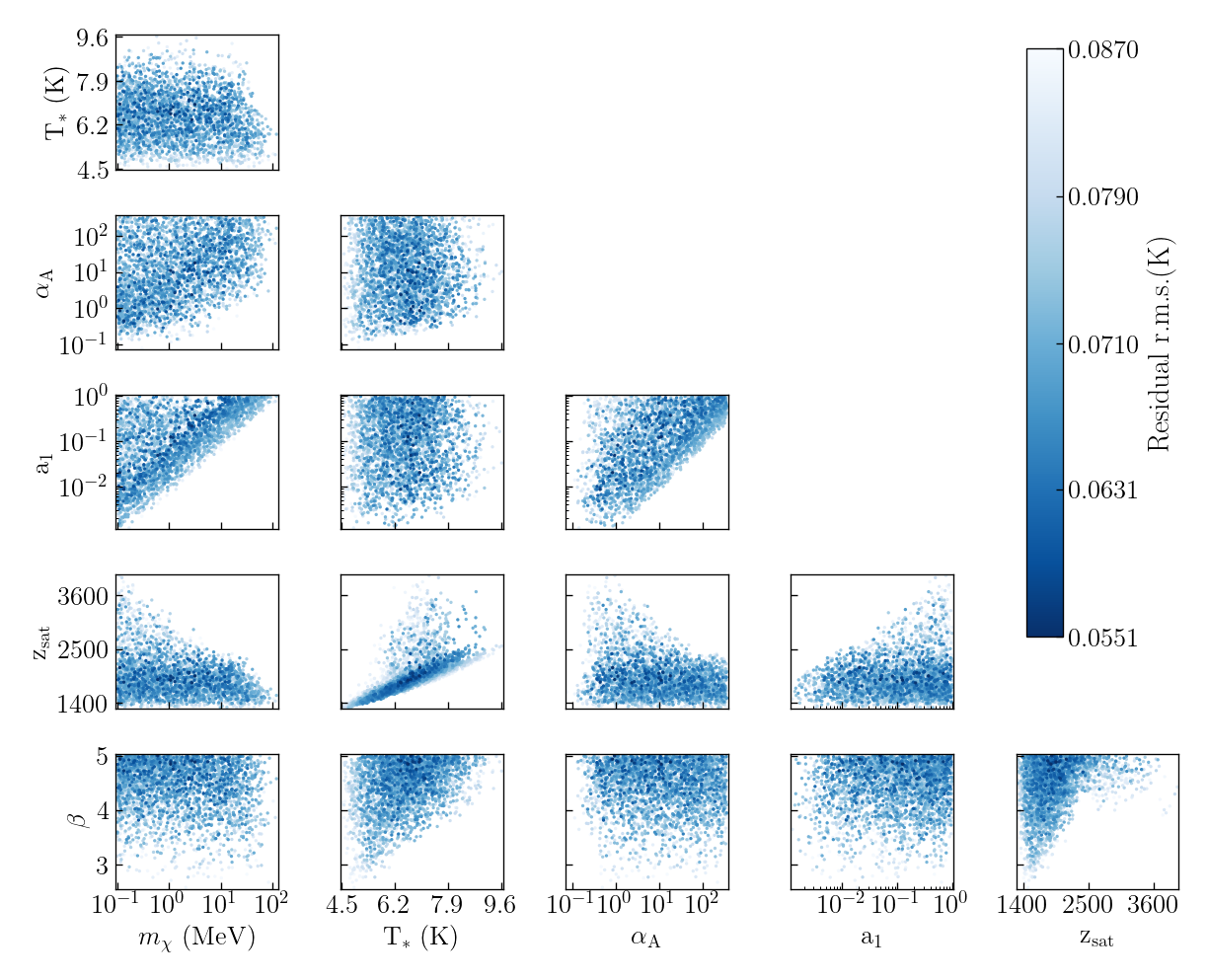}
	\caption{The viable parameter space of the dark matter model shown as points in the 2-D plots for fifteen combinations of different model parameters. The color shade of each sample point is represented by the r.m.s. value of the residual when the EDGES data is fitted with the the EDGES foreground model + dark matter signal of the sample.}
	\label{fig:cp}
\end{figure}
	\begin{figure}[t]
	\centering
	\includegraphics[width=0.65\textwidth]{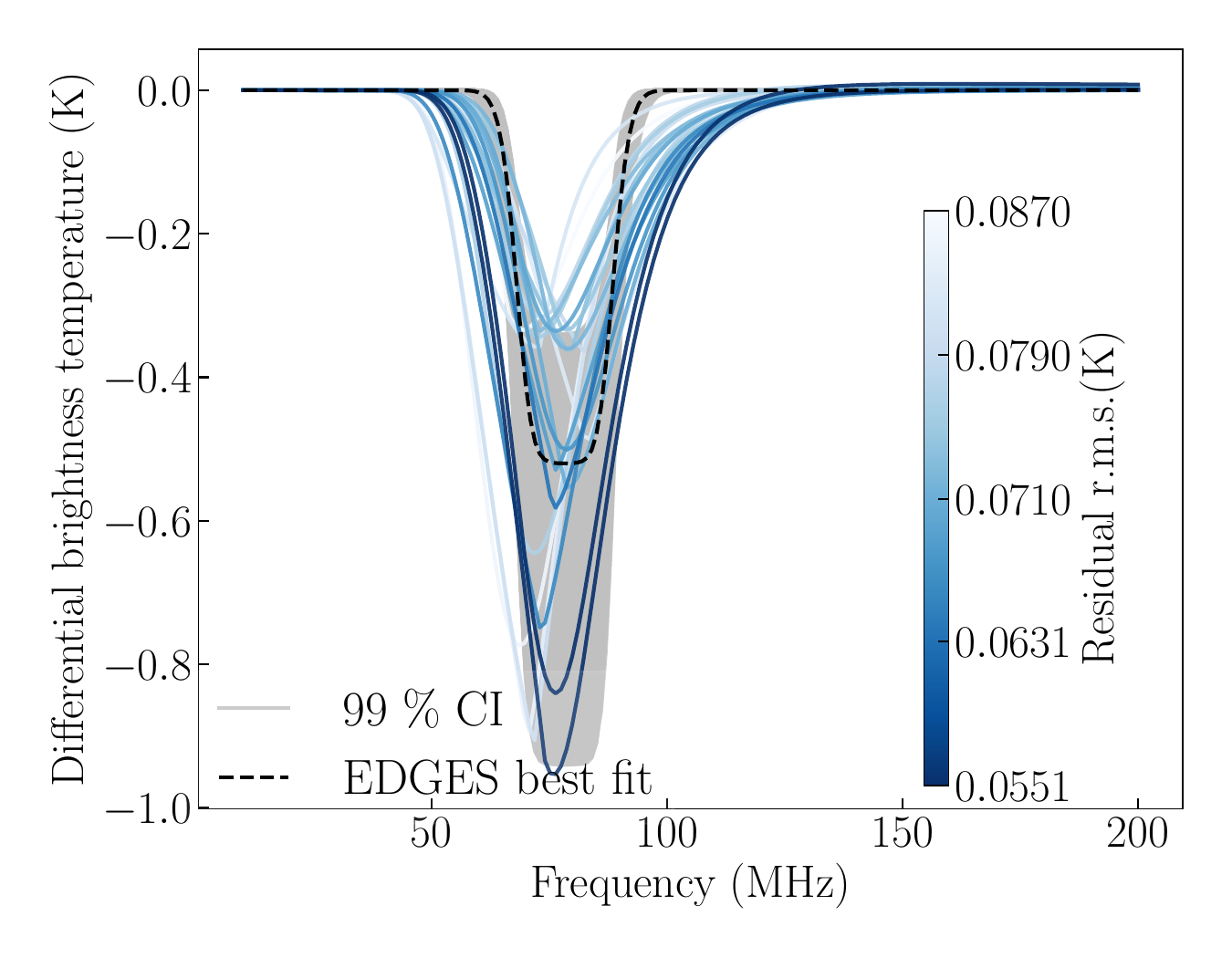}
	\caption{Twenty dark matter global absorption signals drawn from the viable parameter space shown in figure \ref{fig:param_e}. The curves are color coded according to the r.m.s. values of the residual. The gray band shows the 99\% confidence interval (CI) for the EDGES best fit signal, where the signal itself is denoted by the black dashed curve.}
	\label{fig:edges}
\end{figure}

	We repeat this for two million sample points in the six dimensional dark matter parameter space. We then fit the dark matter signal ($\mathrm{T}_{\mathrm{signal}}$)  from each sample + the EDGES foreground model ($\mathrm{T}_{\mathrm{F}}$) to the observed sky temperature ($\mathrm{T}_{\mathrm{sky}}$) and obtain the best fit parameter values for the EDGES foreground model by minimizing the residual given by, 
	\begin{equation}
		\mathrm{residual} = \sqrt{\frac{\sum_{i=1}^{n}\left[\mathrm{T}_{\mathrm{sky}}^i - \left(\mathrm{T}_{\mathrm{F}}(f_i)+\mathrm{T}_{\mathrm{signal}}(f_i)\right)\right]^2}{n}},
	\end{equation}
	where $f_i$ is the frequency of $i^{\text{th}}$ data point and $n$ is the total number of data points. 
	The EDGES foreground model is given by \cite{2018Natur.555...67B},
	\begin{align}
		T_{\text{F}}(f) &= a_0\left(\frac{f}{f_c}\right)^{-2.5} + a_1\left(\frac{f}{f_c}\right)^{-2.5}\log\left(\frac{f}{f_c}\right) + a_2\left(\frac{f}{f_c}\right)^{-2.5}\left[\log\left(\frac{f}{f_c}\right)\right]^2\nonumber\\ &+ a_3\left(\frac{f}{f_c}\right)^{-4.5} + a_4\left(\frac{f}{f_c}\right)^{-2},
	\end{align}
	where $f_c = 74.8$ MHz is the central frequency of the EDGES band and the coefficients $a_k$, where $k$ runs from $0$ to $4$ are free parameters of the foreground model.

	When just the foreground model is fitted to the observed sky temperature, the residual r.m.s. value is 0.087 K. We note that we are doing a random search of the dark matter parameter space. During the fitting procedure, the dark matter parameters remain fixed and only the foreground parameters are varied. Thus for some points in the dark matter parameter space we will find residuals which are $>$ 0.087 K while for some the residual would be smaller. We take the criteria for the presence of a dark matter signal as: residual $<$ 0.087 K. Further, we consider a sample to be viable if it produces a global signal which has a maximum amplitude of $0.5^{+0.5}_{-0.2}$ K at a frequency location $78^{+10}_{-10}$ MHz and a FWHM of $19^{+4}_{-2}$ MHz. We show the viable parameter space in the form of 2-D plots for fifteen combinations of different model parameters in figure \ref{fig:cp} where the color shade of each point represents the value of the residual. We also show twenty such global signals along with their residuals in figure \ref{fig:edges}. The smallest residual we get is 0.055 K. We see that there is a large volume of the dark matter parameter space that is consistent with the EDGES data. 
	We discuss the final allowed parameter space of our model taking into account both astrophysical and direct detection constraints in figure \ref{fig:dd_edges} in section \ref{sec:paramspace2}. 
	Even though our residuals are larger compared to the EDGES flattened Gaussian profile (having a residual of 0.025 K), we note that we fit a physical signal calculated for a specific dark matter model while the flattened Gaussian is a mathematical function specially chosen to fit the pattern seen in the data but without a physical basis. In particular, in such scenarios, look elsewhere effect must be taken into account. Having only residual as the qualitative criterion makes it difficult to have a fruitful comparison between different models.
	
	We would like to emphasize a few important points regarding the shape of the global signal:
	\begin{itemize}[leftmargin=15pt]
		\item[(i)] The left (low frequency) edge of the signal is entirely decided by the bremsstrahlung process which erases the absorption by dark matter at high redshifts. This also fixes the location of the maximum absorption which happens around the redshift of recombination. In particular, the rapid decrease in bremsstrahlung efficiency around recombination provides a sharp edge to the signal at the low frequency end. 
		\item[(ii)] The right (high frequency) edge of the signal is decided by the strength of inelastic collision cross-section relative to the radiative coupling of dark matter. The shape of the high frequency right edge is a strong function of the temperature dependence of the collision cross-section which can be tuned by having the collision cross-section to depend weakly or strongly on the dark matter temperature. 
	\end{itemize}
	\subsubsection{General predictions for the shape of the dark absorption feature}
	\label{subsubsec:par_e}
\begin{figure}[h!]
		\centering
		\begin{subfigure}[b]{7.7cm}
			\includegraphics[width=\textwidth]{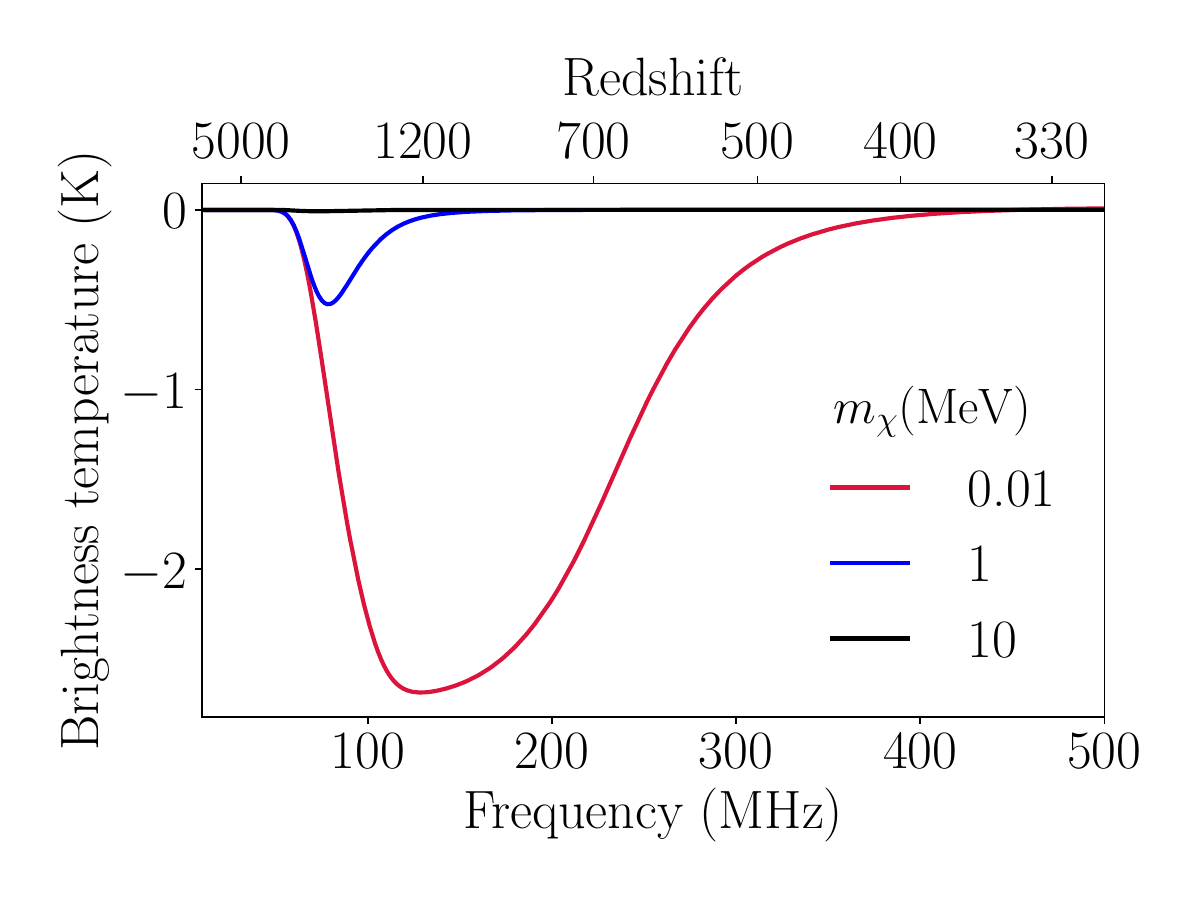}
			\caption{Dark matter mass}
			\label{fig:m_e}
		\end{subfigure}
		\begin{subfigure}[b]{7.7cm}
			\includegraphics[width=\textwidth]{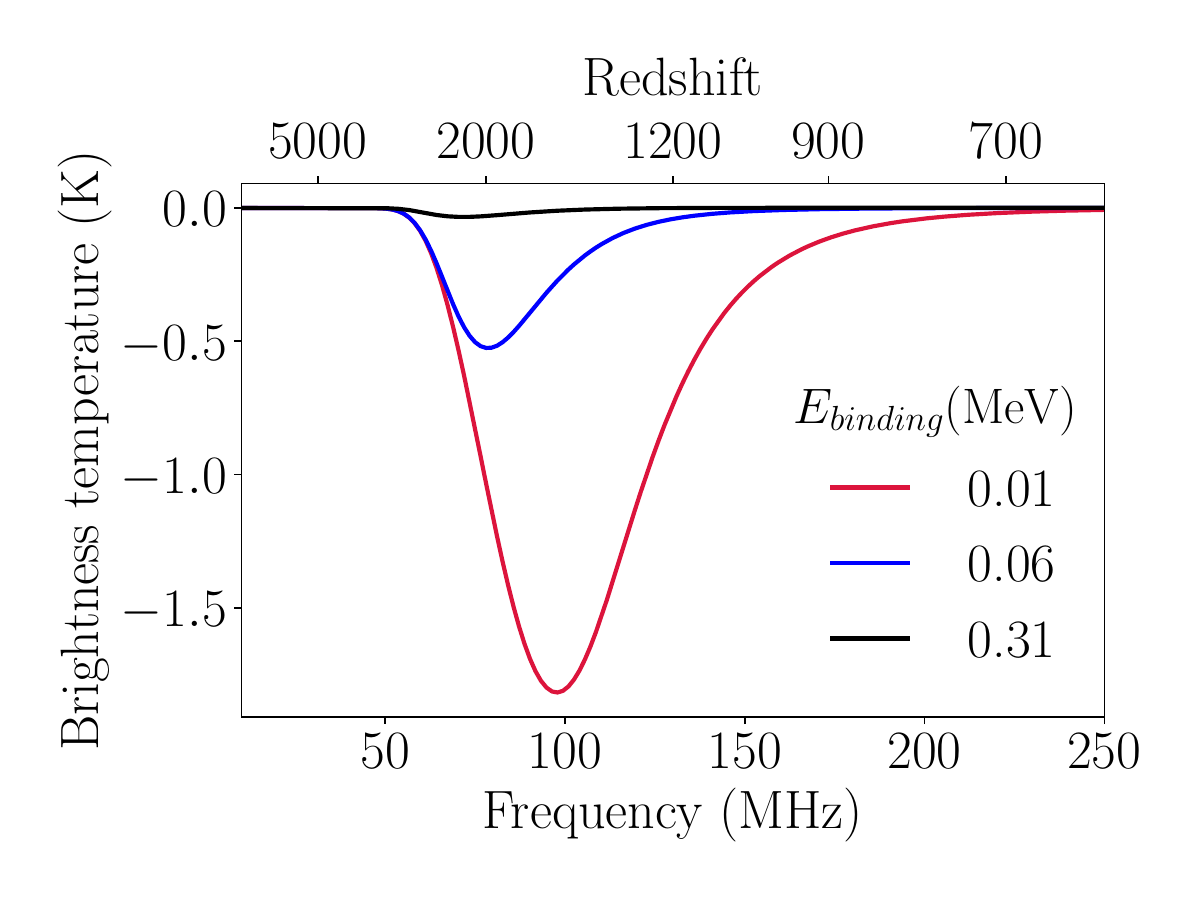}
			\caption{Binding energy}
			\label{fig:be_e}
		\end{subfigure}
		\begin{subfigure}[b]{7.7cm}
			\includegraphics[width=\textwidth]{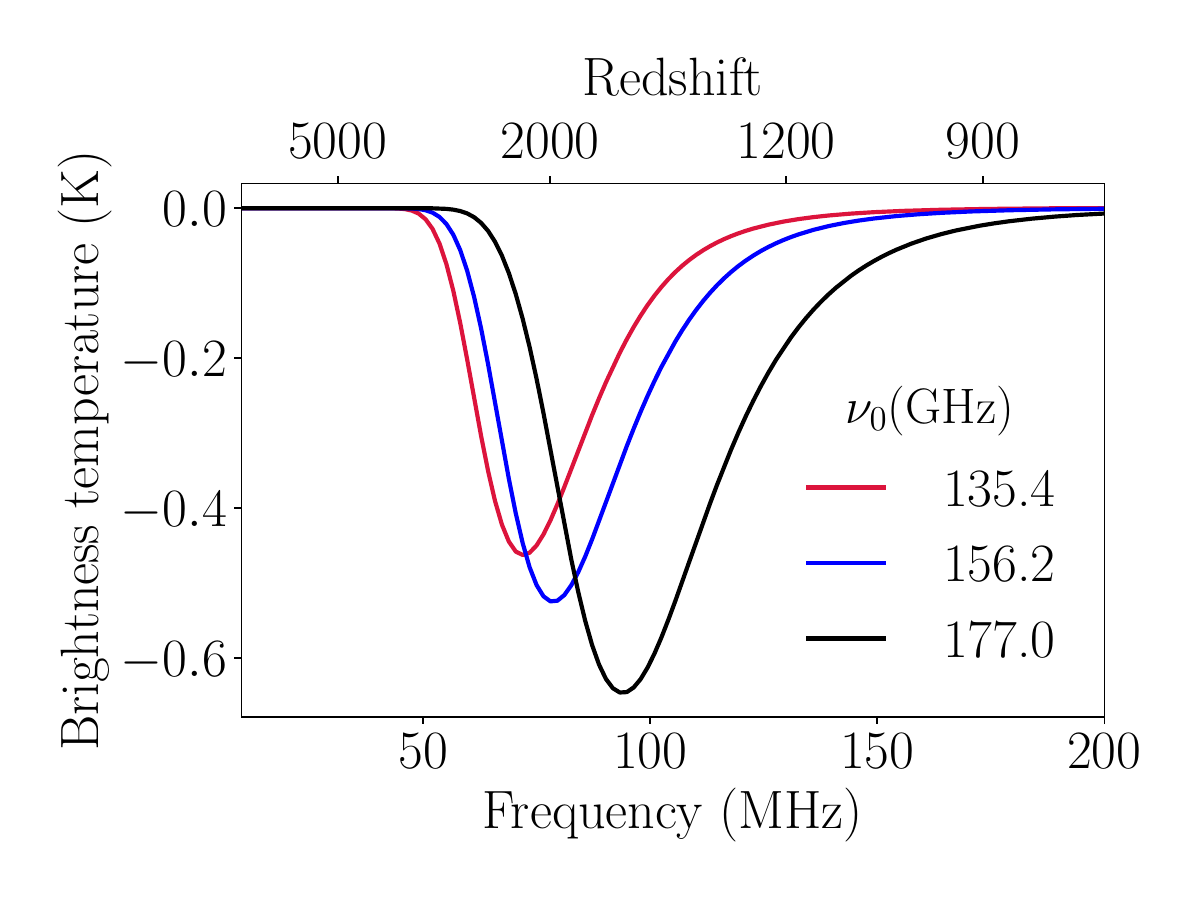}
			\caption{Transition frequency}
			\label{fig:t*_e}
		\end{subfigure}
		\begin{subfigure}[b]{7.7cm}
			\includegraphics[width=\textwidth]{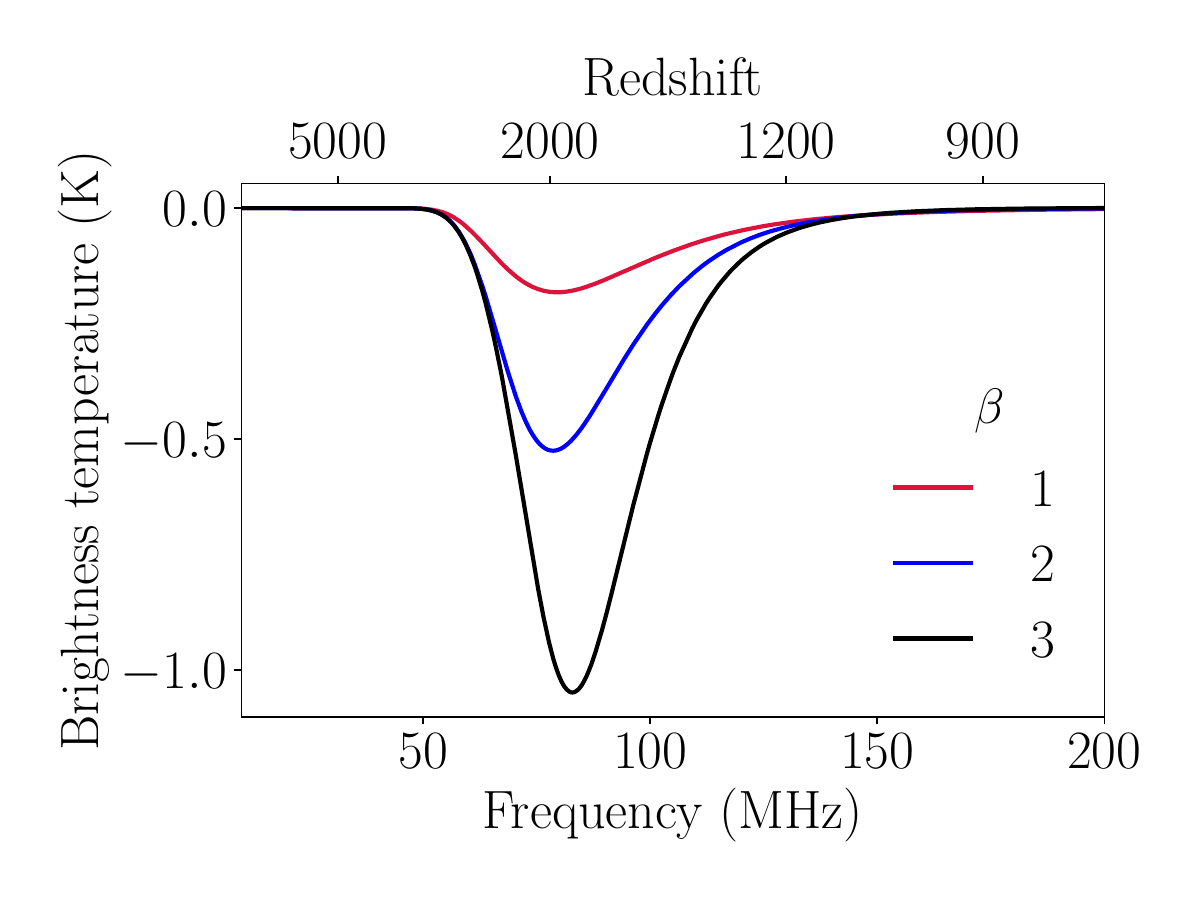}
			\caption{Inelastic collision cross-section}
			\label{fig:cc_e}
		\end{subfigure}
		\begin{subfigure}[b]{7.7cm}
			\includegraphics[width=\textwidth]{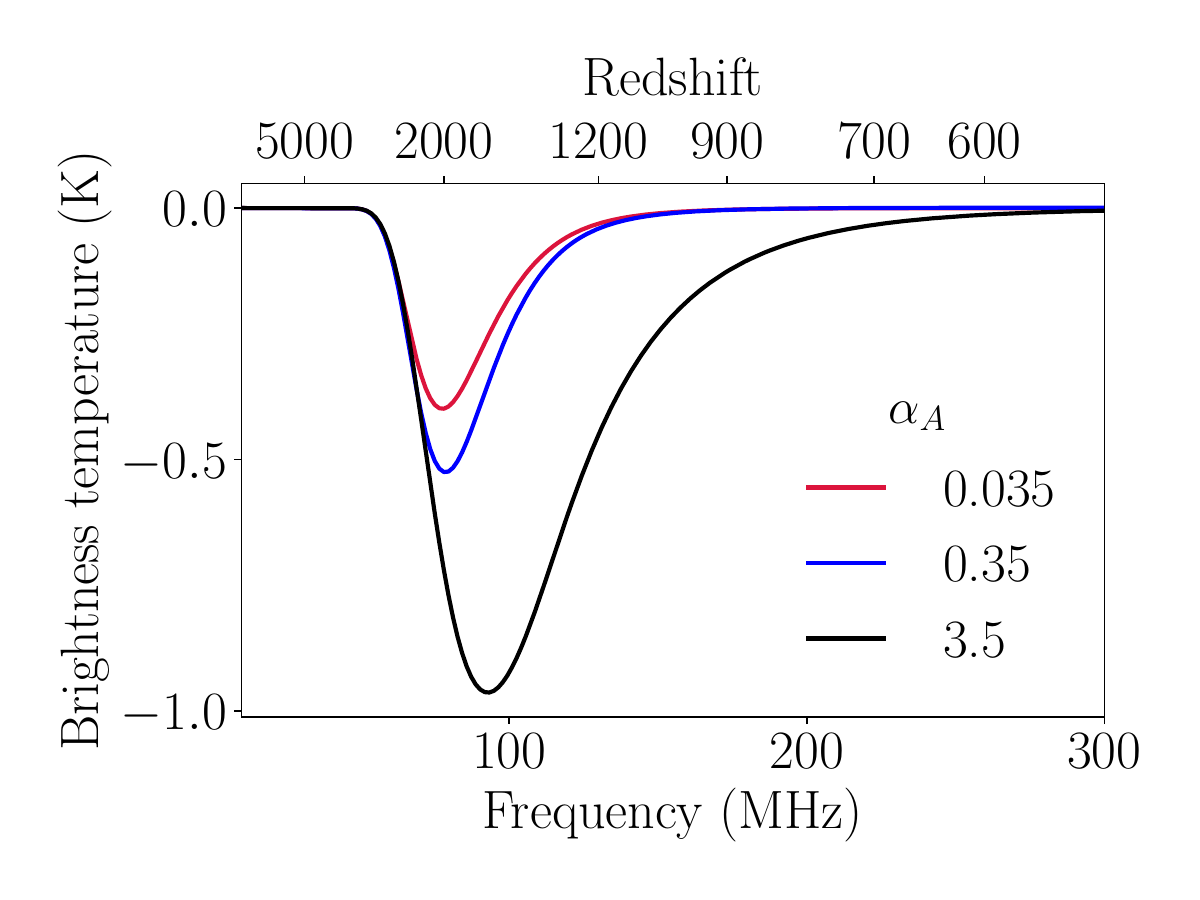}
			\caption{Radiative coupling}
			\label{fig:a10_e}
		\end{subfigure}
		\caption{Role of different model parameters in deciding the shape of the global absorption signal from dark matter. We vary each parameter at a given time keeping all the other parameters fixed.}
		\label{fig:param_e}
	\end{figure}

	To understand the role of different model parameters in determining the shape of the global absorption signal, we vary each model parameter one by one keeping all the other parameters fixed. 
	Even though our plots are restricted to the EDGES frequency band, the qualitative behavior and results are valid for any $\nu_0$ in the Rayleigh Jeans part of the CMB spectrum at recombination. The variation of the dark matter absorption feature in the CMB for different dark matter model parameters is shown in figure \ref{fig:param_e}.
		
	\begin{itemize}[leftmargin=10pt]
		\item \textbf{Dark matter mass:} Given the abundance of dark matter, smaller mass of dark matter implies a higher dark matter number density which increases the strength of the absorption signal.
		\item \textbf{Binding energy:} Dark matter with a higher binding energy decouples earlier (see eq.\eqref{3.3} of section \ref{sec:paramspace}), resulting in a lower dark matter temperature $T_\chi$ in eq.\eqref{4.21}. Since the dark matter inelastic collisional coupling $\propto T_{\chi}^{\beta}$, where $\beta>0$  in eq.\eqref{4.23}, smaller dark matter temperature implies weaker collisional coupling. A weaker collisional coupling compared to radiative coupling drives $T_{\text{ex}}\rightarrow T_\gamma$ earlier, resulting in smaller amplitude of the signal for higher binding energies.
		\item \textbf{Transition frequency:} When the transition frequency is varied, there is an overall shift in the position in frequency of the absorption signal. Moreover the bremsstrahlung rate eq.\eqref{4.17} is a sensitive function of the transition frequency. As we go to lower frequencies, the bremsstrahlung is more efficient in erasing the absorption signal resulting in a smaller amplitude.
		\item \textbf{Inelastic collision cross-section:} The high frequency or right edge of the signal is decided by the relative strength of collisional versus radiative coupling. If we change the power law index ($\beta$) of the collision cross-section while keeping the amplitude at $z_{\text{rec}}=1100$ fixed, a higher $\beta$ would mean stronger collisional coupling at $z>z_{\text{rec}}$. Thus we see stronger absorption as we increase $\beta$.
		\item \textbf{Radiative coupling:} A higher spontaneous emission rate compared to collisional transition rate couples $T_{\text{ex}} \rightarrow T_{\gamma}$ earlier, shifting the absorption signal to higher redshifts. The bremsstrahlung process is more efficient in erasing the signal at higher redshifts. Since the optical depth $\propto \alpha_{A}$, the maximum amplitude of the absorption signal increases for higher values of $\alpha_A$ (see eq.\eqref{a10} for the definition of $\alpha_{A}$).
	\end{itemize}
	
	\subsubsection{General predictions for the dark absorption feature in different parts of the CMB spectrum}
	Irrespective of EDGES, composite dark matter predicts an absorption feature in the CMB spectrum. Our choice of transition frequency was motivated by the EDGES observation. In general for a different transition frequency $\nu_{0}$ and absorption redshift $z_0$, the absorption feature will be appear in a different part of the CMB spectrum. The upcoming experiment called the Array of Precision Spectrometers for the Epoch of RecombinAtion (APSERA) \cite{2015ApJ...810....3S}, which aims to detect the recombination lines in the CMB spectrum will also be sensitive to the dark absorption feature in the 2-6 GHz frequency range.
	
	Any dark absorption feature originating at $z_0 \gtrsim2\times 10^6$ in the CMB cannot be observed. This is because Compton scattering \cite{1957JETP....4..730K} along with the photon number changing processes like bremsstrahlung and double Compton scattering \cite{1969Ap&SS...4..301Z,1981ApJ...244..392L,1981MNRAS.194..439T,1982A&A...107...39D} are efficient in erasing any deviations from the black body spectrum till $z\sim2\times 10^6$. As the bremsstrahlung and double Compton scattering rates fall $\propto \nu^{-2}$ with frequency, they decouple at $z\sim2\times 10^6$ for photons having $x\sim0.01$ \cite{1969Ap&SS...4..301Z,1970Ap&SS...7...20S,1991A&A...246...49B,1993PhRvD..48..485H,2009A&A...507.1243P,2012MNRAS.419.1294C,2012JCAP...06..038K}. With only Compton scattering efficient in $z\sim 2\times 10^6-10^5$ range, the equilibrium spectrum is the Bose-Einstein spectrum and the resulting deviations from blackbody are created in the form of $\mu$-type spectral distortions. If the absorption happens at $z \lesssim 10^5$, we will have a broad absorption feature in the CMB spectrum. The COsmic Background Explorer/Far-InfraRed Absolute Spectrophotometer (COBE/FIRAS) \cite{1996ApJ...473..576F} experiment strongly limits the CMB spectral distortions in $60-600$ GHz band (see Appendix \ref{app:cobe}). Thus any absorption happening at $z_0<2\times 10^6$ corresponding to $60<\nu_{0} (\text{GHz})<1.2\times 10^{9}$ range will be strongly constrained by COBE. 
	
	In addition, CMB photons having $x\sim 50$ correspond to different energy states of hydrogen and helium in $10.2 - 50$ eV range at recombination ($z \sim 1100$). If dark matter absorbs these photons, there would be fewer CMB photons that can excite and ionize hydrogen and helium speeding up recombination. An early recombination would modify the position and amplitude of angular peaks and the Silk-damping tail of the CMB anisotropy power spectrum which is strongly constrained by Planck \cite{2020A&A...641A...5P}. We leave the detailed analysis of the constraining power of different CMB observations on our dark matter model to future work.

	\section{Astrophysical constraints}
	\label{sec:paramspace}
		For a dark matter model to be viable, it should be consistent with the existing cosmological and astrophysical observations. This requirement motivated our choices for the allowed ranges of dark matter model parameters in the previous sections. In this section we justify the these choices. The parameter space for composite dark matter is huge and it is not possible to make a complete study in a single paper. We therefore focus on the parameter range related to observations in the radio band. Note that the goal of this section is not to derive the best constraints on our model but rather to show that a significant and interesting part of the parameter space is allowed and has unique experimental signatures. These constraints should therefore be taken as an order of magnitude estimate which provide a rough guidance for the viable parameter space of the model.
	\begin{figure}[t]
		\centering \hspace{-0.5cm}
		\includegraphics[width=0.33\textwidth]{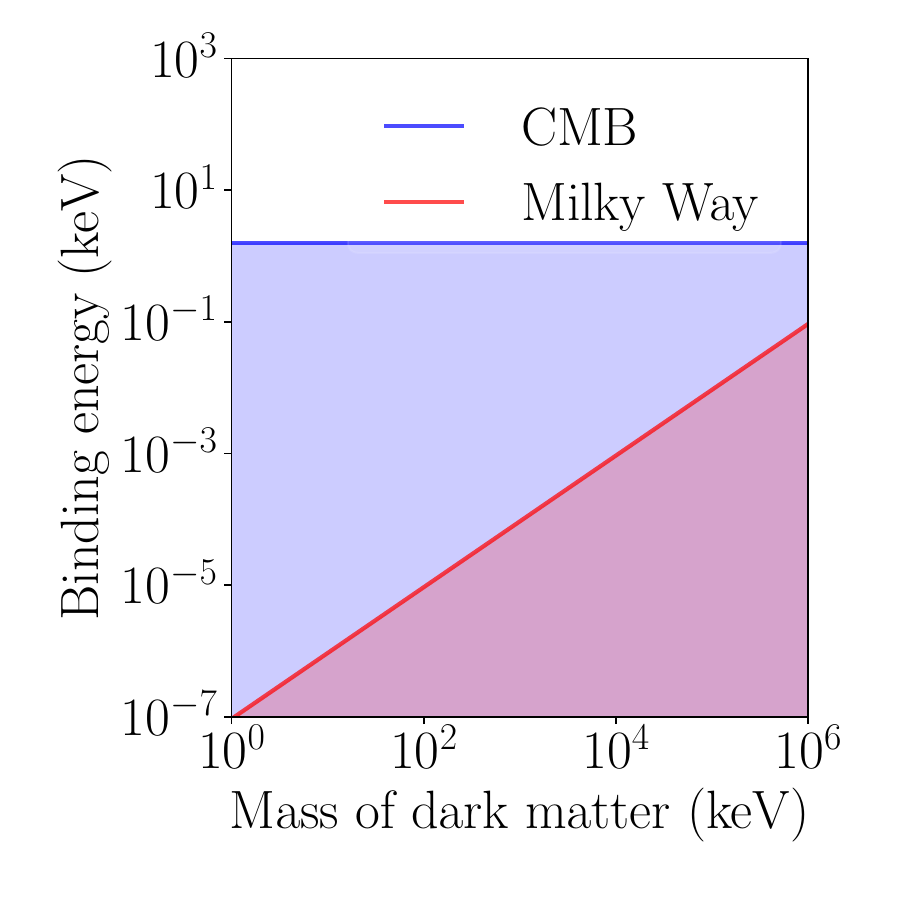}\hspace{-0.5cm}
		\includegraphics[width=0.33\textwidth]{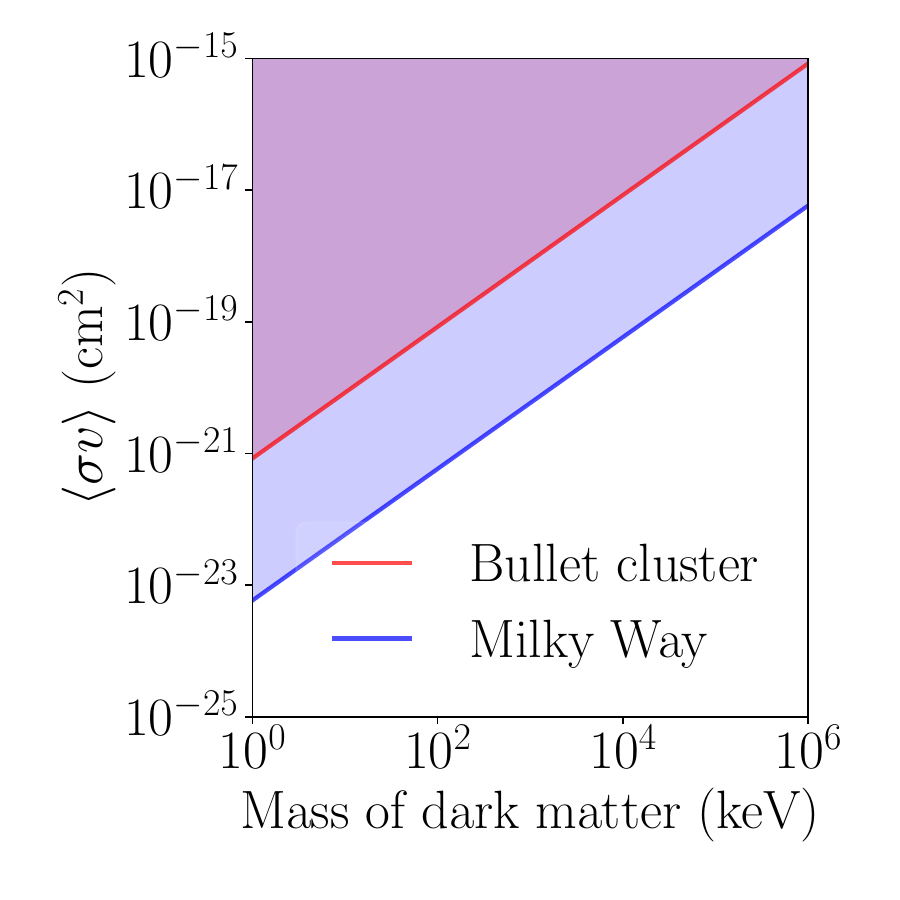}\hspace{-0.5cm}
		\includegraphics[width=0.33\textwidth]{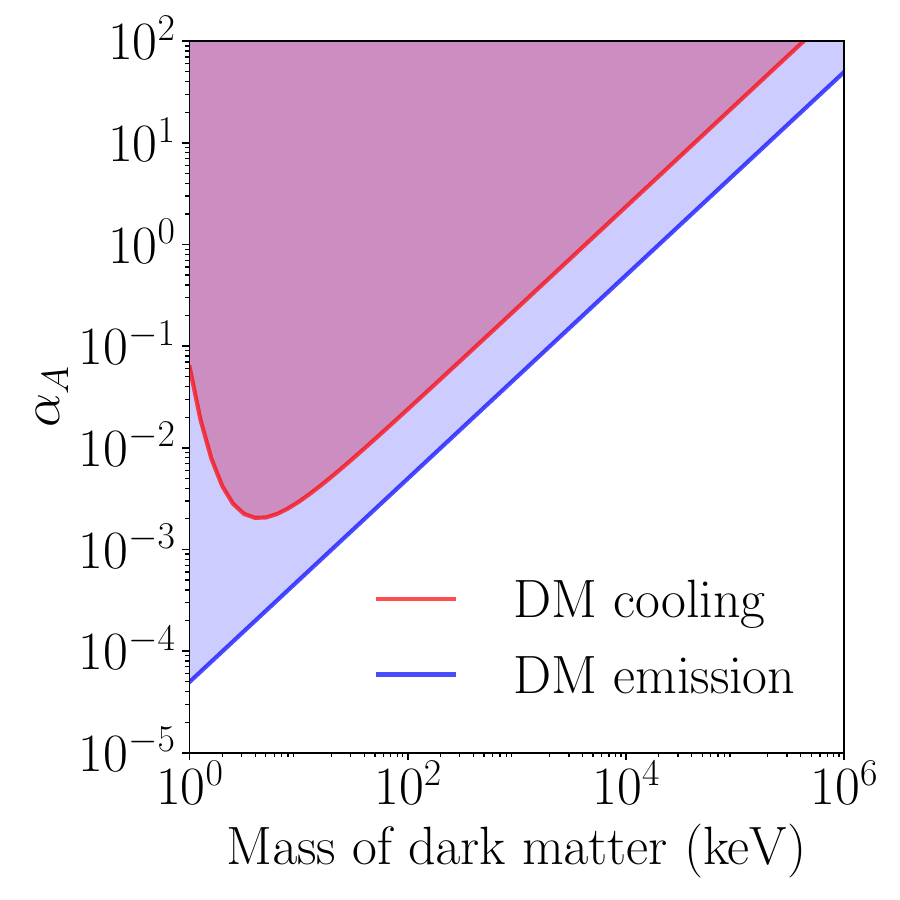}
		\caption{We show the absolutely allowed parameter space for the different astrophysical parameters (left: binding energy, center: collision cross-section, and right: radiative coupling) for our composite dark matter model. The colored region in the plots is ruled out. We choose the transition frequency $\nu_0 = 156.2$ GHz which is motivated to solve the EDGES anomaly.}
		\label{fig:c}
	\end{figure}
	\begin{itemize}[leftmargin=10pt]
		\item \textbf{Mass of dark matter:}	
		At early times when the dark sector was in thermal equilibrium with the Standard Model sector, the Universe was characterized by a single temperature which was equal to the CMB temperature $T_{\gamma}$. At temperatures above the scale at which the dark sector becomes strongly coupled i.e. $k_BT_{\gamma} > \Lambda_N$, the the dark sector in the UV was characterized by electromagnetically charged dark quarks as explained in section \ref{subsec:2.2}.
		
		The presence of electrically charged dark quarks also implies that the dark sector can interact with the Standard Model sector via number changing processes such as annihilation and pair production, and momentum exchange processes such as Coulomb and Compton scattering. Below we list the different scenarios where these interactions can be important:
			\begin{itemize}
				\item[(i)] If the number changing processes are important at scales $\lesssim$ 100 keV, the dark sector can modify the expansion history which can in-turn alter the standard processes in the thermal history of the Universe such as BBN, etc. 
				\item[(ii)] Even if the number changing processes are unimportant, the momentum exchange processes can keep the dark sector in thermal equilibrium with the Standard Model sector. This can result in an excess radiation pressure during recombination which can modify the acoustic peaks in the CMB power spectrum. Such interactions are strongly constrained by the precise observations of Planck \cite{2020A&A...641A...5P}. 
				\item[(iii)] Moreover, for dark matter in our own Milky Way (MW) halo (with a virial temperature $T_{\text{vir}}$) to not ionize, the energy exchange in each inelastic collision ($\sim k_{B}T_{\text{vir}}$) between two dark matter particles must be less than the binding energy of dark matter particles. The lower bound on dark matter binding energy from MW is shown in the first panel of figure \ref{fig:c}. 
		\end{itemize}
		 One way to evade the constraint from (ii) is to assume that the dark sector is characterized by electrically neutral bound states which do not interact with the Standard Model sector at scales (with a wave number $k \lesssim 0.22\, \text{Mpc}^{-1}$) sensitive to Planck. The redshift at which the smallest scale observed by Planck entered the horizon is $z_{*} \simeq10^{5}$. We assume that dark matter recombines into a stable and neutral composite state by $z\geq z_*$. As the Universe expands and the CMB temperature falls, the peak of the CMB blackbody spectrum shifts towards lower energies. There are fewer high energy photons left in the Wein tail of the CMB spectrum that can ionize the composite dark matter. Assuming the recombination history of hydrogen and dark matter to be similar,
		\begin{eqnarray}
			\frac{\text{Binding Energy}}{k_BT_{\gamma}(z = 1100)}\bigg|_{\text{hydrogen}} \approx 50 \,\,\approx\,\, \frac{\text{Binding Energy}}{k_BT_{\gamma}(z = z_*)}\bigg|_{\text{dark matter}}.\label{3.3}
		\end{eqnarray}
		This imposes a lower bound on the binding energy of dark matter, $E_{\text{binding}}\geq 1.6 \, \text{keV}$, as shown in the first panel of figure \ref{fig:c}. 
		
		Even if dark matter becomes an electrically neutral bound state, it can still interact with the baryons owing to higher electromagnetic moment described by operators in eq. \eqref{C28} and \eqref{C32} of Appendix \ref{app:C}. Therefore in order to evade constraints from (ii) we will also assume that the dark matter kinetically decouples from the baryon photon plasma at $z=z_*$. A conservative estimate can be obtained by assuming that the timescale of dark matter electron scattering becomes greater than the age of the Universe at redshift $z_*\simeq10^{5}$, 
		\begin{eqnarray}
			\frac{n_{e}\langle \sigma v\rangle_{\chi e}}{H}\bigg|_{z=z_*} \approx 1, \label{cmb}
		\end{eqnarray}
		where $\langle \sigma v\rangle_{\chi e}$ is the velocity averaged cross section of dark matter electron scattering which includes contribution from both elastic (charge radius operator in eq. \eqref{C32}) and inelastic ($\chi-\chi^*$ transition operator in eq. \eqref{C28}) processes. The velocity averaged cross-sections for elastic (ES) and inelastic (IS) scattering are given by, 
		\begin{align}
			\langle \sigma v\rangle_{\chi e}^{\mathrm{ES}} &= \frac{16\pi\epsilon^2\alpha^2\mu_{\chi e}^2}{3 m_\chi^4}\sqrt{\frac{8T_{\gamma}}{\pi \mu_{\chi e}}}, \label{4.3} \\
			\langle \sigma v\rangle_{\chi e}^{\mathrm{IS}} &= \frac{8\pi\epsilon^2\alpha^2\mu_{\chi e}^2}{m_\chi^2m_e^2}\sqrt{\frac{8T_{\gamma}}{\pi \mu_{\chi e}}} .
			\label{4.4a}
		\end{align}
		Note that at $z_* \approx 10^5$, the kinetic energy of electrons is $\sim  20$ eV. Therefore $\chi$ to $\chi^*$ inelastic scattering is kinematically allowed for dark matter energy splittings $\Delta E \lesssim 20$ eV. For energy splittings $\Delta E \gtrsim 20$ eV, the scattering between dark matter and electron is entirely elastic. These constraints are shown as a black dashed line in figures \ref{fig:dd} and \ref{fig:dd_edges} respectively. We represent the regions dominated by inelastic scattering by dark gray color and regions dominated by elastic scattering by light gray color.

	\item \textbf{Radiative coupling:} 
		In this section we discuss two scenarios in the Milky Way halo that can be used to constrain the radiative coupling of dark matter. We will be assume the dark matter to be collisional such that the two transitioning states are in kinetic equilibrium with the virial temperature of the MW halo i.e. $T_{\text{ex}} = T_{\text{vir}}$.
		\begin{itemize}[leftmargin=10pt]
		\item[(i)] When dark matter is collisionally excited in the MW, its kinetic energy is converted into internal energy. The excited dark matter particles can then de-excite by spontaneously emitting a photon, converting the internal energy of dark matter into radiation. This would lead to a gravitationally unstable dark matter halo, which cools and starts collapsing into a disk similar to baryons which is ruled out by the observations from GAIA \cite{2018PhRvL.121h1101S}. The cooling timescale $t_{\text{cooling}}$ for this process is given by the ratio of thermal energy density $U$ in dark matter to the radiative cooling rate $\mathcal{C}$,
		\begin{align}
			t_{\text{cooling}} \approx \frac{U}{\mathcal{C}} \ \approx \frac{\frac{3}{2}n_\chi k_{B}T_{\text{vir}}}{\mathcal{C}}, \hspace{4pt}
			\text{where}\hspace{10 pt}\mathcal{C} = n_{1}A_{10}k_{B}T_*.
		\end{align}
		The MW dark matter halo remains gravitationally stable if the cooling timescale is longer than the age of MW \cite{2022Natur.603..599X}. This puts an upper limit on the radiative coupling of dark matter (see eq.\eqref{a10} for the definition of $\alpha_{A}$),
		\begin{align}
			\alpha_A < \frac{3}{2}\frac{T_{\text{vir}}}{T_{*}}\frac{1}{t_{\text{MW}}A_{10}^{\text{HI}}}\left(1 + \frac{g_{0}}{g_{1}}e^{\frac{T_{*}}{T_{\text{vir}}}}\right).\label{3.11}
		\end{align}
		In this analysis, we have assumed that dark matter only cools via the dark hyper-fine transitions. In principle, if the states with higher energies can be collisionally excited and have a higher spontaneous emission rate, they will contribute to the cooling process. As discussed in section \ref{sec:model}, in this model we assume the other states have energies comparable to the binding energy of dark matter ($\sim$ MeV) which would not be excited collisionally in most astrophysical scenarios. We plot the limit on $\alpha_A$ in the third panel of figure \ref{fig:c} for transition frequency $156.2$ GHz or $T_{*} = $ 7.5 K. This choice of parameters was motivated by the EDGES observation.
		
		\item[(ii)] The collisionally excited dark matter particles in the MW halo can de-excite by spontaneously emitting a photon, creating a background radiation with a dipole anisotropy owing to our off-center position 8.3 kpc \cite{2017ApJ...837...30G} away from the MW halo center. 
		The experiments like WMAP and Planck \cite{2013ApJS..208...20B,2020A&A...641A...5P} measure the fluctuations in the sky temperature in broad frequency bands in the frequency range of 10-500 GHz. 
		
		Considering $j$ to be the integrated specific intensity and $j_{\text{avg}}$ be the average integrated specific intensity due to dark matter emission (see Appendix \ref{app:cons} for details), the temperature fluctuations along a given LoS in a particular frequency band ($\nu_{\text{min}}$ to $ \nu_{\text{max}}$ with a central frequency $\nu_c$) is given by,
		\begin{align}
			\delta T_{\nu_{c}} = \left(\frac{|j-j_{\text{avg}}|}{\nu_{\text{max}}-\nu_{\text{min}}}\right)\frac{c^{2}}{2k_{B}\nu_{c}^{2}} \label{3.17}.
		\end{align}
		We note that this dipole will be aligned with the galactic plane and thus obscured by Galactic emission. Also this dipole will appear in only one channel that includes $\nu_0$ and would be absent in other channels. Without doing a detailed analysis, we can just require very conservatively that this dipole must be smaller than the cosmological dipole $\delta T_{\nu_{c}}< 3\, $mK, imposing an upper bound on $\alpha_A$. The constraints on $\alpha_{A}$ from  eq.\eqref{3.11} and eq.\eqref{3.17} are shown in the third panel of figure \ref{fig:c}. In our analysis, we have chosen $\nu_0 = 156.2$ GHz which falls in the frequency band of Planck centered at $\nu_c$ = 143 GHz with a bandwidth $\Delta\nu/\nu = 0.33$. We note that CMB is the most precisely measured part of the electromagnetic spectrum over the whole sky and constraints from other parts of the spectrum (for other values of $\nu_{0}$) would be much weaker. 
	\end{itemize}
		
	\item \textbf{Inelastic collision cross-section:} A conservative way to constrain the collision cross-section of dark matter is to assume the timescale of dark matter collisions $t_{\text{collision}}=(n_{\chi}\langle\sigma v\rangle)^{-1}$ in our local neighborhood ($\rho_{\chi}\sim0.4$ GeV cm$^{-3}$) to be greater than the age of the Milky Way ($t_{\text{MW}} \approx$ 13 billion years \cite{2022Natur.603..599X}). This puts an upper limit on the dark matter collision cross-section (elastic + inelastic). For $v\approx 200$ km/s,
	\begin{align}
		\langle\sigma v\rangle_{\chi\chi}<\frac{1}{t_{\text{MW}}}\left(\frac{m_{\chi}}{\rho_{\chi}}\right).\label{3.6}
	\end{align}
	We show this constraint in the second panel of figure \ref{fig:c}. We also show the upper limit on the collision cross-section from Bullet cluster \cite{2004ApJ...606..819M} ($\sigma_{\text{BC}}/m_{\chi}\,<\,2\,\text{cm}^{2}\text{g}^{-1}$) assuming the relative velocity between the two clusters $v_{\text{BC}} \approx $ 4700 km/s. This indirectly puts constraints on dark matter inelastic collision cross-section.
	
	We note that the collision cross-sections usually have a strong temperature dependence which suggests that dark matter collisions may be insignificant in MW but maybe important in other dark matter halos. Our back of the envelope constraints from MW are very conservative. A more careful analysis taking into account the non-negligible dark matter self-interactions which maybe preferred by data, would relax these constraints \cite{2010AdAst2010E...5D,2000PhRvL..84.3760S}. We therefore use the Bullet cluster constraint as a reference in our analysis.
	\end{itemize}

\subsection{Discussion of pre-existing constraints on dark matter with electromagnetic
  couplings in context of our model}
\subsubsection{Millicharged dark matter constraints}
	We note that there are stringent constraints on millicharged dark
        matter in literature \cite{2000JHEP...05..003D,2004JETPL..79....1D,2013PhRvD..88k7701D,2018JHEP...09..051C}. However, in our model the dark matter is
        neutral and composite in most astrophysical environments. The
        existing millicharged dark matter constraints therefore do not
        apply to us. The exception would be extreme environments such as
        cores of collapsing supernovae \cite{1987PhRvL..58.1490H}. The supernovae constraints however
        rely on dark matter particles radiating  away energy from the
        core. In our model, millicharged dark quarks cannot exist as free
        particles and therefore cannot be radiated away, only the
        electrically neutral composites exist as free (asymptotic states) particles. The lightest dark 
        particles that can be radiated away are the dark pions, however
        their production would be a higher order process that would be
        suppressed compared to the direct production of a millicharged
        particles usually considered in deriving the supernova
        constraints. For an analogy, consider the pair-production of pions in $e^{+}e^{-}$ collisions. At short distances, the production goes via pair-production of quarks $\left(e^+e^- \rightarrow \gamma^* \rightarrow q\bar{q}\right)$, since the production of pions are suppressed. Further, just the production of dark pions does not  immediately imply the presence of a new channel for energy loss. The dark quarks produced give rise to further reactions of dark gluons and quarks (shower similar to QCD), yielding a large number of dark pions at the end of the day. This procedure results in the energy of the original dark quarks getting divided into a large number of softer states. Such distribution of energy is inevitable as taught by QCD. The fraction of these particles (and the energy) that can finally escape the core is therefore expected to be small. The computation of this quantity depends on the number of colors and flavors of the gauge group, as well as on the details of hadronization. A quantitative study of these effects is beyond the scope of this paper. Thus, we cannot directly apply existing constraints
        from literature to our model. We expect that there would be
        interesting constraints from supernovae on our model for masses of
        order MeV or smaller but it would require a more complicated
        analysis than is usually done for millicharged dark particles. 

\subsubsection{ BBN constraints}\label{BBN}
There are strong limits on light dark matter models with unspecified but strong electromagnetic coupling or neutrino coupling from BBN and CMB \cite{1986PhRvD..34.2197K,2004PhRvD..70d3526S,2013JCAP...08..041B,2014PhRvD..89h3508N,2015PhRvD..91h3505N,2020JCAP...01..004S}. However, we would like to emphasize that these constraints cannot be directly applied to our model because the thermal history of our model differs significantly from these simple scenarios considered in the literature. These works generically make the assumption that dark matter is produced thermally through annihilation into Standard Model neutrinos or electron-positron pairs and photons, or both. This changes the expansion rate, the photon to neutrino ratio, and the baryon to photon ratio affecting the successful predictions of the BBN and the CMB power spectrum. 

The thermal history of our model is highly non-trivial. At temperatures above the dark confinement scale $\Lambda_{N}$, we expect a dark quark gluon plasma which would be in thermal equilibrium with the visible sector because of $\epsilon$ electromagnetic charge of dark quarks. As the temperature falls below $\Lambda_{N}$, the dark quarks and dark gluons quickly hadronize into bound states. Because of effective strong interactions, the heavier bound states annihilate quickly into the lowest lying states of the dark sector i.e. the lightest dark glueballs and/or dark pions, which can further annihilate/decay to both $\nu\bar{\nu}$ pairs as well as photons. The anniilation/decay channel to neutrinos becomes evident if one considers the origin of electromagnetism, which arises below the scale of electroweak symmetry breaking from electroweak forces $SU(2)\times U(1)_Y$. Once Higgs gets a vev,  $SU(2)\times U(1)_Y$  is spontaneously broken to $U(1)_\text{em}$.  One linear combination of $Y_\mu$ and $W^3_{\mu}$ gives rise to a massless gauge boson which is the photon and orthogonal linear combination gives rise to the massive Z boson. Therefore, unless substantial additional constructions are invoked,  one expects that $\epsilon$ $U(1)_\text{em}$ charge arises from $\epsilon$ $U(1)_Y$ charge, which, in turn, predicts dark quark-$Z$ coupling of strength of order $\epsilon$ as well. The annihilation rate to neutrinos is $K_1\epsilon^2G_F^2$ and to photons is $K_2\epsilon^4\alpha^2$, where $K_1$ and $K_2$ are unknown non-perturbative factors that arise due to dark gluo-dynamics. Due to the non-perturbative nature of dark QCD, the exact calculation of these rates is highly complicated and beyond the scope of this work. However, if the rate of annihilation/decay of dark sector particles to both neutrinos and photons is comparable, there would be no significant change in the baryon to photon and photon to neutrino ratios post-annihilation/decay compared to the Standard Model. 

We note that in references \cite{1986PhRvD..34.2197K,2004PhRvD..70d3526S,2013JCAP...08..041B,2014PhRvD..89h3508N,2015PhRvD..91h3505N}, the effect on the deuterium abundance of dark matter annihilation into photons is opposite to the case where dark matter annihilates into neutrinos. Moreover, the neutrinos produced due to dark sector annihilation/decay would have a non-thermal distribution with average energies higher than the pre-existing thermal neutrinos. The weak interactions of these neutrinos would delay the freezeout of the neutron-to-proton ratio whereas the increased expansion rate of the Universe due to extra dark degrees of freedom will advance the freeze-out. We therefore have two effects that cancel each other. The effect of non-thermal neutrinos from electron-positron annihilation has been studied for standard cosmological model, where the distortion is of order 1\%  \cite{1992PhRvD..46.3372D,1992PZETF..56..129D,1993PhRvD..47.4309F,1997NuPhB.503..426D,2002PhR...370..333D}. In view of the above mentioned many competing effects and significant difference in our model versus the models used in literature, it is not possible to directly use the pre-existing constraints from BBN in the literature and put a hard lower limit on the mass of dark matter in our model. A systematic way to approach this problem would be to take the branching ratios to photons and neutrinos to be a free parameter and fit it to the data i.e. the BBN abundances. One can then use this preferred ratio to understand particulars of the confining dynamics itself. 
The reference \cite{2020JCAP...01..004S} studies sub-MeV thermally produced dark matter with the branching ratio to photons and neutrinos as an additional free parameter. They rule out dark matter masses $m_\chi < 0.8-1.4$ MeV at $2\sigma$ using CMB and BBN observations, where the smaller upper limit is for scalar dark matter and the larger upper limit is for Dirac fermion dark matter.  In our model, the dark matter abundance arises from asymmetry in the dark sector and not from thermal production. Therefore, we have an additional free parameter since the cross-sections are not fixed by requiring the correct dark matter abundance, which could help relax the mass bound. In order to get the exact bound for our model, a full analysis needs to be done which we leave for future work. Extrapolating from current literature, we expect that even in the worst case scenario the mass bound on our dark matter model would also be of order $\sim$ 1 MeV. For reference, we show the 1.4 MeV bound from \cite{2020JCAP...01..004S} in figure \ref{fig:dd2}. 

\color{black}
We also note that the existing measurements from colliders would also place constraints on the coupling of dark matter to Standard Model particles. However, these limits would be weaker than the cosmological and direct detection constraints considered in this work. For instance, if the dark sector particles have weak interactions, they can open a new channel for $Z$ boson decay. However, this rate would be $\epsilon^2$ suppressed. Since the decay width of $Z$ is measured to a precision of few parts in $10^4$, $\epsilon \lesssim 10^{-2}$ would be allowed. There can also be other channels such as if a Standard Model particle $X\rightarrow Y+\gamma$, where $Y$ is a collection of Standard Model particles, it can also decay to $X\rightarrow Y +$ pair of dark sector particles. These processes would make corrections of the order of $(\epsilon/16\pi^2)^2 \lesssim 10^{-10}$ for $\epsilon \sim 10^{-3}$ (compared to $X \rightarrow Y$ rates), where $(1/16\pi^2)^2$ reflects the phase space suppression. As far as we know no channels in the Standard Model are measured to an accuracy of one part in $10^{10}$.

	\section{Direct detection constraints}
	\label{sec:paramspace2} 
	The presence of electromagnetic coupling of dark matter with sub-GeV masses will give rise to signals in the direct detection experiments like XENON 10 \cite{2011PhRvL.107e1301A}, XENON 100 \cite{2016PhRvD..94i2001A, 2017PhRvD..96d3017E}, and Dark-Side \cite{2018PhRvL.121k1303A} which measure the atomic ionization rate due to energy transfer from dark matter to electron. Other experiments such as SENSEI (protoSENSEI@surface \cite{2018PhRvL.121f1803C} and protoSENSEI@MINOS \cite{2019PhRvL.122p1801A}) and CDMS-HVeV \cite{2018PhRvL.121e1301A} search for signals due to transition of electrons from the valence band to the conduction band of the semiconductor targets. We derive constraints on the parameters of our model using the existing limits on dark matter electron scattering cross-section from the direct detection experiments mentioned above, in particular from figure 4 of \cite{2019JCAP...09..070E}. 


	Even though we consider dark matter to be electrically neutral, it can still interact with electrically charged particles via different electromagnetic form factors arising from higher order electric and magnetic multipole moments \cite{2000PhLB..480..181P}. We begin by discussing the case of inelastic scattering between dark matter and electron that can cause $\chi$ to $\chi^*$ hyper-fine transition. Since the initial and final states are parity even (orbital quantum number $l = 0$), the leading order contribution comes from the magnetic dipole transition caused by the magnetic field generated by the electron. The relevant operator for the magnetic dipole interaction is given by eq. \eqref{C28} of Appendix \ref{app:C}. In order to derive the bounds in the parameter in our model we closely follow the formalism developed in \cite{2012PhRvD..85g6007E}, where the relevant cross-section is given in terms of a form factor $F_{\mathrm{DM}}$ and the cross-section $\bar{\sigma}$ evaluated at momentum transfer $q = \alpha m_e$.
	When the magnetic field is generated by the magnetic moment of the electron, the resultant form factor for the interaction is $F_{\mathrm{DM}} = 1$ and the cross-section $\bar{\sigma}$ is given by, 
	\begin{equation} 
		\bar{\sigma}_{1}^{\mathrm{IS}} = \frac{2\pi\epsilon^2\alpha^2\mu_{\chi e}^2}{\,m_\chi^2\,m_e^2},\label{5.1}
	\end{equation}
	where $\mu_{\chi e}$ denotes the reduced mass of dark matter electron system.
	On the other hand if the magnetic field is generated due to the motion of the electron, the resulting form factor is $F_{\mathrm{DM}} = \alpha m_e/q$ and the cross-section $\bar{\sigma}$ is given by,
	\begin{equation}
		\bar{\sigma}_{2}^{\mathrm{IS}} = \frac{8\pi\epsilon^2v_{e\chi}^2\mu_{\chi e}^2}{\,m_\chi^2\,m_e^2},\label{5.2}
	\end{equation}
	where $v_{e\chi}$ is the relative velocity of electron w.r.t. dark matter. Typical dark matter velocity in our solar neighborhood is $v_\chi \sim$ 220 km/s $\sim 10^{-3}$c. In case of Xenon, the electron velocity in the outermost shell is related to the ionization energy of the atom $E_{i}$ by the relation  $v_e = \sqrt{E_{i}/m_e} \sim \alpha c$. In cooled semiconductors, typical kinetic energy of electrons is due to their thermal motion $\sim$ 0.01 eV \cite{2018PhRvL.121f1803C}, resulting in $v_e \sim 10^{-4}\text{c}$. Thus relative velocity between dark matter and electron can be approximated by the dark matter velocity $v_{e\chi} \sim v_\chi \sim 10^{-3}$c. Inelastic scattering is kinematically allowed when the kinetic energy of initial dark matter state is greater than the electron transition energy + energy splitting between $\chi$ and $\chi^*$. Thus we assume that inelastic scattering only takes place when the $\chi-\chi^*$ mass splitting is less than the dark matter kinetic energy. In case of underground detectors like XENON 10 \cite{2011PhRvL.107e1301A}, XENON 100 \cite{2016PhRvD..94i2001A, 2017PhRvD..96d3017E}, and Dark-Side \cite{2018PhRvL.121k1303A}, the electron transition energy is equal to the ionization energy of Xenon or Neon $\sim$ 12 eV. On the other hand for detectors with semi-conductor targets such as SENSEI (protoSENSEI@surface \cite{2018PhRvL.121f1803C} and protoSENSEI@MINOS \cite{2019PhRvL.122p1801A}), the electron transition energy is equal to the bandgap of the semiconductor $\sim$ 1 eV. 

	For dark matter transition energies greater than its kinetic energy, the dominant interaction between dark matter and electron comes from elastic scattering due to the charge radius operator given in eq. \eqref{C32} of Appendix \ref{app:C}. This interaction also has a form factor $F_{\mathrm{DM}} = 1$ and the cross-section $\bar{\sigma}$ is given by,
	\begin{equation}
		\bar{\sigma}_e^{\mathrm{ES}} = \frac{4\pi\epsilon^2\alpha^2\mu_{\chi e}^2}{3 m_\chi^4}.\label{5.3}
	\end{equation}

\begin{figure}
	\centering
	\includegraphics[width=0.65\linewidth]{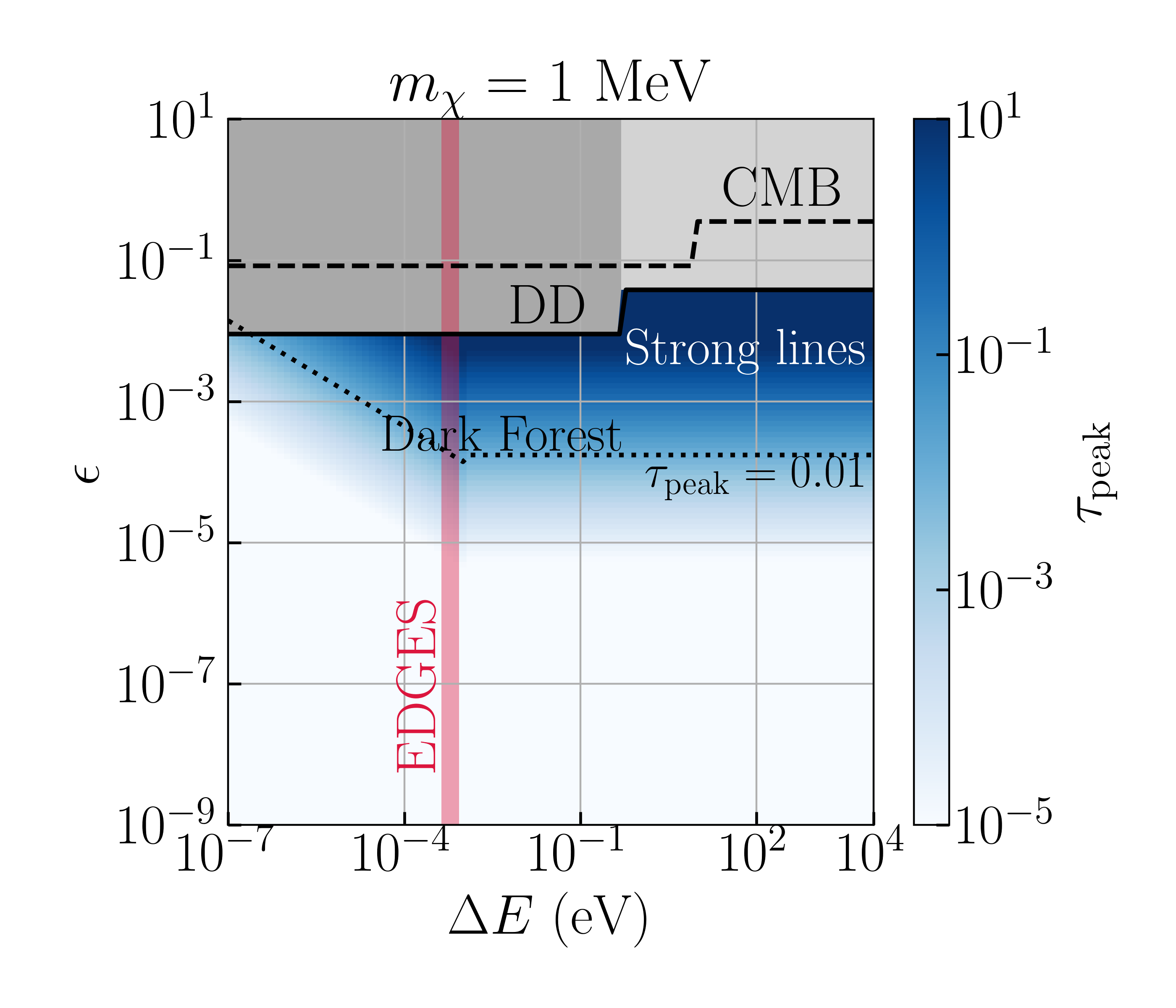}
	\caption{
		The allowed parameter space for $\epsilon$ for dark matter transition energy ranging from radio ($10^{-7}$ eV) to X-rays ($10$ keV) and for dark matter mass $m_\chi = 1$ MeV. The black solid and dashed lines refer to direct detection (DD) \cite{2011PhRvL.107e1301A,2016PhRvD..94i2001A, 2017PhRvD..96d3017E, 2018PhRvL.121k1303A,2018PhRvL.121f1803C, 2019PhRvL.122p1801A, 2018PhRvL.121e1301A} and CMB constraints from dark matter electron scattering. The regions where inelastic scattering ($\chi-\chi^*$ transition) dominates for direct detection is shown by dark gray color and the regions where elastic scattering (charge radius) dominates for direct detection is shown by light gray color. While the dark forest is allowed in the whole electromagnetic spectrum, the region above the red dotted line ($\tau_{\text{peak}}=0.01$) will give rise to significant fraction of strong absorption lines (optical depths $\gtrsim 0.01$) in the dark forest (see figure \ref{fig:tau_min}). The color shade represents the value of optical depth $\tau_{\text{peak}}$ (see eq. \eqref{5.5o} for definition) which becomes darker for larger energy splittings and larger electric charge of dark quarks (see eq. \eqref{C29}). 
		The overlap of crimson band with $\tau_{\text{peak}}\gtrsim0.01$ region corresponds to the parameter space where the absorption of the CMB by dark matter is consistent with the EDGES data (see figure \ref{fig:dd_edges} for more details). }
	\label{fig:dd}
\end{figure}
\begin{figure}[t]
	\centering
	\includegraphics[width=0.48\textwidth]{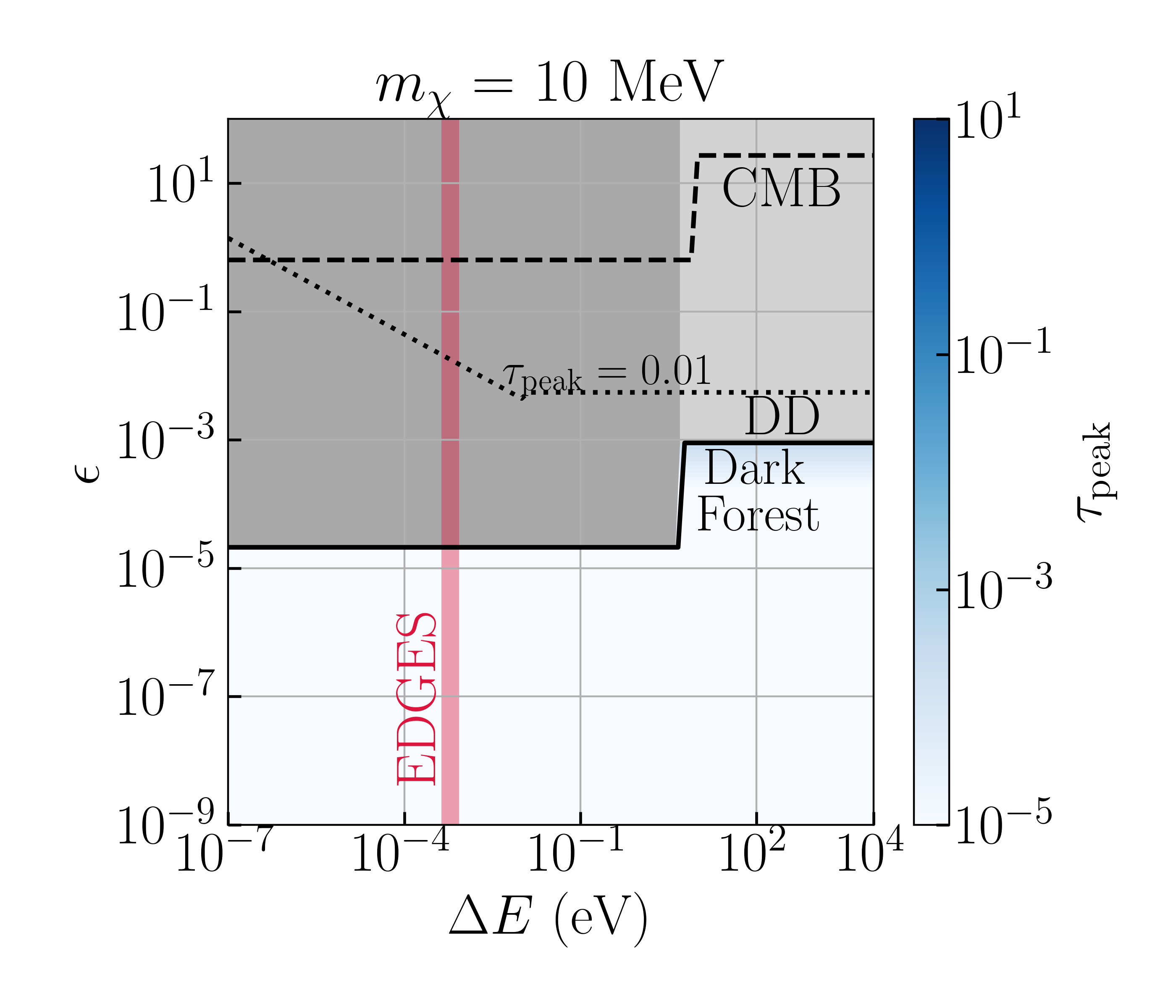}
	\includegraphics[width=0.48\textwidth]{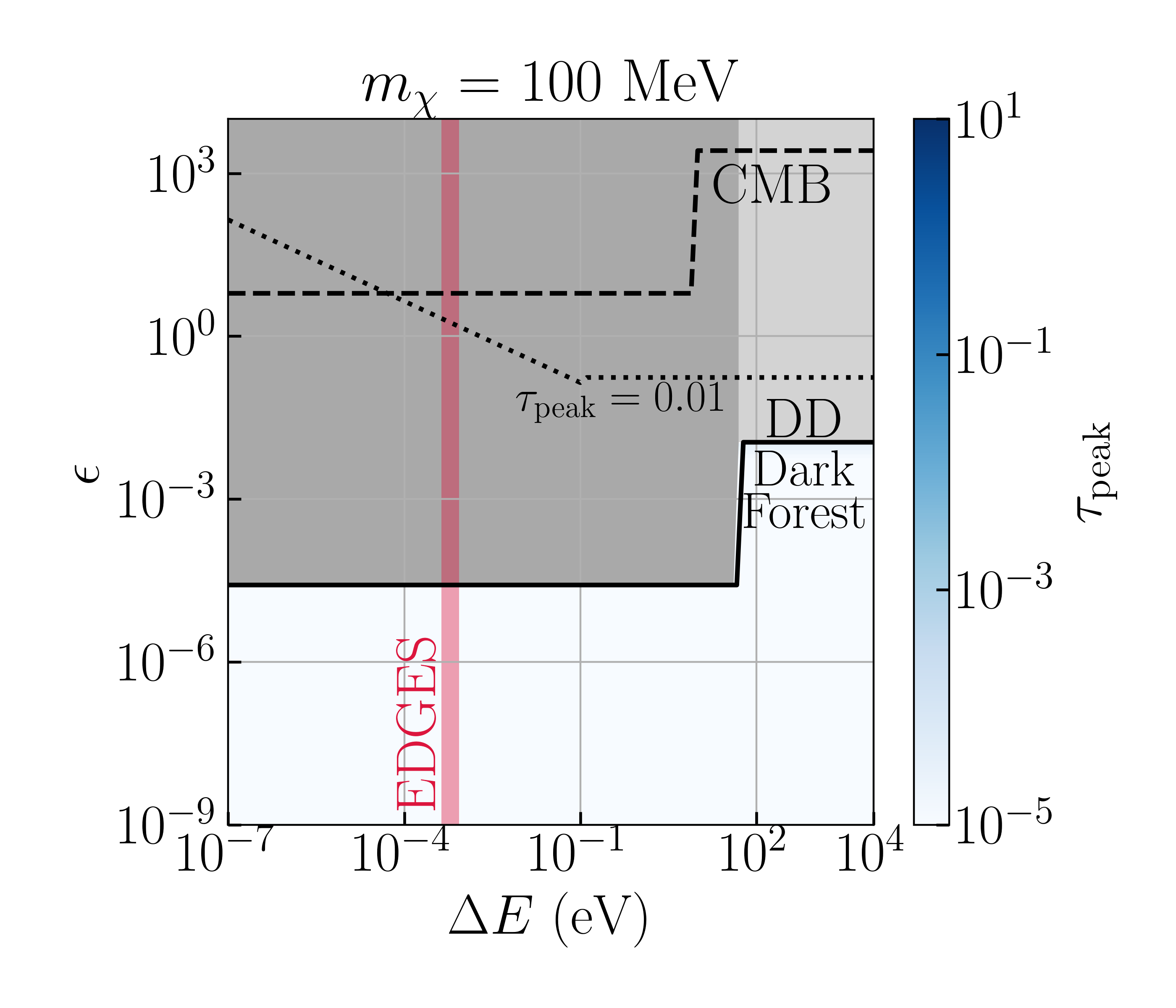}
	\caption{
		Constraints for dark matter masses $m_\chi = 10$ MeV, and $m_\chi = 100$ MeV respectively from left to right. The other details of the figure are same as mentioned in the caption of figure \ref{fig:dd}.}
	\label{fig:dd2}
\end{figure}
\begin{figure}[t]
	\centering
	\includegraphics[width=0.7\textwidth]{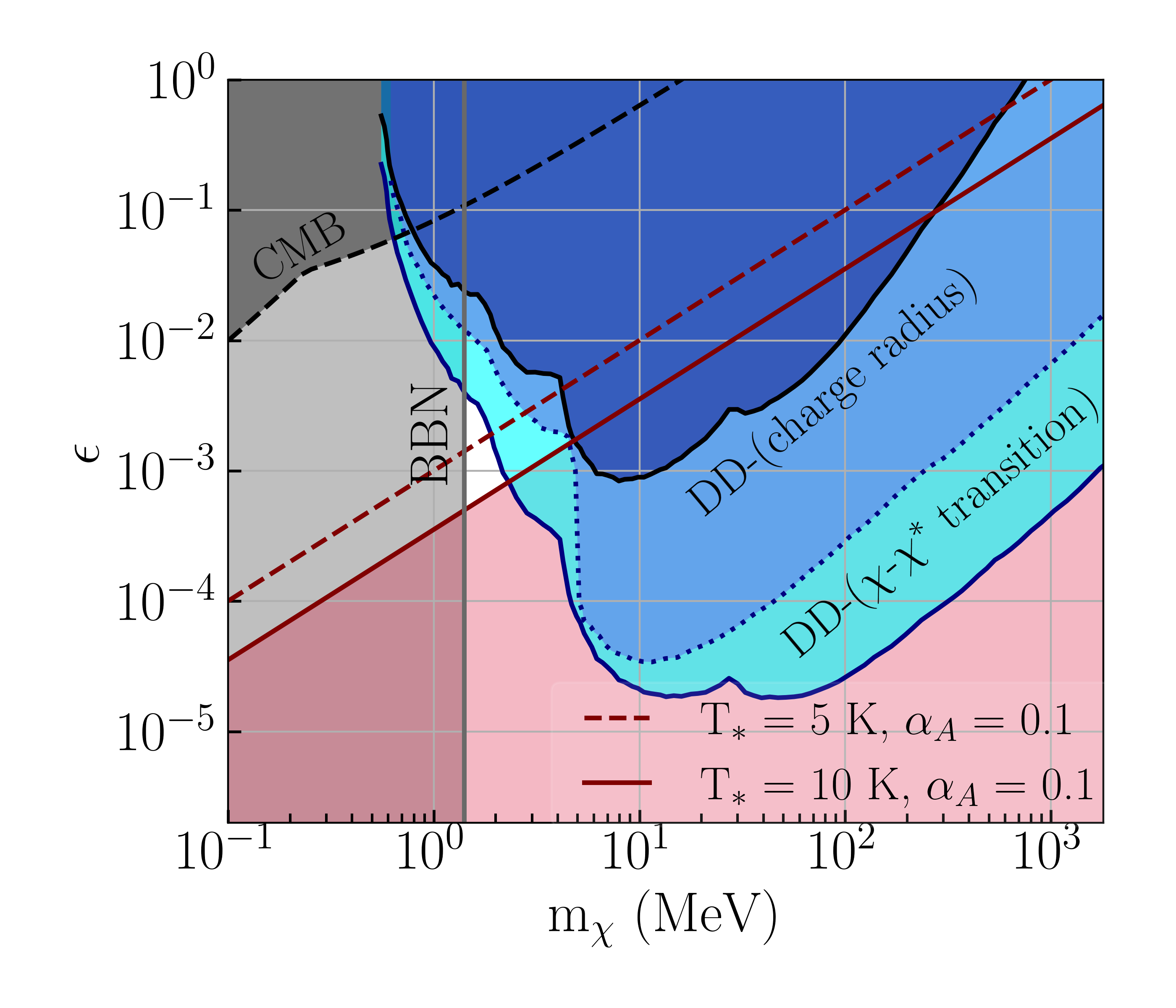}\hfill
	\caption{The allowed parameter space for $\epsilon$ for dark matter masses ranging from 100 keV to 2 GeV. The blue colored regions are disallowed by the direct detection (DD) experiments \cite{2011PhRvL.107e1301A,2016PhRvD..94i2001A, 2017PhRvD..96d3017E, 2018PhRvL.121k1303A,2018PhRvL.121f1803C, 2019PhRvL.122p1801A, 2018PhRvL.121e1301A}. The dark blue solid and dotted lines represent the direct detection constraints from dark matter inelastic scattering ($\chi-\chi^*$ transition) given in eq. \eqref{5.1} and \eqref{5.2} respectively. The elastic scattering (charge radius interaction from eq. \eqref{5.3}) is shown by the solid black line. The black dashed line and dark grey region above is disallowed by CMB constraints (see eq. \eqref{cmb}) from dark matter electron scattering. We also show the BBN constraint on thermal dark matter at $\sim$ 1.4 MeV from \cite{2020JCAP...01..004S} (for detailed discussion see \ref{BBN}). The white and crimson regions are the allowed regions of the parameter space by the CMB and direct detection experiments. The solid red and dashed red lines refer to $\alpha_A=0.1$ for dark matter transition temperatures $T_{*} = 5$ K and $T_{*} = 10$ K respectively. The white region above $\alpha_A=0.1$ is the allowed parameter space for our dark matter model which can explain the EDGES anomaly. The crimson region corresponds to $\alpha_{A}<0.1$ where the global absorption signal becomes too weak to be detected with EDGES.}
	\label{fig:dd_edges}
\end{figure}
We show the direct detection and CMB limits (as described by eq. \eqref{cmb}, \eqref{4.3}, and \eqref{4.4a} in section \ref{sec:paramspace}) on $\epsilon$ from dark matter electron scattering over a wide range of dark matter transition energies ($\chi-\chi^*$ mass splitting). This constraint is shown for dark matter mass $m_\chi = 1$ MeV in figure \ref{fig:dd} and for $m_\chi = 10$ MeV, and $100$ MeV in figure \ref{fig:dd2}. We represent the regions dominated by elastic scattering using light gray and regions dominated by inelastic scattering using dark gray color.

The optical depth represents the fraction of the background intensity absorbed in an dark line. To understand the implications of these constraints on the strength of absorption lines in the dark forest as a function of dark matter model parameters: millicharge $\epsilon$, mass $m_\chi$, and transition energy $\Delta E$, we use the peak optical depth of the absorption line (defined as the maxima of the optical depth profile i.e. $\tau(\nu_h=\nu_0, p)$ in eq. \eqref{4.6o}), which is given by,
\begin{equation}
	\tau_{\text{peak}}
	=\int_{-d_{s}}^{\,d_{o}}\frac{g_1}{g_0}\frac{\alpha_A A_{10}^{\text{HI}}\,c^{2}}{8\pi\nu_{0}^{2}}\frac{\rho\left(r\right)}{m_{\chi}}\frac{1}{\sqrt{\pi}\Delta \nu_D}\left(\frac{1-e^{-\frac{T_{*}}{T_{\text{ex}}\left(r\right)}}}{1 + \left(g_{1}/g_{0}\right)e^{-\frac{T_*}{T_{\text{ex}}\left(r\right)}}}\right)ds,\label{5.4o}
\end{equation}
where the factor $1/\left(\sqrt{\pi}\Delta \nu_D\right)$ comes from the amplitude of the line profile defined in eq. \eqref{3.2o}. 

The dependence of $\tau_{\text{peak}}$ on millicharge $\epsilon$ comes from the spontaneous emission coefficient $\tau_{\text{peak}}\propto \alpha_{A} \propto \epsilon^2$ (see eq. \eqref{C29}).
The peak optical depth is sensitive to the ratio of transition temperature to the excitation temperature ($T_*/T_{\text{ex}}$). Thus we have two cases, when $T_*\gg T_{\text{ex}}$, we can approximate the term in the numerator $\left(1 - e^{-T_*/T_{\text{ex}}}\right) \approx 1$. Since $\Delta \nu_D \propto \nu_0$ and $\alpha_{A}\propto \nu_0^3$, they cancel the $\nu_0^2$ in the denominator, making the peak optical depth independent of dark matter transition energy. Further, since $\alpha_{A}\propto 1/m_\chi^2$, the dependence of peak optical depth on the dark matter mass goes as $\tau_{\text{peak}}\propto \alpha_{A}/m_\chi \propto 1/m_\chi^3$. In the second case, $T_*\ll T_{\text{ex}}$, we can approximate the term in the numerator $\left(1 - e^{-T_*/T_{\text{ex}}}\right) \approx T_*/T_{\text{ex}}$. Since $T_*\propto \Delta E$, the peak optical depth scales linearly with transition energy, $\tau_{\text{peak}}\propto \Delta E$. In case of collisional dark matter, $T_\text{ex} = T_h\propto m_\chi$ (see eq. \ref{4.4oo}), the dependence of peak optical depth on dark matter mass goes as $\tau_{\text{peak}}\propto 1/m_\chi^4$. Putting everything together, we get,
\begin{equation}
	\tau_\text{peak} \propto
	\begin{cases}
		\frac{\epsilon^2\Delta E}{m_\chi^4} & T_*<T_{h},\\
		\frac{\epsilon^2}{m_\chi^3}   & T_*\geq T_{h}.
	\end{cases}
	\label{5.5o}
\end{equation}
In the mock dark forest spectra discussed in section \ref{sssec:df}, we took a specific case for dark matter transition energy $\Delta E = 5\times 10^{-4}\,$eV. We found that at least one strong line having $\tau_{\text{peak}} \approx 0.01$ is expected for $\alpha_{A}/m_{\chi} \gtrsim 0.01\,\text{MeV}^{-1}$ as shown in figure \ref{fig:tau_min}. We use eq. \eqref{5.5o} to scale the result in figure \ref{fig:tau_min} to different sets of dark matter model parameters. 

We represent the value of $\tau_{\text{peak}}$ in figures \ref{fig:dd} and \ref{fig:dd2} by blue color, darker shade highlighting the regions of stronger absorption lines.  The region above the black dotted line is where we expect strong lines in the dark forest with optical depth $\tau_{\text{peak}} \gtrsim 0.01$.
We should be able to probe the strong line region with current experiments and either detect a signature of dark forest or put much stronger bounds on dark matter parameters than what is expected from direct detection experiments. We find that a wide range of parameter space of our model can give rise to dark forest features within the sensitivity of existing experiments from radio to X-rays. In particular, strongest lines are expected from X-ray observations and the existing data will allow us to put much stronger constraints on $\epsilon$ compared to direct detection experiments. We leave a detailed analysis for future work.

In figure \ref{fig:dd_edges}, we show the limits on $\epsilon$ for transition energies $5\lesssim T_*\mathrm{(K)} \lesssim 10$, where our model can produce global signals consistent with EDGES (see figure \ref{fig:cp}). The constraints from direct detection are shown in blue. The solid red and dashed red lines refer to $\alpha_A=0.1$ for dark matter transition temperatures $T_{*} = 5$ K and $T_{*} = 10$ K respectively. Since we require $\alpha_{A}>0.1$ to be consistent with EDGES  (see figure \ref{fig:cp}), we shade the disallowed regions  ($\alpha_{A}<0.1$ for $T_* = 10$ K) crimson.  We find that dark matter masses $m_\chi \lesssim$ 3 MeV are allowed by direct detection and CMB constraints as shown by the white region represented by the label EDGES.

	\color{black}
	\section{Conclusions}
	\label{sec:conclusion}
	In this work, we explore the unique experimental signature of a class of composite dark matter models having electromagnetic transitions: the absorption lines in the spectrum of a background source. Such absorption signatures occur as distinct dark lines in the spectrum of a quasar, and as a global absorption feature or spectral distortion in the spectrum of the CMB.
	\begin{itemize}[leftmargin=10pt]
		\item The absorption of light by multiple dark matter halos along the LoS to a quasar gives rise to a dark forest, similar to the Lyman-$\alpha$ forest from neutral hydrogen. In contrast to the Lyman-$\alpha$ or 21 cm forest, the dark forest is less prone to the uncertainties related to the non-linear baryonic physics.
		
		\item The dark forest is sensitive to the self-interactions of dark matter. The dark forest is also sensitive to the minimum halo mass and therefore the primordial  power spectrum on small scales. 
		
		\item Absorption of CMB photons by dark matter gives rise to $\mu$-type spectral distortions at $z\gtrsim 10^5$ and a global absorption feature in the CMB spectrum at $z\lesssim 10^5$. 
	
		\item The bremsstrahlung process plays an important role if the dark matter absorption frequency falls in the Rayleigh-Jeans tail of the CMB spectrum. In this case, bremsstrahlung determines the high redshift edge of the global absorption feature. The low redshift shape of the global signal is determined by the relative strength of inelastic collisional coupling with respect to the radiative coupling of dark matter.
		
		\item For dark matter masses of $\mathcal{O}$(MeV), our model can produce a global absorption feature which is consistent with the EDGES observation \cite{2018Natur.555...67B}.
             
        \item If the EDGES signal is indeed produced by dark matter absorption, then our model predicts that a dark forest must exist in a frequency band ten times higher than the 21 cm forest frequency band.    
		
		\item A global absorption feature or spectral distortion due to dark matter is a general prediction of our model which can even occur in a different frequency band of the CMB  spectrum. Such signals could be detected in future experiments such as Primordial Inflation
        Explorer (PIXIE) \cite{2011JCAP...07..025K} or APSERA \cite{2015ApJ...810....3S}. 
		
		\item The absorption of CMB photons in the UV band can affect the recombination history and this part of the parameter space can be probed by the CMB anisotropy measurements. 
	
		\item There is a large volume of parameter space where dark forest is more sensitive compared to the best current and planned direct detection experiments.

	\end{itemize}
	We have ignored the clustering of halos in this work. One needs to perform N-body simulations to take into account the two point correlations of the dark forest which we leave for future work. 

	Our results open up a new and unexpected interesting window into the nature of dark matter and a potential new probe of the cosmic web. Our work motivates the search for dark forest and global absorption features across the full electromagnetic spectrum, especially in the desert i.e. the part of the electromagnetic spectrum where there are no strong lines expected from baryons in the Standard Model of cosmology. With the current sensitivity of spectroscopic surveys in different frequency bands, it should already be possible to probe the dark forest at optical and X-ray frequencies. 

	\acknowledgments
	We acknowledge the computational facilities offered by the Department of Theoretical Physics at TIFR, Mumbai. This work is supported by Max Planck Partner Group for cosmology of Max Planck Institute for Astrophysics Garching at TIFR funded by Max-Planck-Gesellschaft. We acknowledge the support from Department of Atomic Energy (DAE) Government of India, under Project Identification No. RTI 4002. We are thankful to Yidong Xu, Steven Furlanetto, Brian Fields, Girish Kulkarni, Shadab Alam, Nissim Kanekar, Aritra Kumar Gon and Shikhar Mittal for useful discussions. 
	
	\appendix
	\section{Dark matter model}\label{app:C}
	By construction, the dark sector model is described by an asymptotically free non-abelian SU(N) gauge theory. At UV energies, the theory is described by weakly coupled dark quarks. The necessary d.o.f. are given in table \ref{tab:ppm} in section \ref{sec:model}. At low energies, SU(N) is strong and spectra of the dark sector is given in terms dark color neutral bound states. The formulation we discuss here closely follows the chiral Lagrangian techniques developed in particle physics \cite{Wise:1992hn,Yan:1992gz,1997PhR...281..145C}. One essential difference though is that there is no analogous weak force present in the dark sector which allows the heavy-light bound dark matter states to decay in comparison to the heavy mesons of the visible sector. We begin this section with the spontaneous breaking of the chiral symmetry of the dark sector. At low energies, the strong dark QCD generates the flavor diagonal condensate $\langle q_{Da}\,q_{Db}^{c} \rangle = \delta_{ab}\,\Lambda_{N}^3$. The flavor symmetry gets spontaneously broken in the following fashion,
	\begin{equation}
		SU(2)^D_{L}\times SU(2)^D_{R} \rightarrow SU(2)^D_{V}.
	\end{equation}
	This symmetry breaking results in three pNGBs/dark pions $\pi_D$ parameterized in eq.\eqref{25} in section \ref{sec:model}. In order to discuss the physics of heavy-light bound states, it is convenient to introduce a field $\xi_D$ defined via,
		\begin{align}
			\Sigma_D(x) = \xi_D^{2}(x);\hspace{2pt}\text{with}\hspace{5pt}
			\xi_D(x) \xrightarrow{SU(2)^D_{L}\times SU(2)^D_{R}} L_D\,\xi_D(x)\,V_D^{\dagger}(x) =  V_D(x)\,\xi_D(x)\,R_D^{\dagger},
		\end{align}
		where $V_D(x)$ is the transformation operator for the unbroken vectorial symmetry $SU(2)^D_{V}$. 
	
	We choose our dark matter candidate to be a heavy-light bound state.  To do this, we first redefine the heavy and light dark quark Dirac spinors in the following way,
		\begin{align}
			Q_{D} \equiv 
			\begin{pmatrix}
				Q_{Dw} \\ 
				Q^{c\,\dagger}_{Dw}
			\end{pmatrix},\,\,\,\,
			q_{D} & \equiv 
			\begin{pmatrix}
				\xi_D^{\dagger}q_{Dw} \\ 
				\xi_D\,q_{Dw}^{c\,\dagger}
			\end{pmatrix},
		\end{align}
		where $Q_{Dw}$ and $q_{Dw}$ are the Weyl spinors for the heavy and light dark quarks respectively.
	The dark matter candidate is defined by the matrix field $\mathcal{X}$ as,
	\begin{align}
		\mathcal{X}_{a} &\equiv \bar{Q}_{D}\,q_{Da} \xrightarrow{SU(2)^D_{L}\times SU(2)^D_{R}} V_{Dab}(x)\mathcal{X}_{b}.
	\end{align}
	In order to construct the operators responsible for interactions between dark bound state and dark pions, we define the vector current $\mathcal{V}_{D\mu}$ and the axial current $\mathcal{A}_{D\mu}$ which incorporate the dark pions $\pi_D$,
	\begin{align}
		\mathcal{V}_{D\mu} &= \frac{1}{2}(\xi_D^{\dagger}\partial_{\mu}\xi_D + \xi_D\partial_{\mu}\xi_D^{\dagger}),\nonumber\\
		\mathcal{A}_{D\mu} &= \frac{1}{2}(\xi_D^{\dagger}\partial_{\mu}\xi_D - \xi_D\partial_{\mu}\xi_D^{\dagger}).
	\end{align}
	Since the dark state $\mathcal{X}$ is a non-relativistic heavy-light bound state, its velocity $v$ is conserved. Using the framework of HQET, we can project the small scale fluctuations \cite{Isgur:1989vq, Georgi:1990um} and write the Lagrangian for $\mathcal{X}_v$ invariant under the full flavor symmetry $SU(2)_{L}^D\times SU(2)_{R}^D$, 
	\begin{align}
		\mathcal{L}_{\mathcal{X}_v} = i\,\text{tr}\big(\bar{\mathcal{X}}_{v}^{a}\,v^{\mu}D_{\mu}^{ab}\,\mathcal{X}_{v}^{b}\big) + ig_1\,\text{tr}\big(\bar{\mathcal{X}}_{v}^{a}\,\gamma^{\mu}\gamma^{5}\,\mathcal{A}_{\mu}^{ab}\mathcal{X}_{v}^{b}\big) + \frac{1}{4}f^{2}_{D}\,\partial_{\mu}\Sigma_{ab}\partial^{\mu}\Sigma^{\dagger}_{ba}\nonumber \\ 
		+ g_2\,f_{D}^2\Lambda_{N}m_{q}\delta_{ab}\,\Sigma_{ab} + g_{3}\,m_q\delta_{ab}\,\text{tr}\big(\bar{\mathcal{X}}_{v}^{a}\mathcal{X}_{v}^{b}\big)\,\xi_D\,\xi_D +\text{h.c.} + ... \hspace{3pt}, \label{ldm}
	\end{align}
	where the covariant derivative is defined as $D_{\mu} = \partial_{\mu} + \mathcal{V}_{D\mu}$ and the trace is taken over the gamma matrices.
	
	The Lagrangian in eq.\eqref{ldm} does not distinguish the spin 1 state $\chi^*$ from the spin 0 state $\chi$. The terms that violate the spin symmetry occur as powers in $1/M_{Q}$ and contribute to the mass gap between $\chi$ and $\chi*$ resulting in the hyper-fine splitting of the dark sector. The tree level operators that generate this mass gap are given by,
	\begin{equation}
		\frac{\lambda}{M_{Q}}\text{tr}\big(\bar{\mathcal{X}}_{v}^{a}\,\sigma^{\mu\nu}\,\mathcal{X}_{v}^{a}\,\sigma_{\mu\nu}\big); \hspace{3pt} \lambda'\frac{m_q}{M_{Q}}\text{tr}\big(\bar{\mathcal{X}}_{v}^{a}\,\sigma^{\mu\nu}\,\mathcal{X}_{v}^{a}\,\sigma_{\mu\nu}\big)\,\xi_D\xi_D \hspace{3pt}... \hspace{3pt}\text{etc.}\label{C27}
	\end{equation}
	We can decompose the matrix field $\mathcal{X}_v$ into a pseudo-scalar field and a vector field using eq.\eqref{dm} and compute the tree level hyper-fine splitting from eq.\eqref{C27}. Further, the axial current operator in eq.\eqref{ldm} gives rise to $\chi^*\chi^*\pi_D$ and $\chi\chi^*\pi_D$ couplings which also contribute to the hyper-fine splitting at the loop level.
	
	The operator that gives rise to photon transitions between $\chi$ and $\chi^*$ is given by,
	\begin{equation}
	\lambda''\frac{\bar{\epsilon} e}{4\tilde{m}_q}\text{tr}\big(\bar{\mathcal{X}}_{v}^{a}\,\sigma^{\mu\nu}\,\mathcal{X}_{v}^{a}\,F_{\mu\nu}\big),\label{C28}
	\end{equation}
	where $\lambda''$ is an unknown coefficient. The strength of this operator is proportional to the electric charge ($\epsilon e$) of the dark quarks and is suppressed by the constituent dark quark mass (say $\tilde{m}_q$). 

	In this work, for simplicity we rewrite the scale $\tilde{m}_q$ as $m_\chi/k$, so that the effective charge in this case would be replaced by $\epsilon e \rightarrow \epsilon e k$, where $k=m_\chi/\tilde{m}_q$. Further, for all phenomenological quantitative analysis in the main text, we refer to the effective strength of the operator, namely $\lambda''k\epsilon$, as $\epsilon$.
	
	An estimate for the spontaneous emission rate $\alpha_{A}$ (for definition see eq. \eqref{a10}) in terms of $\epsilon$ and the hyper-fine splitting of the dark sector ($\Delta E$) can be obtained by scaling the hyper-fine splitting parameters of the hydrogen atom,
	\begin{equation}
		\alpha_A \approx \epsilon^2 \left(\frac{\Delta E}{\Delta E_{\text{HI}}}\right)^3\left(\frac{m_e}{m_\chi}\right)^2,\label{C29}
	\end{equation}
	where $\Delta E_{\text{HI}}$ is the hyper-fine splitting in the hydrogen atom and $m_e$ is the mass of electron.
	\\	
	The operator (in eq. \eqref{C28}) responsible for $\chi - \chi^*$ transition will also give rise to inelastic scattering between dark matter and electron in direct detection experiments. Quantum mechanically, this scattering can be understood as arising due to the magnetic moment of dark matter interacting with the magnetic field generated by a electrically charged particle.

	Dark matter can also have elastic scattering due to its charge radius and the charge radius operator \cite{2000PhLB..480..181P} is given by,
	\begin{equation}
		\frac{\lambda'''\,\bar{\epsilon} e r_\chi^2}{6m_\chi}\text{tr}\big(\bar{\mathcal{X}}_{v}^{a}\,(\partial_{\mu}\mathcal{X}_{v}^{a})\,\partial_{\nu}F_{\mu\nu}\big) + \mathrm{h.c.},\label{C32}
	\end{equation}
	where $r_\chi$ is the charge radius of dark matter which we approximate here to be $r_\chi = 1/m_\chi$. In order to reduce the number of parameters, we also take the effective strength of the operator, namely $\lambda'''\bar{\epsilon}$, as $\sim\epsilon$.

	\section{Dark matter halo density and temperature profile}\label{app:CA}
	The halo mass density is modeled as an NFW profile \cite{1997ApJ...490..493N} with a concentration model taken from \cite{2001MNRAS.321..559B,2008MNRAS.391.1940M} to describe the halo up to 10 times the virial radius \cite{1998ApJ...495...80B} of the halo.
	The halo mass $M_h$ is defined as the total matter enclosed within a sphere around the halo center that encloses a fixed density equal to 200 times the critical density. Thus $r_{200c}$ is defined as,
	\begin{equation}
		\frac{M_{200c}(r<r_{200c})}{4\pi r_{200c}^3/3} = 200\rho_c(z)
	\end{equation}
	where halo mass $M_h = M_{200c} = \int_{0}^{r_{200c}}d^3r\,\rho(r)$.
	
	For simplicity, we assume the dark matter to follow a Maxwell-Boltzmann velocity distribution with a radially dependent rms width $\sigma$ described by the power law profile in \cite{2004MNRAS.352.1109A}, 
	\begin{equation}
		\frac{\rho}{\sigma^3}(r) = 10^{1.46}\bigg(\frac{\rho_c(z)}{v_{\text{vir}}^{3}}\bigg)\bigg(\frac{r}{r_{\text{vir}}}\bigg)^{-1.9}, \label{4.3oo}
	\end{equation}
	where $r$ is the radius of the halo in physical units, $\rho_c$ is the critical density of the Universe, $\rho(r)$ is the NFW density profile of the halo. $v_{\text{vir}}$ and $r_{\text{vir}}$ are the virial velocity and virial radius of the dark matter halo respectively. One can associate a temperature to the width of the velocity distribution using the relation, $\sigma^2(r) = 3k_BT_{h}(r)/m_\chi$, which gives,
	\begin{equation}
		T_h(r) = \frac{m_\chi}{3k}\bigg(\frac{10^{1.46}}{\rho(r)}\bigg(\frac{\rho_c(z)}{v_{\text{vir}}^{3}}\bigg)\bigg(\frac{r}{r_{\text{vir}}}\bigg)^{-1.9}\bigg)^{-2/3}. \label{4.4oo}
	\end{equation}
	The random motion of dark matter particles along the LoS gives rise to a Doppler line profile whose line width in units of speed is given by,
	\begin{equation}
		b(r) = c\frac{\Delta\nu_{D}}{\nu_0} = (1+v_{\text{LoS}}(r, p)/c)\sqrt{\frac{2k_BT_h(r)}{m_\chi}}.\label{C.4}
	\end{equation}
	\newpage
	\section{Dark forest} \label{app:D}
\begin{figure}[h!]
		\centering
		\includegraphics[width=\textwidth]{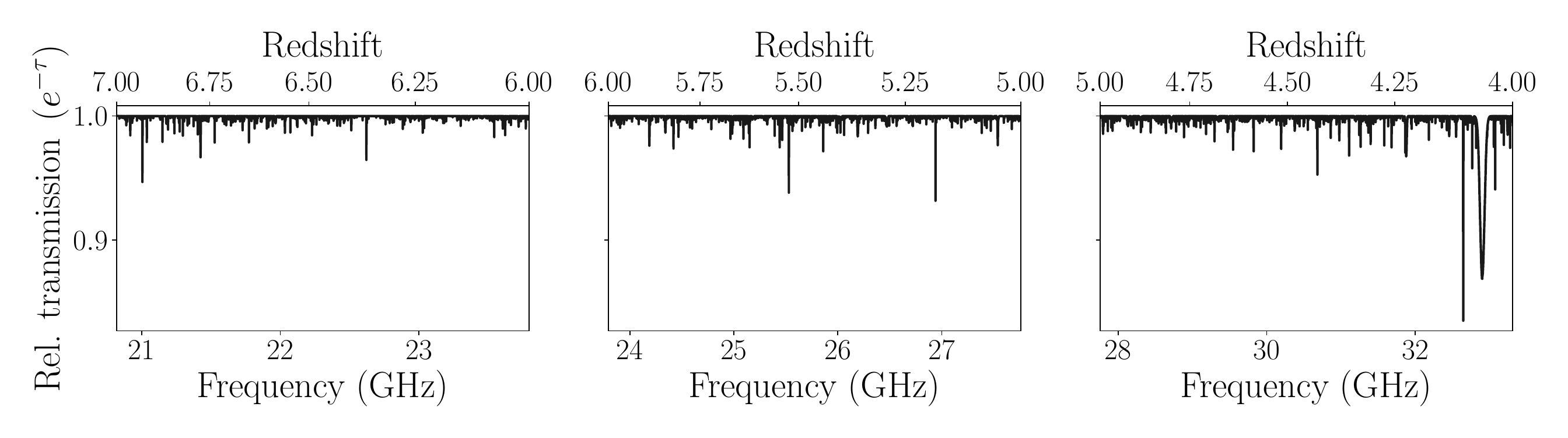}\\
		\includegraphics[width=\textwidth]{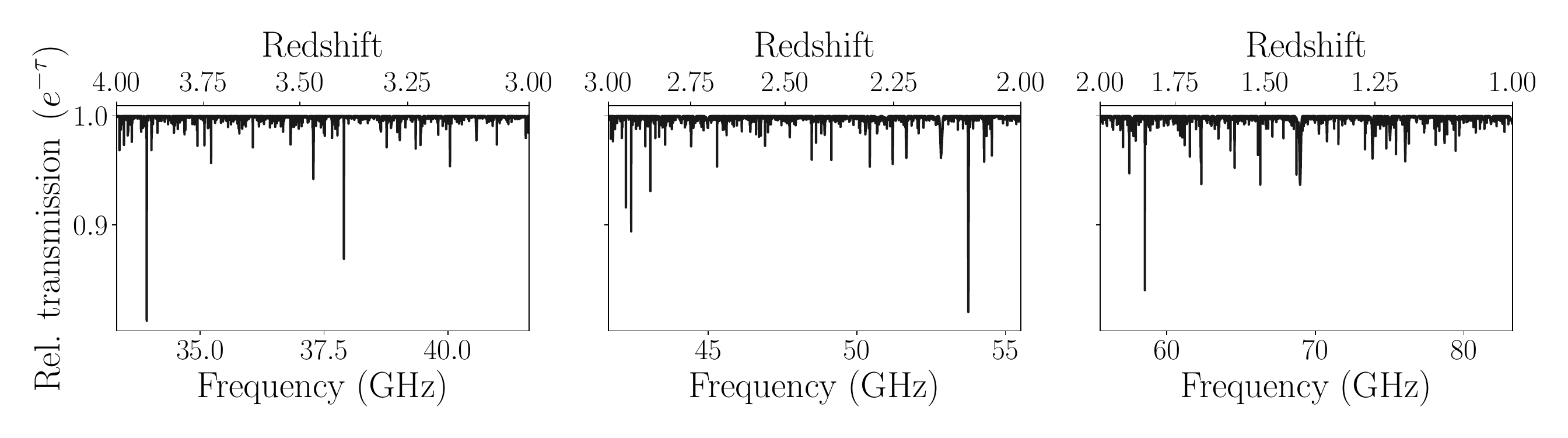}\\
		\includegraphics[width=0.375\textwidth]{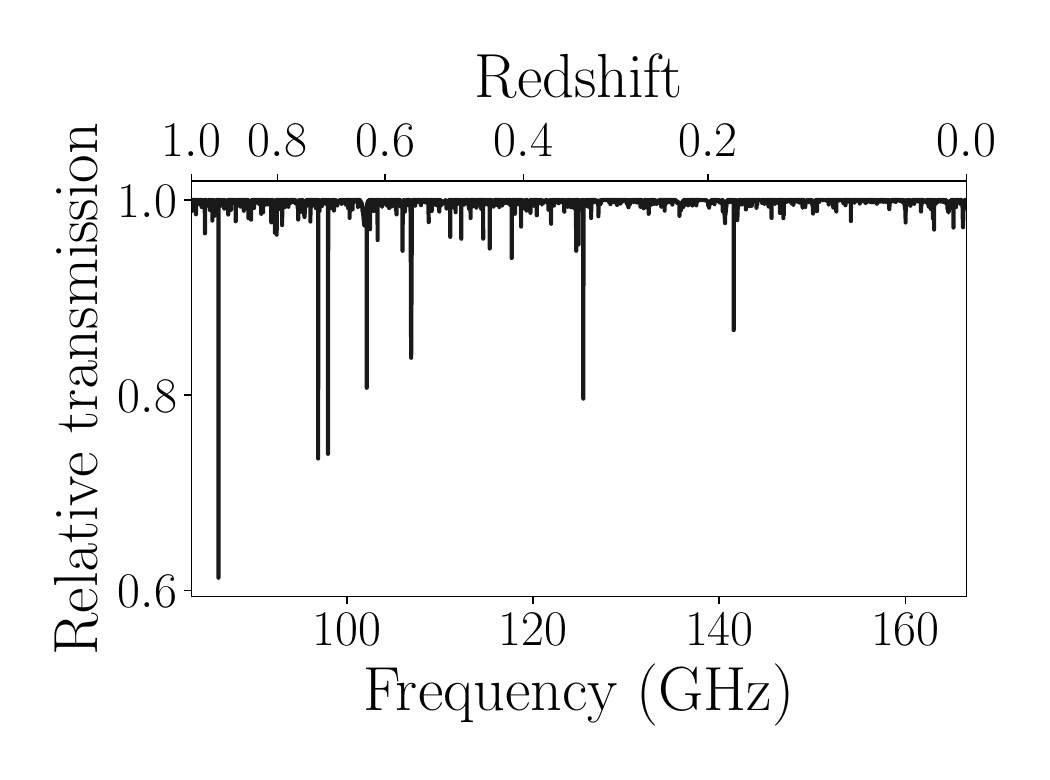}
		\caption{Dark forest spectrum for collisionless DM in for $M_{\text{min}} = 10^6 M_{\odot}/h$.}
		\label{fig:df_6_nc}
	\end{figure}

\begin{figure}[h!]
		\centering
		\includegraphics[width=\textwidth]{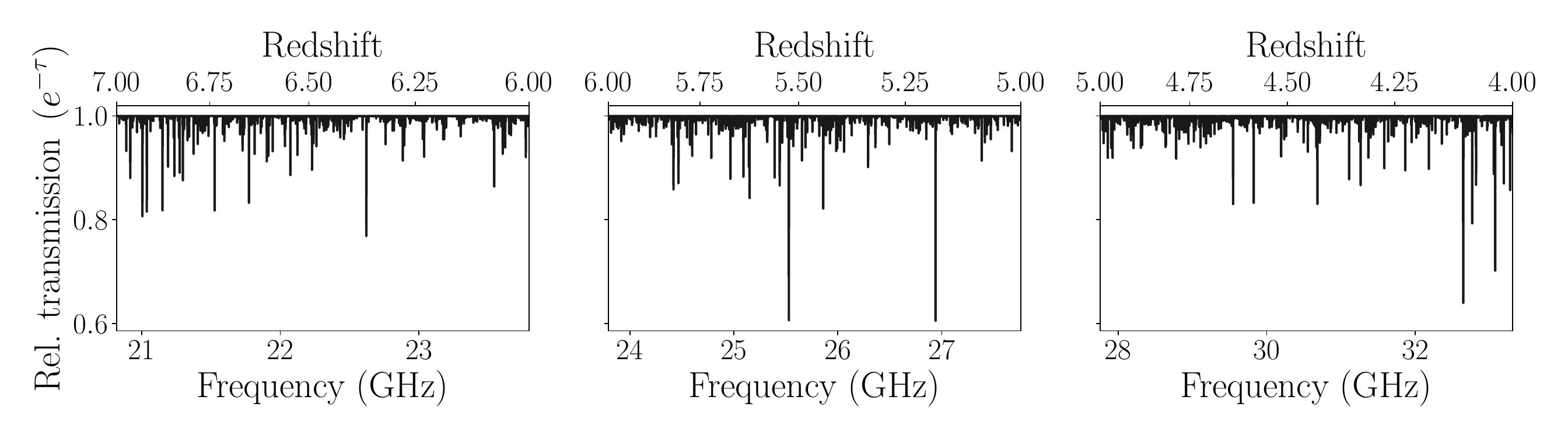}\\
		\includegraphics[width=\textwidth]{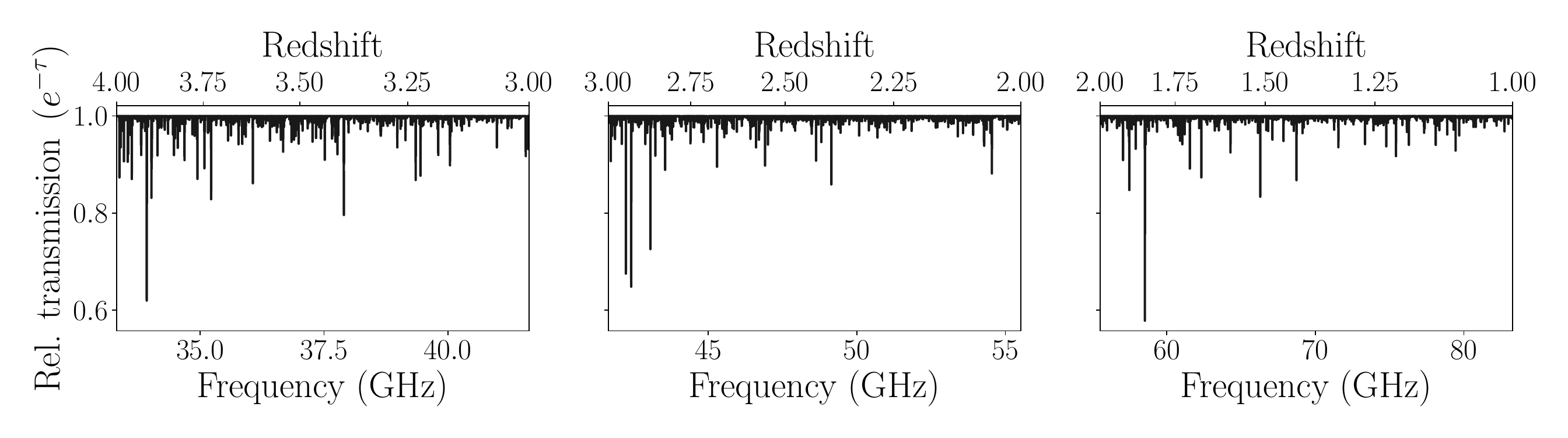}\\
		\includegraphics[width=0.375\textwidth]{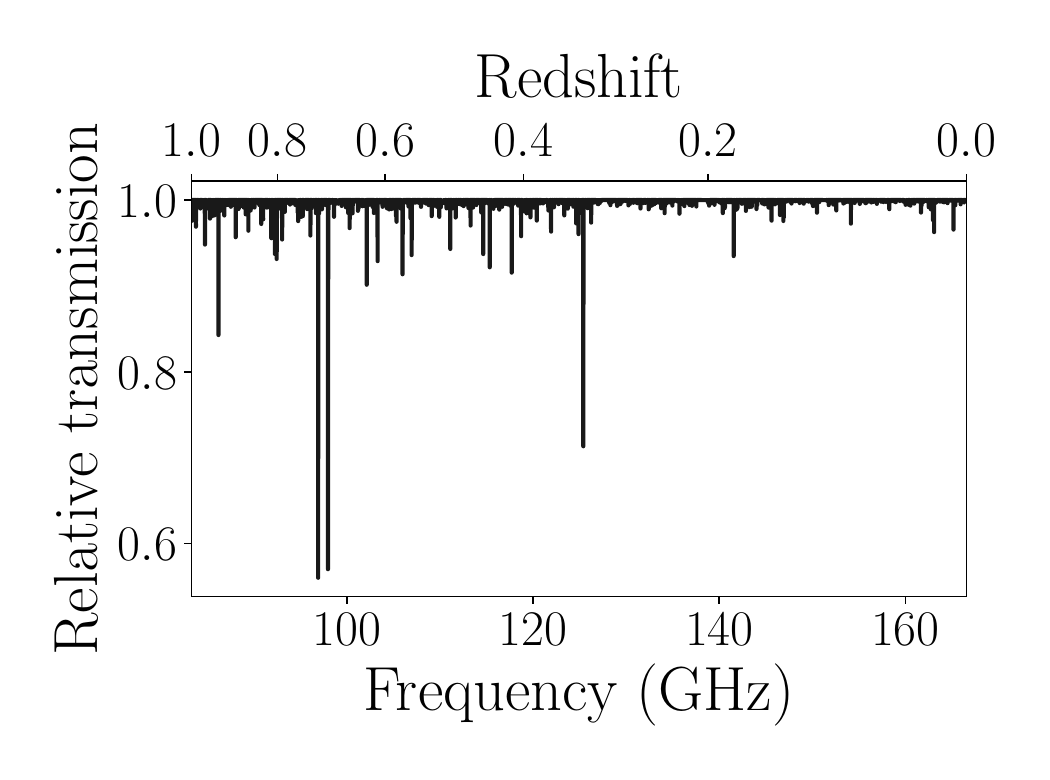}
		\caption{\centering Dark forest spectrum for collisional DM for $M_{\text{min}} = 10^6 M_{\odot}/h$.}
		\label{fig:df_6_c}
	\end{figure}

\begin{figure}[h!]
		\centering
		\includegraphics[width=\textwidth]{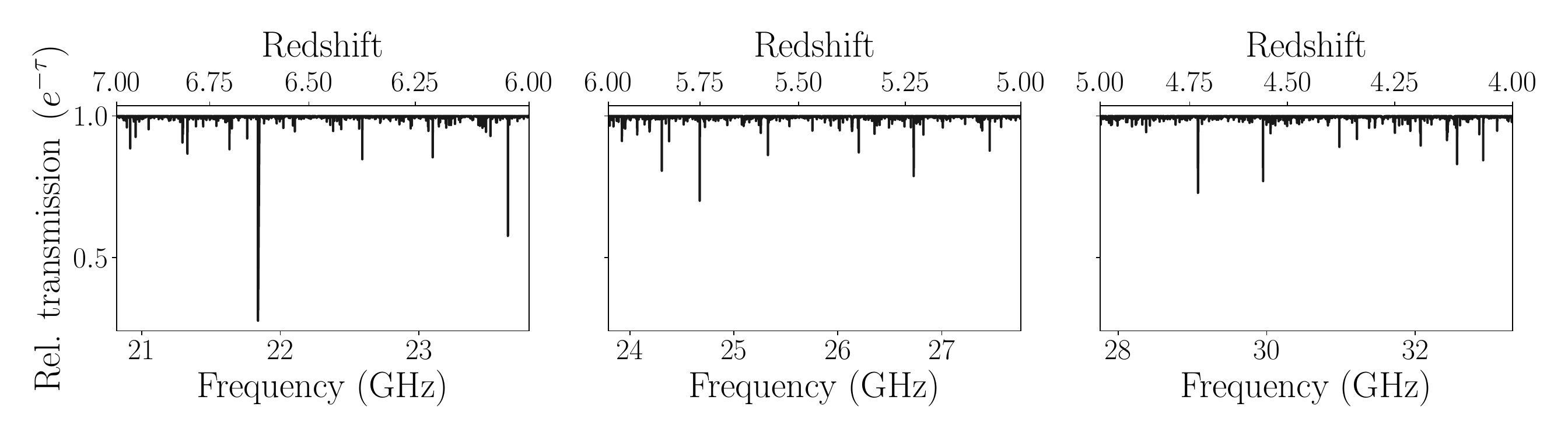}\\
		\includegraphics[width=\textwidth]{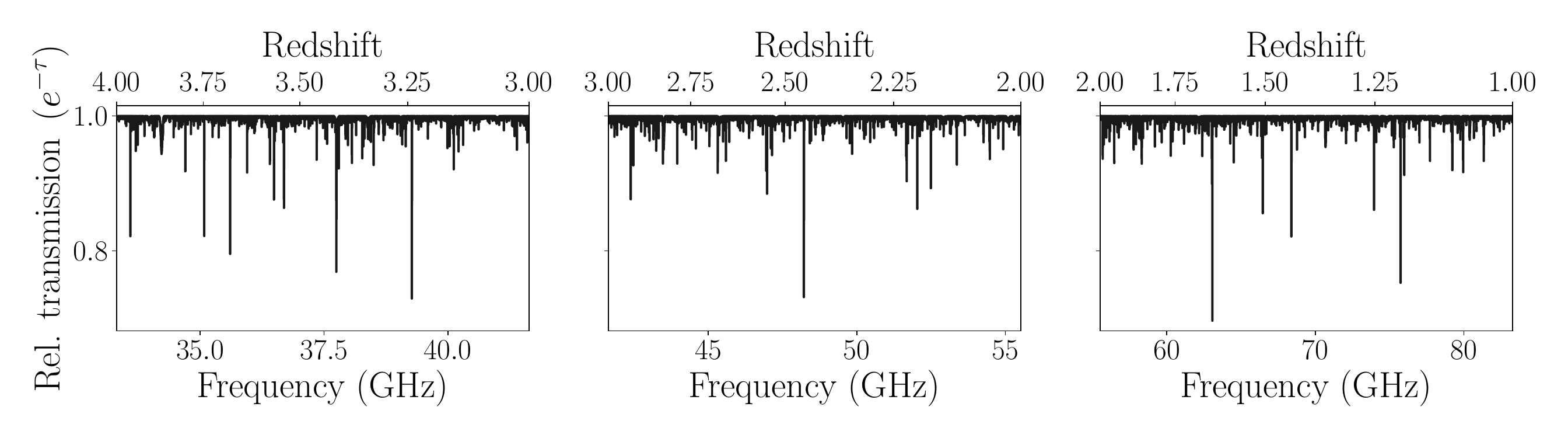}\\
		\includegraphics[width=0.375\textwidth]{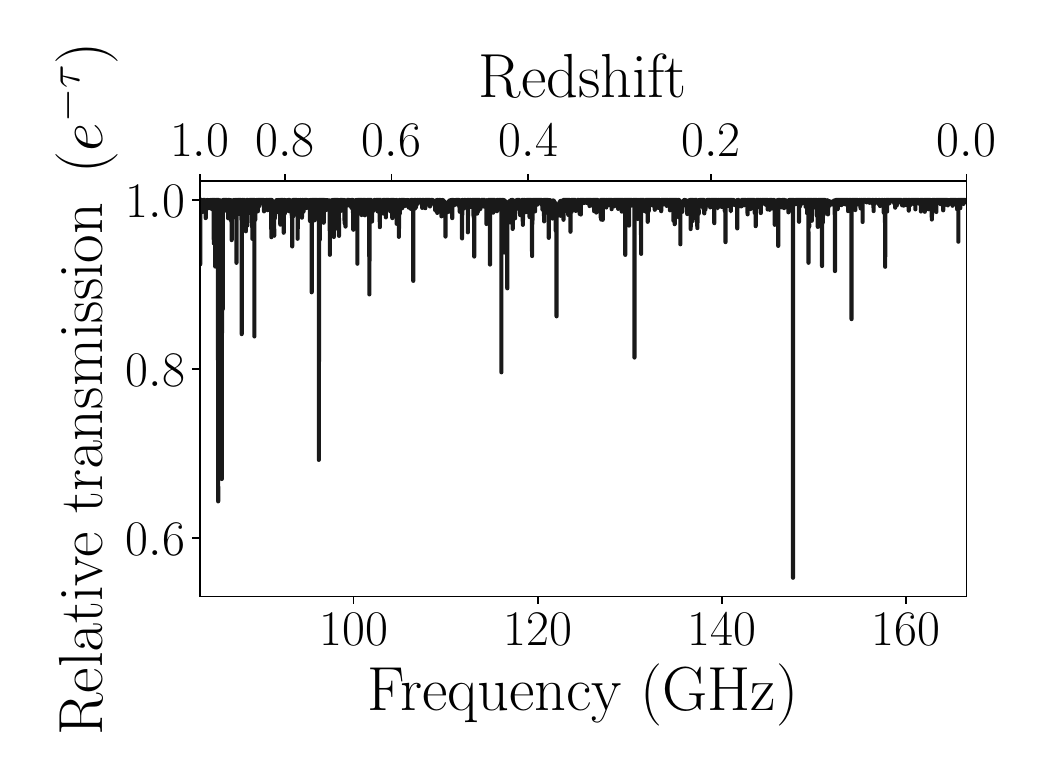}
		\caption{Dark forest spectrum for collisionless DM for $M_{\text{min}} = 10^4 M_{\odot}/h$.}
		\label{fig:df_4_nc}
	\end{figure}

\begin{figure}[t]
		\centering
		\includegraphics[width=\textwidth]{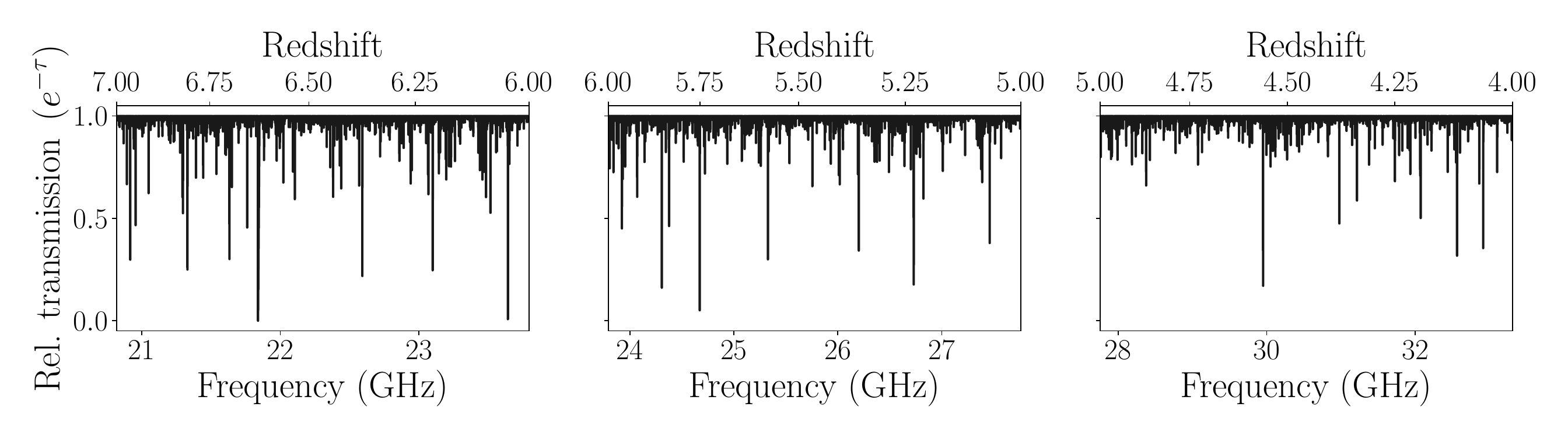}\\
		\includegraphics[width=\textwidth]{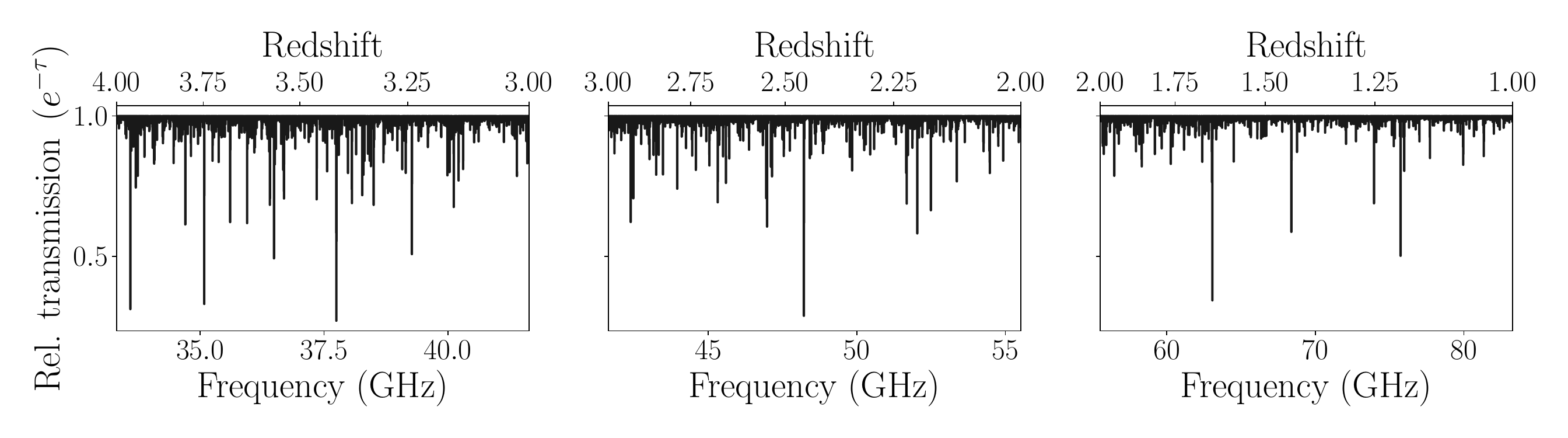}\\
		\includegraphics[width=0.375\textwidth]{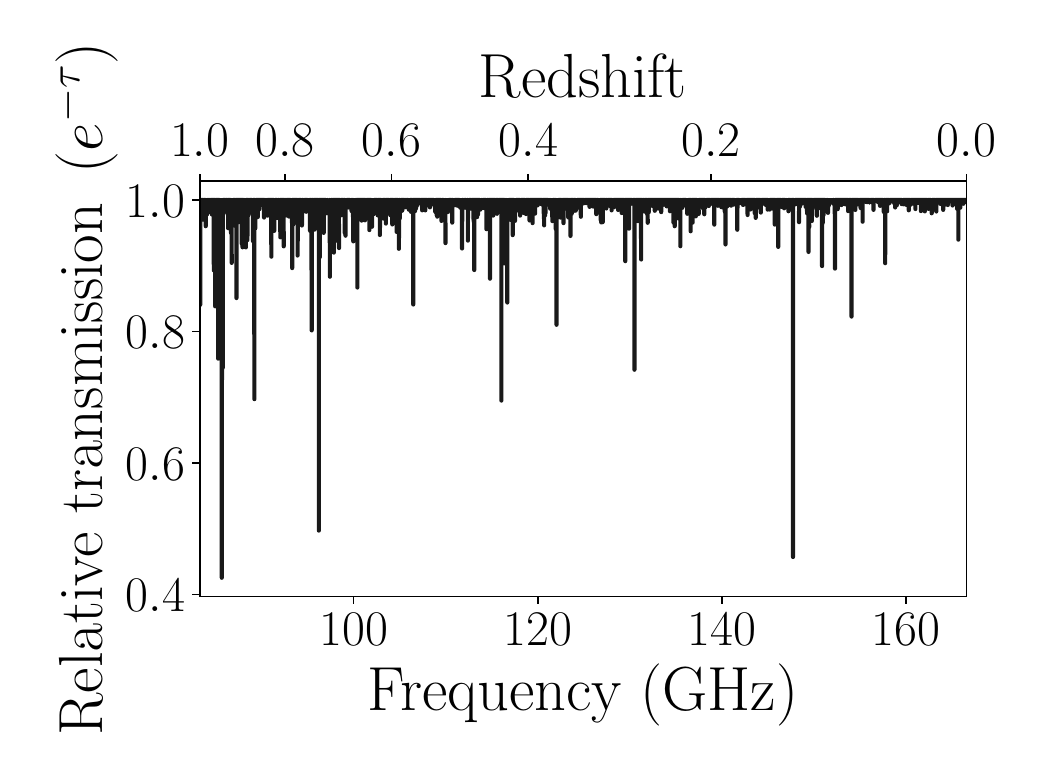}
		\caption{\centering Dark forest spectrum for collisional DM for $M_{\text{min}} = 10^4 M_{\odot}/h$.}
		\label{fig:df_4_c}
	\end{figure}

\begin{figure}[t]
		\centering
		\includegraphics[width=0.5\textwidth]{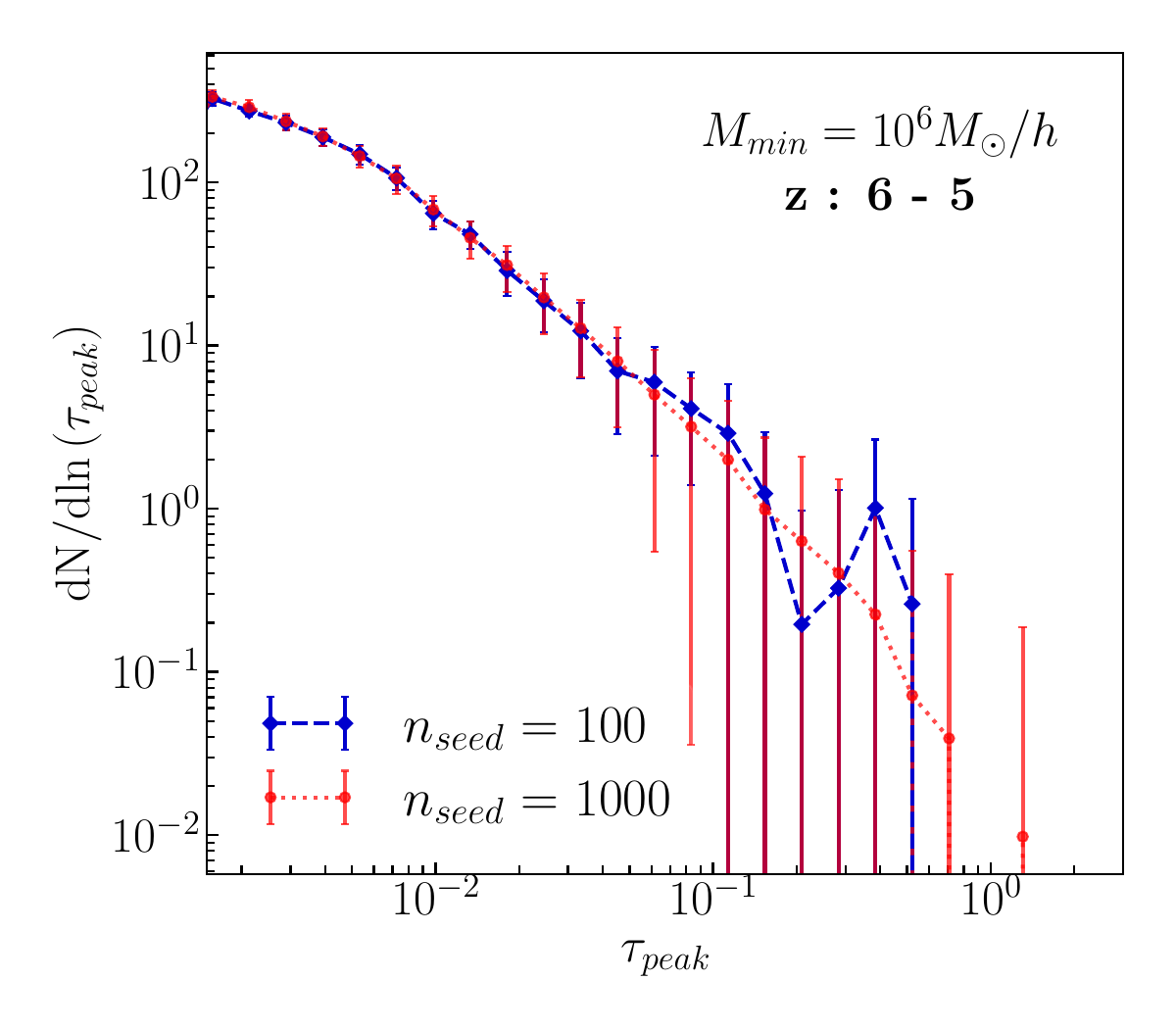}\hfill
		\includegraphics[width=0.5\textwidth]{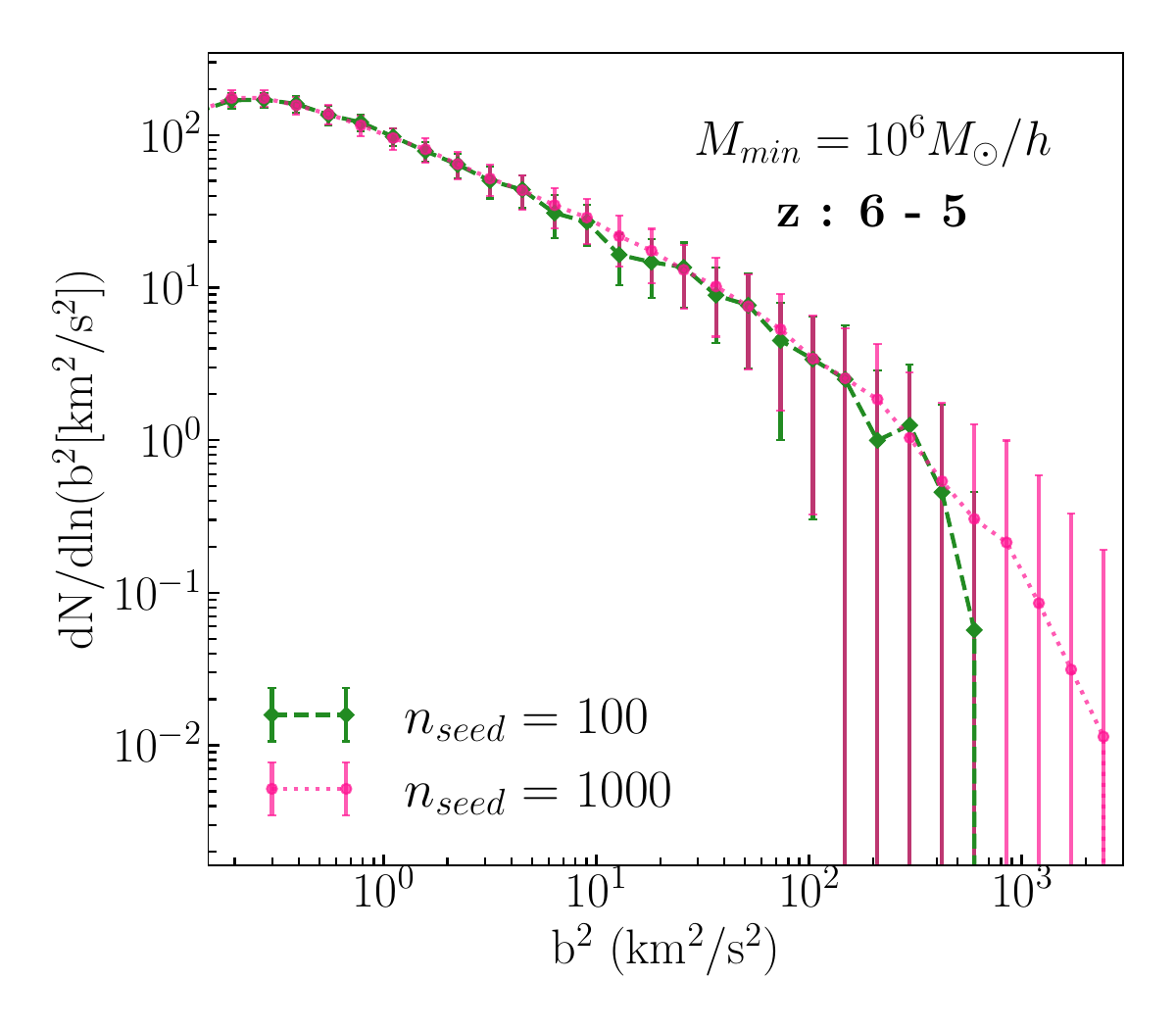}
		\caption{The distribution functions for peak optical depth (left) and line width (right) for collisionless DM for $M_{\text{min}} = 10^6 M_{\odot}/h$ in the redshift range 6-5 for 100 and 1000 LoS directions.}
		\label{fig:df_4v_c}
	\end{figure}

	We show the sample dark forest spectra in the redshift range 7-0 for collisionless and collisional DM for two different choices of $M_{\text{min}}= 10^6$ and $10^4 M_{\odot}/h$ respectively keeping other dark matter model parameters fixed at: $m_\chi = 1$ MeV, $T_{*} = 7.5$ K, and $\alpha_{A}=0.35$ in figures \ref{fig:df_6_nc}~-~\ref{fig:df_4v_c}.
	\newpage
	\newpage
	\newpage
	\newpage
	
	\section{Convergence of distribution functions}\label{app:G}
	We compare the $\tau_{\text{peak}}$ and $b^2$ distribution functions for collisionless DM in the redshift range 6-5 for 100 and 1000 LoS directions in figure \ref{fig:df_4v_c}. The mean count $N$ in a given $\tau_{\text{peak}}$ or $b^2$ bin follows a Poisson distribution with a standard deviation $\propto \sqrt{N}$. When the average is taken over spectra from 1000 different LoS directions, we find that the tail becomes smoother and the error bars on the distribution functions contract as expected.
	\section{Radiative transfer equation in an expanding Universe}\label{app:B}
	Let $u_{\nu}(\Omega)$ be the radiation energy density per unit frequency per unit solid angle. In terms of the photon distribution function $f(\nu)$, 
	\begin{equation}\label{A.1}
		u_{\nu}(\Omega)d\nu\,d\Omega = h\nu f(\nu)(h/c)^{3}\nu^{2}\,d\nu\,d\Omega. 
	\end{equation}
	In an expanding Universe, frequency redshifts as $1/a$, where $a$ as the scale factor, while $f(\nu)$ remains conserved. From eq.\eqref{A.1}, in the absence of a source or sink, $u_{\nu}(\Omega)/\nu^{3}$ or  $u_{\nu}(\Omega)a^{3}$ remains constant with respect to time. 
	\begin{align}\label{A.5}
		\frac{d(u_{\nu}a^{3})}{dt} &= 0,\\
		a^{3}\frac{\partial u_{\nu}}{\partial t} + a^{3}\bigg(\frac{d\nu}{dt}\bigg)\frac{\partial u_{\nu}}{\partial \nu} + 3a^{2}\dot{a}u_{\nu} &= 0,\\
		a^{3}\frac{\partial u_{\nu}}{\partial t} -
		\nu a^{2}\dot{a}\frac{\partial u_{\nu}}{\partial \nu} + 3a^{2}\dot{a}u_{\nu} &= 0,\\
		\frac{\partial u_{\nu}}{\partial t} -
		\nu\frac{\dot{a}}{a}\frac{\partial u_{\nu}}{\partial \nu} + 3\frac{\dot{a}}{a}u_{\nu} &=0.
	\end{align}  
	In the presence of a source having emission coefficient $j_\nu$ and absorption coefficient $\alpha_{\nu}$, the radiative transfer equation becomes \cite{1997ApJ...486..581G},
	\begin{align}
		\frac{\partial u_{\nu}}{\partial t} -
		\nu\frac{\dot{a}}{a}\frac{\partial u_{\nu}}{\partial\nu} + 3\frac{\dot{a}}{a}u_{\nu} &= -c\alpha_{\nu}u_{\nu}+j_{\nu}. \label{A.6}
	\end{align} 
	The specific intensity $I_{\nu}$, defined as the energy per unit area per unit time per unit frequency per unit solid angle is related to the energy density per unit frequency per unit solid angle as $u_{\nu}=I_{\nu}/c$. The specific intensity in temperature units is given by the following relation,
	\begin{eqnarray}
		cu_{\nu} = I_{\nu} =\frac{2\nu^{2}}{c^{2}}k_BT_{b},\label{4.6a}
	\end{eqnarray}
	where $T_b$ is the brightness temperature at a fixed frequency $\nu$. We can use eq.\eqref{4.6a} to cast eq.\eqref{A.6} in terms of the brightness temperature \cite{1986rpa..book.....R,2010asph.book.....C},
	\begin{equation}
		\frac{dT_{b}}{d t} + \frac{\dot{a}}{a}T_{b}=c\alpha_{\nu}\big(-T_{b}+\frac{c^{2}}{2\nu^{2}k_B}\frac{j_{\nu}}{\alpha_{\nu}}\big). \label{4.7}
	\end{equation}
	If the level population defined in eq.\eqref{29} is in kinetic equilibrium at temperature $T_{\text{ex}}$, by using the Einstein relations we can find the ratio between the emission and absorption coefficients which is given by \cite{1986rpa..book.....R},
	\begin{equation}
		\frac{j_{\nu}}{\alpha_{\nu}} = \frac{2h\nu^3}{c^2}\frac{1}{(e^{h\nu/k_BT_{\text{ex}}}-1)}.
	\end{equation}
	This relation following from the principle of detailed balance depends only on the excitation temperature $T_{\text{ex}}$ of the two levels.
	
	\section{Limits on CMB spectral distortions from COBE/FIRAS}\label{app:cobe}
\begin{figure}[t]
		\centering
		\includegraphics[width=0.5\textwidth]{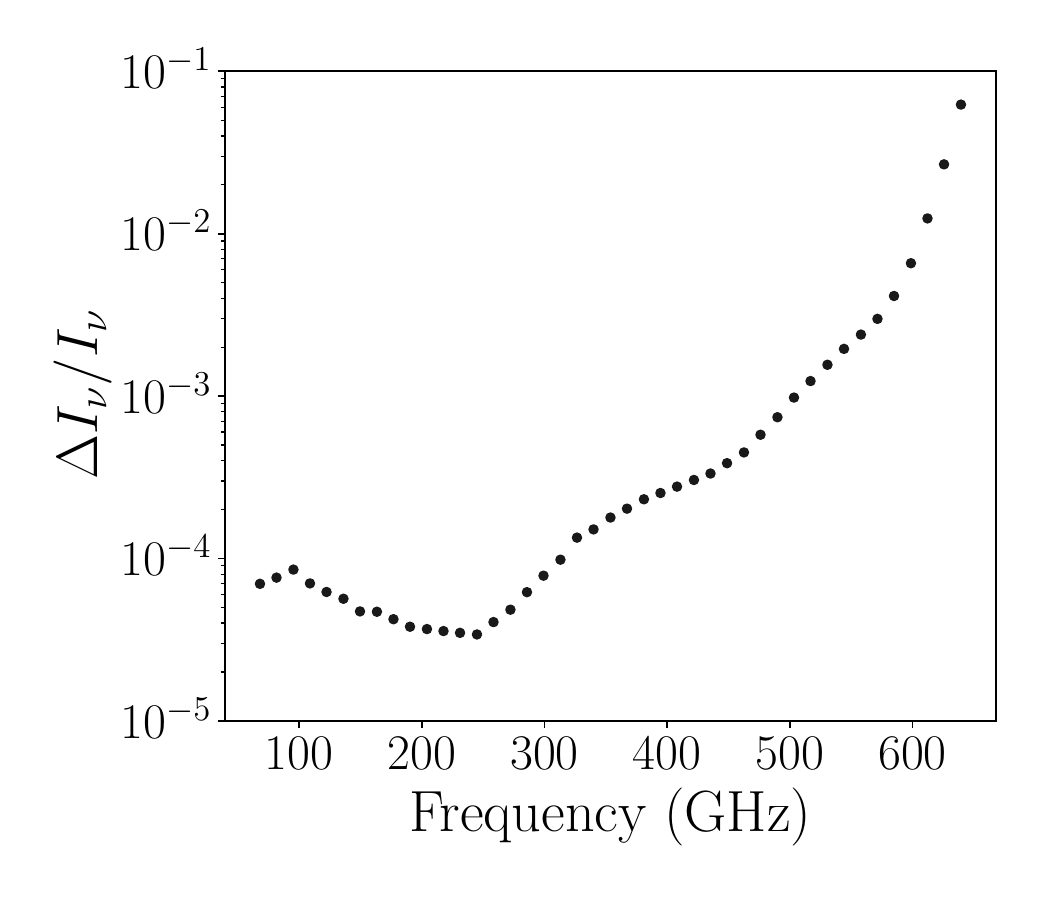}
		\caption{1-$\sigma$ limits on the fractional deviation from the CMB blackbody spectrum ($\Delta I_{\nu}/I_{\nu}$) as a function of frequency from COBE/FIRAS data \cite{1996ApJ...473..576F, cobedata}.}
		\label{fig:cobe_bound}
	\end{figure}

	We show the limits on spectral distortions of CMB from a perfect blackbody from COBE/FIRAS \cite{1996ApJ...473..576F, cobedata} in the 60-600 GHz band in figure \ref{fig:cobe_bound}.
	
	\section{Milky Way model for constraining the radiative coupling from CMB anisotropy maps}\label{app:cons}
	We assume that dark matter in MW is distributed according to the following NFW \cite{1997ApJ...490..493N} density profile,
	\begin{equation}
		\rho_\chi(r) = \rho_{0}\frac{1}{\frac{r}{r_{0}}\big(1 + (\frac{r}{r_{0}})\big)^{2}},
	\end{equation}
	with $\rho_{0}\,(10^{6} \, M_{\odot}\, \text{kpc}^{-3}) = 3.486$ and $r_{0} \,\text{(kpc)} = 26.77$ \cite{2019A&A...621A..56P} and has a virial temperature $T_{\text{vir}}$.
	The emissivity due to dark matter emission from MW along the LoS is given by \cite{1986rpa..book.....R},
	\begin{equation}
		j_{\nu} = \frac{h\nu}{4\pi}\phi(\nu)n_{1}(r)\alpha_AA_{10}^{\text{HI}},
	\end{equation}
	where $\nu$ is the frequency of the emitted photon and $\phi(\nu)$ is the Doppler line profile centered at $\nu_0$ with a line width,
	\begin{align}
		\Delta\nu_{D} &= \frac{\nu_{0}}{c}\sqrt{\frac{2k_{B}T_{\text{vir}}}{m_\chi}}.
	\end{align}
	For a band covering a frequency range from $\nu_{\text{min}}$ to $ \nu_{\text{max}}$, the integrated specific intensity along a particular LoS is given by,
	\begin{eqnarray}
				j = \int_{\nu_{\text{min}}}^{\nu_{\text{max}}}d\nu\,j_{\nu}.
	\end{eqnarray}
	
	\newpage
	\bibliography{reference.bib}

\begin{thebibliography}{100}

\bibitem{2022arXiv220306380B}
Kimberly~K. {Boddy et al.}
\newblock {Astrophysical and Cosmological Probes of Dark Matter}.
\newblock {\em arXiv e-prints}, page arXiv:2203.06380, March 2022.
\newblock \href {http://arxiv.org/abs/2203.06380} {\path{arXiv:2203.06380}},
  {\small[\href{https://ui.adsabs.harvard.edu/abs/2022arXiv220306380B}{ADS}]}.

\bibitem{2022PTEP.2022h3C01W}
R.~L. {Workman et al.} and {Particle Data Group}.
\newblock {Review of Particle Physics}.
\newblock {\em Progress of Theoretical and Experimental Physics},
  2022(8):083C01, August 2022.
\newblock \href {http://dx.doi.org/10.1093/ptep/ptac097} {\path{[DOI]}},
  {\small[\href{https://ui.adsabs.harvard.edu/abs/2022PTEP.2022h3C01W}{ADS}]}.

\bibitem{2019PhRvL.123e1103M}
David J.~E. {Marsh} and Jens~C. {Niemeyer}.
\newblock {Strong Constraints on Fuzzy Dark Matter from Ultrafaint Dwarf Galaxy
  Eridanus II}.
\newblock {\em \prl}, 123(5):051103, August 2019.
\newblock \href {http://arxiv.org/abs/1810.08543} {\path{arXiv:1810.08543}},
  \href {http://dx.doi.org/10.1103/PhysRevLett.123.051103} {\path{[DOI]}},
  {\small[\href{https://ui.adsabs.harvard.edu/abs/2019PhRvL.123e1103M}{ADS}]}.

\bibitem{2022PhRvD.106f3517D}
Neal {Dalal} and Andrey {Kravtsov}.
\newblock {Excluding fuzzy dark matter with sizes and stellar kinematics of
  ultrafaint dwarf galaxies}.
\newblock {\em \prd}, 106(6):063517, September 2022.
\newblock \href {http://dx.doi.org/10.1103/PhysRevD.106.063517} {\path{[DOI]}},
  {\small[\href{https://ui.adsabs.harvard.edu/abs/2022PhRvD.106f3517D}{ADS}]}.

\bibitem{2021PhRvL.126i1101N}
E.~O. {Nadler et al.} and {DES Collaboration}.
\newblock {Constraints on Dark Matter Properties from Observations of Milky Way
  Satellite Galaxies}.
\newblock {\em \prl}, 126(9):091101, March 2021.
\newblock \href {http://arxiv.org/abs/2008.00022} {\path{arXiv:2008.00022}},
  \href {http://dx.doi.org/10.1103/PhysRevLett.126.091101} {\path{[DOI]}},
  {\small[\href{https://ui.adsabs.harvard.edu/abs/2021PhRvL.126i1101N}{ADS}]}.

\bibitem{2010JCAP...05..021K}
David~E. {Kaplan}, Gordan~Z. {Krnjaic}, Keith~R. {Rehermann}, and
  Christopher~M. {Wells}.
\newblock {Atomic dark matter}.
\newblock {\em \jcap}, 2010(5):021, May 2010.
\newblock \href {http://arxiv.org/abs/0909.0753} {\path{arXiv:0909.0753}},
  \href {http://dx.doi.org/10.1088/1475-7516/2010/05/021} {\path{[DOI]}},
  {\small[\href{https://ui.adsabs.harvard.edu/abs/2010JCAP...05..021K}{ADS}]}.

\bibitem{2012PhRvD..86k5013C}
James~M. {Cline}, Guy~D. {Moore}, and Andrew~R. {Frey}.
\newblock {Composite magnetic dark matter and the 130 GeV line}.
\newblock {\em \prd}, 86(11):115013, December 2012.
\newblock \href {http://arxiv.org/abs/1208.2685} {\path{arXiv:1208.2685}},
  \href {http://dx.doi.org/10.1103/PhysRevD.86.115013} {\path{[DOI]}},
  {\small[\href{https://ui.adsabs.harvard.edu/abs/2012PhRvD..86k5013C}{ADS}]}.

\bibitem{2014JCAP...05..033F}
Mads~T. {Frandsen}, Francesco {Sannino}, Ian~M. {Shoemaker}, and Ole
  {Svendsen}.
\newblock {X-ray lines from dark matter: the good, the bad, and the unlikely}.
\newblock {\em \jcap}, 2014(5):033, May 2014.
\newblock \href {http://arxiv.org/abs/1403.1570} {\path{arXiv:1403.1570}},
  \href {http://dx.doi.org/10.1088/1475-7516/2014/05/033} {\path{[DOI]}},
  {\small[\href{https://ui.adsabs.harvard.edu/abs/2014JCAP...05..033F}{ADS}]}.

\bibitem{2014PhRvD..89l1302C}
James~M. {Cline}, Zuowei {Liu}, Guy~D. {Moore}, Yasaman {Farzan}, and Wei
  {Xue}.
\newblock {3.5 keV x rays as the ``21 cm line'' of dark atoms, and a link to
  light sterile neutrinos}.
\newblock {\em \prd}, 89(12):121302, June 2014.
\newblock \href {http://arxiv.org/abs/1404.3729} {\path{arXiv:1404.3729}},
  \href {http://dx.doi.org/10.1103/PhysRevD.89.121302} {\path{[DOI]}},
  {\small[\href{https://ui.adsabs.harvard.edu/abs/2014PhRvD..89l1302C}{ADS}]}.

\bibitem{2022JHEP...06..047B}
Masha {Baryakhtar}, Asher {Berlin}, Hongwan {Liu}, and Neal {Weiner}.
\newblock {Electromagnetic signals of inelastic dark matter scattering}.
\newblock {\em Journal of High Energy Physics}, 2022(6):47, June 2022.
\newblock \href {http://dx.doi.org/10.1007/JHEP06(2022)047} {\path{[DOI]}},
  {\small[\href{https://ui.adsabs.harvard.edu/abs/2022JHEP...06..047B}{ADS}]}.

\bibitem{2010PhRvD..82g5019F}
Brian {Feldstein}, Peter~W. {Graham}, and Surjeet {Rajendran}.
\newblock {Luminous dark matter}.
\newblock {\em \prd}, 82(7):075019, October 2010.
\newblock \href {http://arxiv.org/abs/1008.1988} {\path{arXiv:1008.1988}},
  \href {http://dx.doi.org/10.1103/PhysRevD.82.075019} {\path{[DOI]}},
  {\small[\href{https://ui.adsabs.harvard.edu/abs/2010PhRvD..82g5019F}{ADS}]}.

\bibitem{2007PhRvD..75b3521P}
Stefano {Profumo} and Kris {Sigurdson}.
\newblock {Shadow of dark matter}.
\newblock {\em \prd}, 75(2):023521, January 2007.
\newblock \href {http://arxiv.org/abs/astro-ph/0611129}
  {\path{arXiv:astro-ph/0611129}}, \href
  {http://dx.doi.org/10.1103/PhysRevD.75.023521} {\path{[DOI]}},
  {\small[\href{https://ui.adsabs.harvard.edu/abs/2007PhRvD..75b3521P}{ADS}]}.

\bibitem{2010PhRvD..81i5001K}
Graham~D. {Kribs}, Tuhin~S. {Roy}, John {Terning}, and Kathryn~M. {Zurek}.
\newblock {Quirky composite dark matter}.
\newblock {\em \prd}, 81(9):095001, May 2010.
\newblock \href {http://arxiv.org/abs/0909.2034} {\path{arXiv:0909.2034}},
  \href {http://dx.doi.org/10.1103/PhysRevD.81.095001} {\path{[DOI]}},
  {\small[\href{https://ui.adsabs.harvard.edu/abs/2010PhRvD..81i5001K}{ADS}]}.

\bibitem{1994Natur.370..629M}
Ben {Moore}.
\newblock {Evidence against dissipation-less dark matter from observations of
  galaxy haloes}.
\newblock {\em \nat}, 370(6491):629--631, August 1994.
\newblock \href {http://dx.doi.org/10.1038/370629a0} {\path{[DOI]}},
  {\small[\href{https://ui.adsabs.harvard.edu/abs/1994Natur.370..629M}{ADS}]}.

\bibitem{2008ApJ...676..920K}
Rachel {Kuzio de Naray}, Stacy~S. {McGaugh}, and W.~J.~G. {de Blok}.
\newblock {Mass Models for Low Surface Brightness Galaxies with High-Resolution
  Optical Velocity Fields}.
\newblock {\em \apj}, 676(2):920--943, April 2008.
\newblock \href {http://arxiv.org/abs/0712.0860} {\path{arXiv:0712.0860}},
  \href {http://dx.doi.org/10.1086/527543} {\path{[DOI]}},
  {\small[\href{https://ui.adsabs.harvard.edu/abs/2008ApJ...676..920K}{ADS}]}.

\bibitem{2008AJ....136.2648D}
W.~J.~G. {de Blok}, F.~{Walter}, E.~{Brinks}, C.~{Trachternach}, S.~H. {Oh},
  and Jr. {Kennicutt}, R.~C.
\newblock {High-Resolution Rotation Curves and Galaxy Mass Models from THINGS}.
\newblock {\em \aj}, 136(6):2648--2719, December 2008.
\newblock \href {http://arxiv.org/abs/0810.2100} {\path{arXiv:0810.2100}},
  \href {http://dx.doi.org/10.1088/0004-6256/136/6/2648} {\path{[DOI]}},
  {\small[\href{https://ui.adsabs.harvard.edu/abs/2008AJ....136.2648D}{ADS}]}.

\bibitem{2011ApJ...742...20W}
Matthew~G. {Walker} and Jorge {Pe{\~n}arrubia}.
\newblock {A Method for Measuring (Slopes of) the Mass Profiles of Dwarf
  Spheroidal Galaxies}.
\newblock {\em \apj}, 742(1):20, November 2011.
\newblock \href {http://arxiv.org/abs/1108.2404} {\path{arXiv:1108.2404}},
  \href {http://dx.doi.org/10.1088/0004-637X/742/1/20} {\path{[DOI]}},
  {\small[\href{https://ui.adsabs.harvard.edu/abs/2011ApJ...742...20W}{ADS}]}.

\bibitem{1996ApJ...462..563N}
Julio~F. {Navarro}, Carlos~S. {Frenk}, and Simon D.~M. {White}.
\newblock {The Structure of Cold Dark Matter Halos}.
\newblock {\em \apj}, 462:563, May 1996.
\newblock \href {http://arxiv.org/abs/astro-ph/9508025}
  {\path{arXiv:astro-ph/9508025}}, \href {http://dx.doi.org/10.1086/177173}
  {\path{[DOI]}},
  {\small[\href{https://ui.adsabs.harvard.edu/abs/1996ApJ...462..563N}{ADS}]}.

\bibitem{1997ApJ...490..493N}
Julio~F. {Navarro}, Carlos~S. {Frenk}, and Simon D.~M. {White}.
\newblock {A Universal Density Profile from Hierarchical Clustering}.
\newblock {\em \apj}, 490(2):493--508, December 1997.
\newblock \href {http://arxiv.org/abs/astro-ph/9611107}
  {\path{arXiv:astro-ph/9611107}}, \href {http://dx.doi.org/10.1086/304888}
  {\path{[DOI]}},
  {\small[\href{https://ui.adsabs.harvard.edu/abs/1997ApJ...490..493N}{ADS}]}.

\bibitem{1999MNRAS.310.1147M}
B.~{Moore}, T.~{Quinn}, F.~{Governato}, J.~{Stadel}, and G.~{Lake}.
\newblock {Cold collapse and the core catastrophe}.
\newblock {\em \mnras}, 310(4):1147--1152, December 1999.
\newblock \href {http://arxiv.org/abs/astro-ph/9903164}
  {\path{arXiv:astro-ph/9903164}}, \href
  {http://dx.doi.org/10.1046/j.1365-8711.1999.03039.x} {\path{[DOI]}},
  {\small[\href{https://ui.adsabs.harvard.edu/abs/1999MNRAS.310.1147M}{ADS}]}.

\bibitem{2013PhRvD..87j3515C}
Francis-Yan {Cyr-Racine} and Kris {Sigurdson}.
\newblock {Cosmology of atomic dark matter}.
\newblock {\em \prd}, 87(10):103515, May 2013.
\newblock \href {http://arxiv.org/abs/1209.5752} {\path{arXiv:1209.5752}},
  \href {http://dx.doi.org/10.1103/PhysRevD.87.103515} {\path{[DOI]}},
  {\small[\href{https://ui.adsabs.harvard.edu/abs/2013PhRvD..87j3515C}{ADS}]}.

\bibitem{2014PhRvD..89d3514C}
James~M. {Cline}, Zuowei {Liu}, Guy~D. {Moore}, and Wei {Xue}.
\newblock {Scattering properties of dark atoms and molecules}.
\newblock {\em \prd}, 89(4):043514, February 2014.
\newblock \href {http://arxiv.org/abs/1311.6468} {\path{arXiv:1311.6468}},
  \href {http://dx.doi.org/10.1103/PhysRevD.89.043514} {\path{[DOI]}},
  {\small[\href{https://ui.adsabs.harvard.edu/abs/2014PhRvD..89d3514C}{ADS}]}.

\bibitem{2016PhRvD..94l3017B}
Kimberly~K. {Boddy}, Manoj {Kaplinghat}, Anna {Kwa}, and Annika H.~G. {Peter}.
\newblock {Hidden sector hydrogen as dark matter: Small-scale structure
  formation predictions and the importance of hyperfine interactions}.
\newblock {\em \prd}, 94(12):123017, December 2016.
\newblock \href {http://arxiv.org/abs/1609.03592} {\path{arXiv:1609.03592}},
  \href {http://dx.doi.org/10.1103/PhysRevD.94.123017} {\path{[DOI]}},
  {\small[\href{https://ui.adsabs.harvard.edu/abs/2016PhRvD..94l3017B}{ADS}]}.

\bibitem{1999ApJ...522...82K}
Anatoly {Klypin}, Andrey~V. {Kravtsov}, Octavio {Valenzuela}, and Francisco
  {Prada}.
\newblock {Where Are the Missing Galactic Satellites?}
\newblock {\em \apj}, 522(1):82--92, September 1999.
\newblock \href {http://arxiv.org/abs/astro-ph/9901240}
  {\path{arXiv:astro-ph/9901240}}, \href {http://dx.doi.org/10.1086/307643}
  {\path{[DOI]}},
  {\small[\href{https://ui.adsabs.harvard.edu/abs/1999ApJ...522...82K}{ADS}]}.

\bibitem{1999ApJ...524L..19M}
Ben {Moore}, Sebastiano {Ghigna}, Fabio {Governato}, George {Lake}, Thomas
  {Quinn}, Joachim {Stadel}, and Paolo {Tozzi}.
\newblock {Dark Matter Substructure within Galactic Halos}.
\newblock {\em \apjl}, 524(1):L19--L22, October 1999.
\newblock \href {http://arxiv.org/abs/astro-ph/9907411}
  {\path{arXiv:astro-ph/9907411}}, \href {http://dx.doi.org/10.1086/312287}
  {\path{[DOI]}},
  {\small[\href{https://ui.adsabs.harvard.edu/abs/1999ApJ...524L..19M}{ADS}]}.

\bibitem{2015PhRvD..91b3512F}
R.~{Foot} and S.~{Vagnozzi}.
\newblock {Dissipative hidden sector dark matter}.
\newblock {\em \prd}, 91(2):023512, January 2015.
\newblock \href {http://arxiv.org/abs/1409.7174} {\path{arXiv:1409.7174}},
  \href {http://dx.doi.org/10.1103/PhysRevD.91.023512} {\path{[DOI]}},
  {\small[\href{https://ui.adsabs.harvard.edu/abs/2015PhRvD..91b3512F}{ADS}]}.

\bibitem{2016JCAP...07..013F}
Robert {Foot} and Sunny {Vagnozzi}.
\newblock {Solving the small-scale structure puzzles with dissipative dark
  matter}.
\newblock {\em \jcap}, 2016(7):013, July 2016.
\newblock \href {http://arxiv.org/abs/1602.02467} {\path{arXiv:1602.02467}},
  \href {http://dx.doi.org/10.1088/1475-7516/2016/07/013} {\path{[DOI]}},
  {\small[\href{https://ui.adsabs.harvard.edu/abs/2016JCAP...07..013F}{ADS}]}.

\bibitem{2015JCAP...01..021S}
Katelin {Schutz} and Tracy~R. {Slatyer}.
\newblock {Self-scattering for Dark Matter with an excited state}.
\newblock {\em \jcap}, 2015(1):021--021, January 2015.
\newblock \href {http://arxiv.org/abs/1409.2867} {\path{arXiv:1409.2867}},
  \href {http://dx.doi.org/10.1088/1475-7516/2015/01/021} {\path{[DOI]}},
  {\small[\href{https://ui.adsabs.harvard.edu/abs/2015JCAP...01..021S}{ADS}]}.

\bibitem{2017JCAP...03..048B}
Mattias {Blennow}, Stefan {Clementz}, and Juan {Herrero-Garcia}.
\newblock {Self-interacting inelastic dark matter: a viable solution to the
  small scale structure problems}.
\newblock {\em \jcap}, 2017(3):048, March 2017.
\newblock \href {http://arxiv.org/abs/1612.06681} {\path{arXiv:1612.06681}},
  \href {http://dx.doi.org/10.1088/1475-7516/2017/03/048} {\path{[DOI]}},
  {\small[\href{https://ui.adsabs.harvard.edu/abs/2017JCAP...03..048B}{ADS}]}.

\bibitem{2018PhRvD..97b3002D}
Anirban {Das} and Basudeb {Dasgupta}.
\newblock {New dissipation mechanisms from multilevel dark matter scattering}.
\newblock {\em \prd}, 97(2):023002, January 2018.
\newblock \href {http://arxiv.org/abs/1709.06577} {\path{arXiv:1709.06577}},
  \href {http://dx.doi.org/10.1103/PhysRevD.97.023002} {\path{[DOI]}},
  {\small[\href{https://ui.adsabs.harvard.edu/abs/2018PhRvD..97b3002D}{ADS}]}.

\bibitem{2022PhRvL.128t1301C}
Francis-Yan {Cyr-Racine}, Fei {Ge}, and Lloyd {Knox}.
\newblock {Symmetry of Cosmological Observables, a Mirror World Dark Sector,
  and the Hubble Constant}.
\newblock {\em \prl}, 128(20):201301, May 2022.
\newblock \href {http://arxiv.org/abs/2107.13000} {\path{arXiv:2107.13000}},
  \href {http://dx.doi.org/10.1103/PhysRevLett.128.201301} {\path{[DOI]}},
  {\small[\href{https://ui.adsabs.harvard.edu/abs/2022PhRvL.128t1301C}{ADS}]}.

\bibitem{2022PhRvD.105i5005B}
Nikita {Blinov}, Gordan {Krnjaic}, and Shirley~Weishi {Li}.
\newblock {Realistic model of dark atoms to resolve the Hubble tension}.
\newblock {\em \prd}, 105(9):095005, May 2022.
\newblock \href {http://arxiv.org/abs/2108.11386} {\path{arXiv:2108.11386}},
  \href {http://dx.doi.org/10.1103/PhysRevD.105.095005} {\path{[DOI]}},
  {\small[\href{https://ui.adsabs.harvard.edu/abs/2022PhRvD.105i5005B}{ADS}]}.

\bibitem{2011PhRvL.107e1301A}
J.~{Angle et al.}
\newblock {Search for Light Dark Matter in XENON10 Data}.
\newblock {\em \prl}, 107(5):051301, July 2011.
\newblock \href {http://arxiv.org/abs/1104.3088} {\path{arXiv:1104.3088}},
  \href {http://dx.doi.org/10.1103/PhysRevLett.107.051301} {\path{[DOI]}},
  {\small[\href{https://ui.adsabs.harvard.edu/abs/2011PhRvL.107e1301A}{ADS}]}.

\bibitem{2016PhRvD..94i2001A}
E.~{Aprile et al.}
\newblock {Low-mass dark matter search using ionization signals in XENON100}.
\newblock {\em \prd}, 94(9):092001, November 2016.
\newblock \href {http://arxiv.org/abs/1605.06262} {\path{arXiv:1605.06262}},
  \href {http://dx.doi.org/10.1103/PhysRevD.94.092001} {\path{[DOI]}},
  {\small[\href{https://ui.adsabs.harvard.edu/abs/2016PhRvD..94i2001A}{ADS}]}.

\bibitem{2017PhRvD..96d3017E}
Rouven {Essig}, Tomer {Volansky}, and Tien-Tien {Yu}.
\newblock {New constraints and prospects for sub-GeV dark matter scattering off
  electrons in xenon}.
\newblock {\em \prd}, 96(4):043017, August 2017.
\newblock \href {http://arxiv.org/abs/1703.00910} {\path{arXiv:1703.00910}},
  \href {http://dx.doi.org/10.1103/PhysRevD.96.043017} {\path{[DOI]}},
  {\small[\href{https://ui.adsabs.harvard.edu/abs/2017PhRvD..96d3017E}{ADS}]}.

\bibitem{2018PhRvL.121k1303A}
P.~{Agnes et al.}
\newblock {Constraints on Sub-GeV Dark-Matter-Electron Scattering from the
  DarkSide-50 Experiment}.
\newblock {\em \prl}, 121(11):111303, September 2018.
\newblock \href {http://arxiv.org/abs/1802.06998} {\path{arXiv:1802.06998}},
  \href {http://dx.doi.org/10.1103/PhysRevLett.121.111303} {\path{[DOI]}},
  {\small[\href{https://ui.adsabs.harvard.edu/abs/2018PhRvL.121k1303A}{ADS}]}.

\bibitem{2018PhRvL.121f1803C}
Michael {Crisler}, Rouven {Essig}, Juan {Estrada}, Guillermo {Fernandez},
  Javier {Tiffenberg}, Miguel~Sofo {Haro}, Tomer {Volansky}, Tien-Tien {Yu},
  and {Sensei Collaboration}.
\newblock {SENSEI: First Direct-Detection Constraints on Sub-GeV Dark Matter
  from a Surface Run}.
\newblock {\em \prl}, 121(6):061803, August 2018.
\newblock \href {http://arxiv.org/abs/1804.00088} {\path{arXiv:1804.00088}},
  \href {http://dx.doi.org/10.1103/PhysRevLett.121.061803} {\path{[DOI]}},
  {\small[\href{https://ui.adsabs.harvard.edu/abs/2018PhRvL.121f1803C}{ADS}]}.

\bibitem{2019PhRvL.122p1801A}
Orr {Abramoff et al.}
\newblock {SENSEI: Direct-Detection Constraints on Sub-GeV Dark Matter from a
  Shallow Underground Run Using a Prototype Skipper CCD}.
\newblock {\em \prl}, 122(16):161801, April 2019.
\newblock \href {http://arxiv.org/abs/1901.10478} {\path{arXiv:1901.10478}},
  \href {http://dx.doi.org/10.1103/PhysRevLett.122.161801} {\path{[DOI]}},
  {\small[\href{https://ui.adsabs.harvard.edu/abs/2019PhRvL.122p1801A}{ADS}]}.

\bibitem{2018PhRvL.121e1301A}
R.~et~al. {Agnese}.
\newblock {First Dark Matter Constraints from a SuperCDMS Single-Charge
  Sensitive Detector}.
\newblock {\em \prl}, 121(5):051301, August 2018.
\newblock \href {http://arxiv.org/abs/1804.10697} {\path{arXiv:1804.10697}},
  \href {http://dx.doi.org/10.1103/PhysRevLett.121.051301} {\path{[DOI]}},
  {\small[\href{https://ui.adsabs.harvard.edu/abs/2018PhRvL.121e1301A}{ADS}]}.

\bibitem{1991CSci...60...95S}
G.~{Swarup}, S.~{Ananthakrishnan}, V.~K. {Kapahi}, A.~P. {Rao}, C.~R.
  {Subrahmanya}, and V.~K. {Kulkarni}.
\newblock {The Giant Metre-Wave Radio Telescope}.
\newblock {\em Current Science}, 60:95, January 1991.
\newblock
  {\small[\href{https://ui.adsabs.harvard.edu/abs/1991CSci...60...95S}{ADS}]}.

\bibitem{2017CSci..113..707G}
Y.~{Gupta}, B.~{Ajithkumar}, H.~S. {Kale}, S.~{Nayak}, S.~{Sabhapathy},
  S.~{Sureshkumar}, R.~V. {Swami}, J.~N. {Chengalur}, S.~K. {Ghosh}, C.~H.
  {Ishwara-Chandra}, B.~C. {Joshi}, N.~{Kanekar}, D.~V. {Lal}, and S.~{Roy}.
\newblock {The upgraded GMRT: opening new windows on the radio Universe}.
\newblock {\em Current Science}, 113(4):707--714, August 2017.
\newblock \href {http://dx.doi.org/10.18520/cs/v113/i04/707-714}
  {\path{[DOI]}},
  {\small[\href{https://ui.adsabs.harvard.edu/abs/2017CSci..113..707G}{ADS}]}.

\bibitem{2020PASA...37....2W}
A~{Weltman et. al}.
\newblock {Fundamental physics with the Square Kilometre Array}.
\newblock {\em \pasa}, 37:e002, January 2020.
\newblock \href {http://arxiv.org/abs/1810.02680} {\path{arXiv:1810.02680}},
  \href {http://dx.doi.org/10.1017/pasa.2019.42} {\path{[DOI]}},
  {\small[\href{https://ui.adsabs.harvard.edu/abs/2020PASA...37....2W}{ADS}]}.

\bibitem{2004MNRAS.352..142B}
Somnath {Bharadwaj} and Sk.~Saiyad {Ali}.
\newblock {The cosmic microwave background radiation fluctuations from HI
  perturbations prior to reionization}.
\newblock {\em \mnras}, 352(1):142--146, July 2004.
\newblock \href {http://arxiv.org/abs/astro-ph/0401206}
  {\path{arXiv:astro-ph/0401206}}, \href
  {http://dx.doi.org/10.1111/j.1365-2966.2004.07907.x} {\path{[DOI]}},
  {\small[\href{https://ui.adsabs.harvard.edu/abs/2004MNRAS.352..142B}{ADS}]}.

\bibitem{2004ApJ...608..622Z}
Matias {Zaldarriaga}, Steven~R. {Furlanetto}, and Lars {Hernquist}.
\newblock {21 Centimeter Fluctuations from Cosmic Gas at High Redshifts}.
\newblock {\em \apj}, 608(2):622--635, June 2004.
\newblock \href {http://arxiv.org/abs/astro-ph/0311514}
  {\path{arXiv:astro-ph/0311514}}, \href {http://dx.doi.org/10.1086/386327}
  {\path{[DOI]}},
  {\small[\href{https://ui.adsabs.harvard.edu/abs/2004ApJ...608..622Z}{ADS}]}.

\bibitem{2006PhR...433..181F}
Steven~R. {Furlanetto}, S.~Peng {Oh}, and Frank~H. {Briggs}.
\newblock {Cosmology at low frequencies: The 21 cm transition and the
  high-redshift Universe}.
\newblock {\em \physrep}, 433(4-6):181--301, October 2006.
\newblock \href {http://arxiv.org/abs/astro-ph/0608032}
  {\path{arXiv:astro-ph/0608032}}, \href
  {http://dx.doi.org/10.1016/j.physrep.2006.08.002} {\path{[DOI]}},
  {\small[\href{https://ui.adsabs.harvard.edu/abs/2006PhR...433..181F}{ADS}]}.

\bibitem{2012RPPh...75h6901P}
Jonathan~R. {Pritchard} and Abraham {Loeb}.
\newblock {21 cm cosmology in the 21st century}.
\newblock {\em Reports on Progress in Physics}, 75(8):086901, August 2012.
\newblock \href {http://arxiv.org/abs/1109.6012} {\path{arXiv:1109.6012}},
  \href {http://dx.doi.org/10.1088/0034-4885/75/8/086901} {\path{[DOI]}},
  {\small[\href{https://ui.adsabs.harvard.edu/abs/2012RPPh...75h6901P}{ADS}]}.

\bibitem{2018Natur.555...67B}
Judd~D. {Bowman}, Alan E.~E. {Rogers}, Raul~A. {Monsalve}, Thomas~J. {Mozdzen},
  and Nivedita {Mahesh}.
\newblock {An absorption profile centred at 78 megahertz in the sky-averaged
  spectrum}.
\newblock {\em \nat}, 555(7694):67--70, March 2018.
\newblock \href {http://arxiv.org/abs/1810.05912} {\path{arXiv:1810.05912}},
  \href {http://dx.doi.org/10.1038/nature25792} {\path{[DOI]}},
  {\small[\href{https://ui.adsabs.harvard.edu/abs/2018Natur.555...67B}{ADS}]}.

\bibitem{2021AJ....162...38M}
Nivedita {Mahesh}, Judd~D. {Bowman}, Thomas~J. {Mozdzen}, Alan E.~E. {Rogers},
  Raul~A. {Monsalve}, Steven~G. {Murray}, and David {Lewis}.
\newblock {Validation of the EDGES Low-band Antenna Beam Model}.
\newblock {\em \aj}, 162(2):38, August 2021.
\newblock \href {http://arxiv.org/abs/2103.00423} {\path{arXiv:2103.00423}},
  \href {http://dx.doi.org/10.3847/1538-3881/abfdab} {\path{[DOI]}},
  {\small[\href{https://ui.adsabs.harvard.edu/abs/2021AJ....162...38M}{ADS}]}.

\bibitem{2018Natur.555...71B}
Rennan {Barkana}.
\newblock {Possible interaction between baryons and dark-matter particles
  revealed by the first stars}.
\newblock {\em \nat}, 555(7694):71--74, March 2018.
\newblock \href {http://arxiv.org/abs/1803.06698} {\path{arXiv:1803.06698}},
  \href {http://dx.doi.org/10.1038/nature25791} {\path{[DOI]}},
  {\small[\href{https://ui.adsabs.harvard.edu/abs/2018Natur.555...71B}{ADS}]}.

\bibitem{2018ApJ...858L..17F}
Chang {Feng} and Gilbert {Holder}.
\newblock {Enhanced Global Signal of Neutral Hydrogen Due to Excess Radiation
  at Cosmic Dawn}.
\newblock {\em \apjl}, 858(2):L17, May 2018.
\newblock \href {http://arxiv.org/abs/1802.07432} {\path{arXiv:1802.07432}},
  \href {http://dx.doi.org/10.3847/2041-8213/aac0fe} {\path{[DOI]}},
  {\small[\href{https://ui.adsabs.harvard.edu/abs/2018ApJ...858L..17F}{ADS}]}.

\bibitem{2018PhRvD..98j3529K}
Ely~D. {Kovetz}, Vivian {Poulin}, Vera {Gluscevic}, Kimberly~K. {Boddy}, Rennan
  {Barkana}, and Marc {Kamionkowski}.
\newblock {Tighter limits on dark matter explanations of the anomalous EDGES 21
  cm signal}.
\newblock {\em \prd}, 98(10):103529, November 2018.
\newblock \href {http://arxiv.org/abs/1807.11482} {\path{arXiv:1807.11482}},
  \href {http://dx.doi.org/10.1103/PhysRevD.98.103529} {\path{[DOI]}},
  {\small[\href{https://ui.adsabs.harvard.edu/abs/2018PhRvD..98j3529K}{ADS}]}.

\bibitem{2018PhRvD..98j3005B}
Rennan {Barkana}, Nadav~Joseph {Outmezguine}, Diego {Redigol}, and Tomer
  {Volansky}.
\newblock {Strong constraints on light dark matter interpretation of the EDGES
  signal}.
\newblock {\em \prd}, 98:103005, November 2018.
\newblock \href {http://arxiv.org/abs/1803.03091} {\path{arXiv:1803.03091}},
  \href {http://dx.doi.org/10.1103/PhysRevD.98.103005} {\path{[DOI]}},
  {\small[\href{https://ui.adsabs.harvard.edu/abs/2018PhRvD..98j3005B}{ADS}]}.

\bibitem{2019PhRvD.100b3528C}
Cyril {Creque-Sarbinowski}, Lingyuan {Ji}, Ely~D. {Kovetz}, and Marc
  {Kamionkowski}.
\newblock {Direct millicharged dark matter cannot explain the EDGES signal}.
\newblock {\em \prd}, 100(2):023528, July 2019.
\newblock \href {http://arxiv.org/abs/1903.09154} {\path{arXiv:1903.09154}},
  \href {http://dx.doi.org/10.1103/PhysRevD.100.023528} {\path{[DOI]}},
  {\small[\href{https://ui.adsabs.harvard.edu/abs/2019PhRvD.100b3528C}{ADS}]}.

\bibitem{2018PhRvL.121a1102B}
Asher {Berlin}, Dan {Hooper}, Gordan {Krnjaic}, and Samuel~D. {McDermott}.
\newblock {Severely Constraining Dark-Matter Interpretations of the 21-cm
  Anomaly}.
\newblock {\em \prl}, 121(1):011102, July 2018.
\newblock \href {http://arxiv.org/abs/1803.02804} {\path{arXiv:1803.02804}},
  \href {http://dx.doi.org/10.1103/PhysRevLett.121.011102} {\path{[DOI]}},
  {\small[\href{https://ui.adsabs.harvard.edu/abs/2018PhRvL.121a1102B}{ADS}]}.

\bibitem{2019PhRvD.100l3011L}
Hongwan {Liu}, Nadav~Joseph {Outmezguine}, Diego {Redigolo}, and Tomer
  {Volansky}.
\newblock {Reviving millicharged dark matter for 21-cm cosmology}.
\newblock {\em \prd}, 100(12):123011, December 2019.
\newblock \href {http://arxiv.org/abs/1908.06986} {\path{arXiv:1908.06986}},
  \href {http://dx.doi.org/10.1103/PhysRevD.100.123011} {\path{[DOI]}},
  {\small[\href{https://ui.adsabs.harvard.edu/abs/2019PhRvD.100l3011L}{ADS}]}.

\bibitem{2022ChPhC..46d5102L}
Qiaodan {Li} and Zuowei {Liu}.
\newblock {Two-component millicharged dark matter and the EDGES 21 cm signal}.
\newblock {\em Chinese Physics C}, 46(4):045102, April 2022.
\newblock \href {http://arxiv.org/abs/2110.14996} {\path{arXiv:2110.14996}},
  \href {http://dx.doi.org/10.1088/1674-1137/ac3d2b} {\path{[DOI]}},
  {\small[\href{https://ui.adsabs.harvard.edu/abs/2022ChPhC..46d5102L}{ADS}]}.

\bibitem{2022PhRvD.105g5020M}
Anubhav {Mathur}, Surjeet {Rajendran}, and Harikrishnan {Ramani}.
\newblock {Composite solution to the EDGES anomaly}.
\newblock {\em \prd}, 105(7):075020, April 2022.
\newblock \href {http://arxiv.org/abs/2102.11284} {\path{arXiv:2102.11284}},
  \href {http://dx.doi.org/10.1103/PhysRevD.105.075020} {\path{[DOI]}},
  {\small[\href{https://ui.adsabs.harvard.edu/abs/2022PhRvD.105g5020M}{ADS}]}.

\bibitem{2022NatAs...6..607S}
Saurabh {Singh}, Nambissan~T. {Jishnu}, Ravi {Subrahmanyan}, N.~{Udaya
  Shankar}, B.~S. {Girish}, A.~{Raghunathan}, R.~{Somashekar}, K.~S. {Srivani},
  and Mayuri {Sathyanarayana Rao}.
\newblock {On the detection of a cosmic dawn signal in the radio background}.
\newblock {\em Nature Astronomy}, 6:607--617, February 2022.
\newblock \href {http://arxiv.org/abs/2112.06778} {\path{arXiv:2112.06778}},
  \href {http://dx.doi.org/10.1038/s41550-022-01610-5} {\path{[DOI]}},
  {\small[\href{https://ui.adsabs.harvard.edu/abs/2022NatAs...6..607S}{ADS}]}.

\bibitem{2022MNRAS.tmp.2422M}
Steven~G. {Murray}, Judd~D. {Bowman}, Peter~H. {Sims}, Nivedita {Mahesh}, Alan
  E.~E. {Rogers}, Raul~A. {Monsalve}, Titu {Samson}, and Akshatha~Konakondula
  {Vydula}.
\newblock {A bayesian calibration framework for EDGES}.
\newblock {\em \mnras}, September 2022.
\newblock \href {http://arxiv.org/abs/2209.03459} {\path{arXiv:2209.03459}},
  \href {http://dx.doi.org/10.1093/mnras/stac2600} {\path{[DOI]}},
  {\small[\href{https://ui.adsabs.harvard.edu/abs/2022MNRAS.tmp.2422M}{ADS}]}.

\bibitem{2015ApJ...799...90B}
G.~{Bernardi}, M.~{McQuinn}, and L.~J. {Greenhill}.
\newblock {Foreground Model and Antenna Calibration Errors in the Measurement
  of the Sky-averaged {\ensuremath{\lambda}}21 cm Signal at
  z\raisebox{-0.5ex}\textasciitilde 20}.
\newblock {\em \apj}, 799(1):90, January 2015.
\newblock \href {http://arxiv.org/abs/1404.0887} {\path{arXiv:1404.0887}},
  \href {http://dx.doi.org/10.1088/0004-637X/799/1/90} {\path{[DOI]}},
  {\small[\href{https://ui.adsabs.harvard.edu/abs/2015ApJ...799...90B}{ADS}]}.

\bibitem{2016MNRAS.461.2847B}
G.~{Bernardi et al.}
\newblock {Bayesian constraints on the global 21-cm signal from the Cosmic
  Dawn}.
\newblock {\em \mnras}, 461(3):2847--2855, September 2016.
\newblock \href {http://arxiv.org/abs/1606.06006} {\path{arXiv:1606.06006}},
  \href {http://dx.doi.org/10.1093/mnras/stw1499} {\path{[DOI]}},
  {\small[\href{https://ui.adsabs.harvard.edu/abs/2016MNRAS.461.2847B}{ADS}]}.

\bibitem{2018MNRAS.478.4193P}
D.~C. {Price et al.}
\newblock {Design and characterization of the Large-aperture Experiment to
  Detect the Dark Age (LEDA) radiometer systems}.
\newblock {\em \mnras}, 478(3):4193--4213, August 2018.
\newblock \href {http://arxiv.org/abs/1709.09313} {\path{arXiv:1709.09313}},
  \href {http://dx.doi.org/10.1093/mnras/sty1244} {\path{[DOI]}},
  {\small[\href{https://ui.adsabs.harvard.edu/abs/2018MNRAS.478.4193P}{ADS}]}.

\bibitem{2019JAI.....850004P}
L.~{Philip et al.}
\newblock {Probing Radio Intensity at High-Z from Marion: 2017 Instrument}.
\newblock {\em Journal of Astronomical Instrumentation}, 8(2):1950004, January
  2019.
\newblock \href {http://arxiv.org/abs/1806.09531} {\path{arXiv:1806.09531}},
  \href {http://dx.doi.org/10.1142/S2251171719500041} {\path{[DOI]}},
  {\small[\href{https://ui.adsabs.harvard.edu/abs/2019JAI.....850004P}{ADS}]}.

\bibitem{2022NatAs...6..984D}
E.~{de Lera Acedo et al.}
\newblock {The REACH radiometer for detecting the 21-cm hydrogen signal from
  redshift z {\ensuremath{\approx}} 7.5-28}.
\newblock {\em Nature Astronomy}, 6:984--998, July 2022.
\newblock \href {http://dx.doi.org/10.1038/s41550-022-01709-9} {\path{[DOI]}},
  {\small[\href{https://ui.adsabs.harvard.edu/abs/2022NatAs...6..984D}{ADS}]}.

\bibitem{2014ApJ...782L...9V}
Tabitha~C. {Voytek}, Aravind {Natarajan}, Jos{\'e}~Miguel {J{\'a}uregui
  Garc{\'\i}a}, Jeffrey~B. {Peterson}, and Omar {L{\'o}pez-Cruz}.
\newblock {Probing the Dark Ages at z \raisebox{-0.5ex}\textasciitilde 20: The
  SCI-HI 21 cm All-sky Spectrum Experiment}.
\newblock {\em \apjl}, 782(1):L9, February 2014.
\newblock \href {http://arxiv.org/abs/1311.0014} {\path{arXiv:1311.0014}},
  \href {http://dx.doi.org/10.1088/2041-8205/782/1/L9} {\path{[DOI]}},
  {\small[\href{https://ui.adsabs.harvard.edu/abs/2014ApJ...782L...9V}{ADS}]}.

\bibitem{2019ApJ...883..126N}
Bang~D. {Nhan}, David~D. {Bordenave}, Richard~F. {Bradley}, Jack~O. {Burns},
  Keith {Tauscher}, David {Rapetti}, and Patricia~J. {Klima}.
\newblock {Assessment of the Projection-induced Polarimetry Technique for
  Constraining the Foreground Spectrum in Global 21 cm Cosmology}.
\newblock {\em \apj}, 883(2):126, October 2019.
\newblock \href {http://arxiv.org/abs/1811.04917} {\path{arXiv:1811.04917}},
  \href {http://dx.doi.org/10.3847/1538-4357/ab391b} {\path{[DOI]}},
  {\small[\href{https://ui.adsabs.harvard.edu/abs/2019ApJ...883..126N}{ADS}]}.

\bibitem{2020A&A...641A...6P}
N.~{Aghanim et al}. {Planck Collaboration}.
\newblock {Planck 2018 results. VI. Cosmological parameters}.
\newblock {\em \aap}, 641:A6, September 2020.
\newblock \href {http://arxiv.org/abs/1807.06209} {\path{arXiv:1807.06209}},
  \href {http://dx.doi.org/10.1051/0004-6361/201833910} {\path{[DOI]}},
  {\small[\href{https://ui.adsabs.harvard.edu/abs/2020A&A...641A...6P}{ADS}]}.

\bibitem{2011MNRAS.412..748C}
J.~{Chluba} and R.~M. {Thomas}.
\newblock {Towards a complete treatment of the cosmological recombination
  problem}.
\newblock {\em \mnras}, 412(2):748--764, April 2011.
\newblock \href {http://arxiv.org/abs/1010.3631} {\path{arXiv:1010.3631}},
  \href {http://dx.doi.org/10.1111/j.1365-2966.2010.17940.x} {\path{[DOI]}},
  {\small[\href{https://ui.adsabs.harvard.edu/abs/2011MNRAS.412..748C}{ADS}]}.

\bibitem{2010MNRAS.407..599C}
J.~{Chluba}, G.~M. {Vasil}, and L.~J. {Dursi}.
\newblock {Recombinations to the Rydberg states of hydrogen and their effect
  during the cosmological recombination epoch}.
\newblock {\em \mnras}, 407(1):599--612, September 2010.
\newblock \href {http://arxiv.org/abs/1003.4928} {\path{arXiv:1003.4928}},
  \href {http://dx.doi.org/10.1111/j.1365-2966.2010.16940.x} {\path{[DOI]}},
  {\small[\href{https://ui.adsabs.harvard.edu/abs/2010MNRAS.407..599C}{ADS}]}.

\bibitem{2018ApJS..239...35D}
Benedikt {Diemer}.
\newblock {COLOSSUS: A Python Toolkit for Cosmology, Large-scale Structure, and
  Dark Matter Halos}.
\newblock {\em \apjs}, 239(2):35, December 2018.
\newblock \href {http://arxiv.org/abs/1712.04512} {\path{arXiv:1712.04512}},
  \href {http://dx.doi.org/10.3847/1538-4365/aaee8c} {\path{[DOI]}},
  {\small[\href{https://ui.adsabs.harvard.edu/abs/2018ApJS..239...35D}{ADS}]}.

\bibitem{2020CoPhC.25607478S}
Vladyslav {Shtabovenko}, Rolf {Mertig}, and Frederik {Orellana}.
\newblock {FeynCalc 9.3: New features and improvements}.
\newblock {\em Computer Physics Communications}, 256:107478, November 2020.
\newblock \href {http://arxiv.org/abs/2001.04407} {\path{arXiv:2001.04407}},
  \href {http://dx.doi.org/10.1016/j.cpc.2020.107478} {\path{[DOI]}},
  {\small[\href{https://ui.adsabs.harvard.edu/abs/2020CoPhC.25607478S}{ADS}]}.

\bibitem{2016CoPhC.207..432S}
Vladyslav {Shtabovenko}, Rolf {Mertig}, and Frederik {Orellana}.
\newblock {New developments in FeynCalc 9.0}.
\newblock {\em Computer Physics Communications}, 207:432--444, October 2016.
\newblock \href {http://arxiv.org/abs/1601.01167} {\path{arXiv:1601.01167}},
  \href {http://dx.doi.org/10.1016/j.cpc.2016.06.008} {\path{[DOI]}},
  {\small[\href{https://ui.adsabs.harvard.edu/abs/2016CoPhC.207..432S}{ADS}]}.

\bibitem{Mertig:1990an}
R.~Mertig, M.~Bohm, and Ansgar Denner.
\newblock {FEYN CALC: Computer algebraic calculation of Feynman amplitudes}.
\newblock {\em Comput. Phys. Commun.}, 64:345--359, 1991.
\newblock \href {http://dx.doi.org/10.1016/0010-4655(91)90130-D}
  {\path{[DOI]}}.

\bibitem{2001PhRvL..86.4757A}
Nima {Arkani-Hamed}, Andrew~G. {Cohen}, and Howard {Georgi}.
\newblock {(De)Constructing Dimensions}.
\newblock {\em \prl}, 86(21):4757--4761, May 2001.
\newblock \href {http://arxiv.org/abs/hep-th/0104005}
  {\path{arXiv:hep-th/0104005}}, \href
  {http://dx.doi.org/10.1103/PhysRevLett.86.4757} {\path{[DOI]}},
  {\small[\href{https://ui.adsabs.harvard.edu/abs/2001PhRvL..86.4757A}{ADS}]}.

\bibitem{2006PhRvD..73d5016B}
Brian {Batell} and Tony {Gherghetta}.
\newblock {Localized U(1) gauge fields, millicharged particles, and
  holography}.
\newblock {\em \prd}, 73(4):045016, February 2006.
\newblock \href {http://arxiv.org/abs/hep-ph/0512356}
  {\path{arXiv:hep-ph/0512356}}, \href
  {http://dx.doi.org/10.1103/PhysRevD.73.045016} {\path{[DOI]}},
  {\small[\href{https://ui.adsabs.harvard.edu/abs/2006PhRvD..73d5016B}{ADS}]}.

\bibitem{2016PhRvD..93h5007K}
David~E. {Kaplan} and Riccardo {Rattazzi}.
\newblock {Large field excursions and approximate discrete symmetries from a
  clockwork axion}.
\newblock {\em \prd}, 93(8):085007, April 2016.
\newblock \href {http://arxiv.org/abs/1511.01827} {\path{arXiv:1511.01827}},
  \href {http://dx.doi.org/10.1103/PhysRevD.93.085007} {\path{[DOI]}},
  {\small[\href{https://ui.adsabs.harvard.edu/abs/2016PhRvD..93h5007K}{ADS}]}.

\bibitem{2017JHEP...02..036G}
Gian~F. {Giudice} and Matthew {McCullough}.
\newblock {A clockwork theory}.
\newblock {\em Journal of High Energy Physics}, 2017(2):36, February 2017.
\newblock \href {http://arxiv.org/abs/1610.07962} {\path{arXiv:1610.07962}},
  \href {http://dx.doi.org/10.1007/JHEP02(2017)036} {\path{[DOI]}},
  {\small[\href{https://ui.adsabs.harvard.edu/abs/2017JHEP...02..036G}{ADS}]}.

\bibitem{2004JETPL..79....1D}
S.~L. {Dubovsky}, D.~S. {Gorbunov}, and G.~I. {Rubtsov}.
\newblock {Narrowing the window for millicharged particles by CMB anisotropy}.
\newblock {\em Soviet Journal of Experimental and Theoretical Physics Letters},
  79(1):1--5, January 2004.
\newblock \href {http://arxiv.org/abs/hep-ph/0311189}
  {\path{arXiv:hep-ph/0311189}}, \href {http://dx.doi.org/10.1134/1.1675909}
  {\path{[DOI]}},
  {\small[\href{https://ui.adsabs.harvard.edu/abs/2004JETPL..79....1D}{ADS}]}.

\bibitem{2013PhRvD..88k7701D}
A.~D. {Dolgov}, S.~L. {Dubovsky}, G.~I. {Rubtsov}, and I.~I. {Tkachev}.
\newblock {Constraints on millicharged particles from Planck data}.
\newblock {\em \prd}, 88(11):117701, December 2013.
\newblock \href {http://arxiv.org/abs/1310.2376} {\path{arXiv:1310.2376}},
  \href {http://dx.doi.org/10.1103/PhysRevD.88.117701} {\path{[DOI]}},
  {\small[\href{https://ui.adsabs.harvard.edu/abs/2013PhRvD..88k7701D}{ADS}]}.

\bibitem{2011PhRvD..83f3509M}
Samuel~D. {McDermott}, Hai-Bo {Yu}, and Kathryn~M. {Zurek}.
\newblock {Turning off the lights: How dark is dark matter?}
\newblock {\em \prd}, 83(6):063509, March 2011.
\newblock \href {http://arxiv.org/abs/1011.2907} {\path{arXiv:1011.2907}},
  \href {http://dx.doi.org/10.1103/PhysRevD.83.063509} {\path{[DOI]}},
  {\small[\href{https://ui.adsabs.harvard.edu/abs/2011PhRvD..83f3509M}{ADS}]}.

\bibitem{2004ApJ...606..819M}
M.~{Markevitch}, A.~H. {Gonzalez}, D.~{Clowe}, A.~{Vikhlinin}, W.~{Forman},
  C.~{Jones}, S.~{Murray}, and W.~{Tucker}.
\newblock {Direct Constraints on the Dark Matter Self-Interaction Cross Section
  from the Merging Galaxy Cluster 1E 0657-56}.
\newblock {\em \apj}, 606(2):819--824, May 2004.
\newblock \href {http://arxiv.org/abs/astro-ph/0309303}
  {\path{arXiv:astro-ph/0309303}}, \href {http://dx.doi.org/10.1086/383178}
  {\path{[DOI]}},
  {\small[\href{https://ui.adsabs.harvard.edu/abs/2004ApJ...606..819M}{ADS}]}.

\bibitem{Isgur:1989vq}
Nathan Isgur and Mark~B. Wise.
\newblock {Weak Decays of Heavy Mesons in the Static Quark Approximation}.
\newblock {\em Phys. Lett. B}, 232:113--117, 1989.
\newblock \href {http://dx.doi.org/10.1016/0370-2693(89)90566-2}
  {\path{[DOI]}}.

\bibitem{Georgi:1990um}
Howard Georgi.
\newblock {An Effective Field Theory for Heavy Quarks at Low-energies}.
\newblock {\em Phys. Lett. B}, 240:447--450, 1990.
\newblock \href {http://dx.doi.org/10.1016/0370-2693(90)91128-X}
  {\path{[DOI]}}.

\bibitem{1994NuPhB.412..181J}
Elizabeth {Jenkins}.
\newblock {Heavy meson masses in chiral perturbation theory with heavy quark
  symmetry}.
\newblock {\em Nuclear Physics B}, 412(1):181--200, January 1994.
\newblock \href {http://arxiv.org/abs/hep-ph/9212295}
  {\path{arXiv:hep-ph/9212295}}, \href
  {http://dx.doi.org/10.1016/0550-3213(94)90499-5} {\path{[DOI]}},
  {\small[\href{https://ui.adsabs.harvard.edu/abs/1994NuPhB.412..181J}{ADS}]}.

\bibitem{Mehen:2005hc}
Thomas Mehen and Roxanne~P. Springer.
\newblock {Even- and odd-parity charmed meson masses in heavy hadron chiral
  perturbation theory}.
\newblock {\em Phys. Rev. D}, 72:034006, 2005.
\newblock \href {http://arxiv.org/abs/hep-ph/0503134}
  {\path{arXiv:hep-ph/0503134}}, \href
  {http://dx.doi.org/10.1103/PhysRevD.72.034006} {\path{[DOI]}}.

\bibitem{Wise:1992hn}
Mark~B. Wise.
\newblock {Chiral perturbation theory for hadrons containing a heavy quark}.
\newblock {\em Phys. Rev. D}, 45(7):R2188, 1992.
\newblock \href {http://dx.doi.org/10.1103/PhysRevD.45.R2188} {\path{[DOI]}}.

\bibitem{Yan:1992gz}
Tung-Mow Yan, Hai-Yang Cheng, Chi-Yee Cheung, Guey-Lin Lin, Y.~C. Lin, and
  Hoi-Lai Yu.
\newblock {Heavy quark symmetry and chiral dynamics}.
\newblock {\em Phys. Rev. D}, 46:1148--1164, 1992.
\newblock [Erratum: Phys.Rev.D 55, 5851 (1997)].
\newblock \href {http://dx.doi.org/10.1103/PhysRevD.46.1148} {\path{[DOI]}}.

\bibitem{1997PhR...281..145C}
R.~{Casalbuoni}, A.~{Deandrea}, N.~{Di Bartolomeo}, R.~{Gatto}, F.~{Feruglio},
  and G.~{Nardulli}.
\newblock {Phenomenology of heavy meson chiral lagrangians}.
\newblock {\em \physrep}, 281(3):145--238, March 1997.
\newblock \href {http://arxiv.org/abs/hep-ph/9605342}
  {\path{arXiv:hep-ph/9605342}}, \href
  {http://dx.doi.org/10.1016/S0370-1573(96)00027-0} {\path{[DOI]}},
  {\small[\href{https://ui.adsabs.harvard.edu/abs/1997PhR...281..145C}{ADS}]}.

\bibitem{2002ApJ...579....1F}
Steven~R. {Furlanetto} and Abraham {Loeb}.
\newblock {The 21 Centimeter Forest: Radio Absorption Spectra as Probes of
  Minihalos before Reionization}.
\newblock {\em \apj}, 579(1):1--9, November 2002.
\newblock \href {http://arxiv.org/abs/astro-ph/0206308}
  {\path{arXiv:astro-ph/0206308}}, \href {http://dx.doi.org/10.1086/342757}
  {\path{[DOI]}},
  {\small[\href{https://ui.adsabs.harvard.edu/abs/2002ApJ...579....1F}{ADS}]}.

\bibitem{2022arXiv221006498R}
Elham {Rahimi}, Evan {Vienneau}, Nassim {Bozorgnia}, and Andrew {Robertson}.
\newblock {The local dark matter distribution in self-interacting dark matter
  halos}.
\newblock {\em arXiv e-prints}, page arXiv:2210.06498, October 2022.
\newblock \href {http://arxiv.org/abs/2210.06498} {\path{arXiv:2210.06498}},
  {\small[\href{https://ui.adsabs.harvard.edu/abs/2022arXiv221006498R}{ADS}]}.

\bibitem{2007PhRvD..75b3513C}
Sergio {Colafrancesco}, Stefano {Profumo}, and Piero {Ullio}.
\newblock {Detecting dark matter WIMPs in the Draco dwarf: A multiwavelength
  perspective}.
\newblock {\em \prd}, 75(2):023513, January 2007.
\newblock \href {http://arxiv.org/abs/astro-ph/0607073}
  {\path{arXiv:astro-ph/0607073}}, \href
  {http://dx.doi.org/10.1103/PhysRevD.75.023513} {\path{[DOI]}},
  {\small[\href{https://ui.adsabs.harvard.edu/abs/2007PhRvD..75b3513C}{ADS}]}.

\bibitem{2013MNRAS.431L..20Z}
J.~{Zavala}, M.~{Vogelsberger}, and M.~G. {Walker}.
\newblock {Constraining self-interacting dark matter with the Milky way's dwarf
  spheroidals.}
\newblock {\em \mnras}, 431:L20--L24, April 2013.
\newblock \href {http://arxiv.org/abs/1211.6426} {\path{arXiv:1211.6426}},
  \href {http://dx.doi.org/10.1093/mnrasl/sls053} {\path{[DOI]}},
  {\small[\href{https://ui.adsabs.harvard.edu/abs/2013MNRAS.431L..20Z}{ADS}]}.

\bibitem{2017JCAP...07..025R}
Marco {Regis}, Laura {Richter}, and Sergio {Colafrancesco}.
\newblock {Dark matter in the Reticulum II dSph: a radio search}.
\newblock {\em \jcap}, 2017(7):025, July 2017.
\newblock \href {http://arxiv.org/abs/1703.09921} {\path{arXiv:1703.09921}},
  \href {http://dx.doi.org/10.1088/1475-7516/2017/07/025} {\path{[DOI]}},
  {\small[\href{https://ui.adsabs.harvard.edu/abs/2017JCAP...07..025R}{ADS}]}.

\bibitem{2018PhRvD..98h3024C}
Andrea {Caputo}, Carlos~Pe{\~n}a {Garay}, and Samuel~J. {Witte}.
\newblock {Looking for axion dark matter in dwarf spheroidal galaxies}.
\newblock {\em \prd}, 98(8):083024, October 2018.
\newblock \href {http://arxiv.org/abs/1805.08780} {\path{arXiv:1805.08780}},
  \href {http://dx.doi.org/10.1103/PhysRevD.98.083024} {\path{[DOI]}},
  {\small[\href{https://ui.adsabs.harvard.edu/abs/2018PhRvD..98h3024C}{ADS}]}.

\bibitem{2021PhLB..81436075R}
Marco {Regis}, Marco {Taoso}, Daniel {Vaz}, Jarle {Brinchmann}, Sebastiaan~L.
  {Zoutendijk}, Nicolas~F. {Bouch{\'e}}, and Matthias {Steinmetz}.
\newblock {Searching for light in the darkness: Bounds on ALP dark matter with
  the optical MUSE-faint survey}.
\newblock {\em Physics Letters B}, 814:136075, March 2021.
\newblock \href {http://arxiv.org/abs/2009.01310} {\path{arXiv:2009.01310}},
  \href {http://dx.doi.org/10.1016/j.physletb.2021.136075} {\path{[DOI]}},
  {\small[\href{https://ui.adsabs.harvard.edu/abs/2021PhLB..81436075R}{ADS}]}.

\bibitem{2021PhRvD.103l3028W}
Digvijay {Wadekar} and Glennys~R. {Farrar}.
\newblock {Gas-rich dwarf galaxies as a new probe of dark matter interactions
  with ordinary matter}.
\newblock {\em \prd}, 103(12):123028, June 2021.
\newblock \href {http://arxiv.org/abs/1903.12190} {\path{arXiv:1903.12190}},
  \href {http://dx.doi.org/10.1103/PhysRevD.103.123028} {\path{[DOI]}},
  {\small[\href{https://ui.adsabs.harvard.edu/abs/2021PhRvD.103l3028W}{ADS}]}.

\bibitem{2019IAUS..344..483F}
Yakov {Faerman}, Amiel {Sternberg}, and Christopher~F. {McKee}.
\newblock {Ultra-Compact High Velocity Clouds as Minihalos and Dwarf Galaxies}.
\newblock In Kristen B.~W. {McQuinn} and Sabrina {Stierwalt}, editors, {\em
  Dwarf Galaxies: From the Deep Universe to the Present}, volume 344, pages
  483--487, October 2019.
\newblock \href {http://dx.doi.org/10.1017/S1743921318006269} {\path{[DOI]}},
  {\small[\href{https://ui.adsabs.harvard.edu/abs/2019IAUS..344..483F}{ADS}]}.

\bibitem{2008MNRAS.384..535R}
Emma~V. {Ryan-Weber}, Ayesha {Begum}, Tom {Oosterloo}, Sabyasachi {Pal},
  Michael~J. {Irwin}, Vasily {Belokurov}, N.~Wyn {Evans}, and Daniel~B.
  {Zucker}.
\newblock {The Local Group dwarf Leo T: HI on the brink of star formation}.
\newblock {\em \mnras}, 384(2):535--540, February 2008.
\newblock \href {http://arxiv.org/abs/0711.2979} {\path{arXiv:0711.2979}},
  \href {http://dx.doi.org/10.1111/j.1365-2966.2007.12734.x} {\path{[DOI]}},
  {\small[\href{https://ui.adsabs.harvard.edu/abs/2008MNRAS.384..535R}{ADS}]}.

\bibitem{2006MNRAS.370.1867F}
Steven~R. {Furlanetto}.
\newblock {The 21-cm forest}.
\newblock {\em \mnras}, 370(4):1867--1875, August 2006.
\newblock \href {http://arxiv.org/abs/astro-ph/0604223}
  {\path{arXiv:astro-ph/0604223}}, \href
  {http://dx.doi.org/10.1111/j.1365-2966.2006.10603.x} {\path{[DOI]}},
  {\small[\href{https://ui.adsabs.harvard.edu/abs/2006MNRAS.370.1867F}{ADS}]}.

\bibitem{2011MNRAS.410.2025X}
Yidong {Xu}, Andrea {Ferrara}, and Xuelei {Chen}.
\newblock {The earliest galaxies seen in 21 cm line absorption}.
\newblock {\em \mnras}, 410(3):2025--2042, January 2011.
\newblock \href {http://arxiv.org/abs/1009.1149} {\path{arXiv:1009.1149}},
  \href {http://dx.doi.org/10.1111/j.1365-2966.2010.17579.x} {\path{[DOI]}},
  {\small[\href{https://ui.adsabs.harvard.edu/abs/2011MNRAS.410.2025X}{ADS}]}.

\bibitem{2008ApJ...688..709T}
Jeremy {Tinker}, Andrey~V. {Kravtsov}, Anatoly {Klypin}, Kevork {Abazajian},
  Michael {Warren}, Gustavo {Yepes}, Stefan {Gottl{\"o}ber}, and Daniel~E.
  {Holz}.
\newblock {Toward a Halo Mass Function for Precision Cosmology: The Limits of
  Universality}.
\newblock {\em \apj}, 688(2):709--728, December 2008.
\newblock \href {http://arxiv.org/abs/0803.2706} {\path{arXiv:0803.2706}},
  \href {http://dx.doi.org/10.1086/591439} {\path{[DOI]}},
  {\small[\href{https://ui.adsabs.harvard.edu/abs/2008ApJ...688..709T}{ADS}]}.

\bibitem{2015MNRAS.450.1465G}
Antonella {Garzilli}, Tom {Theuns}, and Joop {Schaye}.
\newblock {The broadening of Lyman-{\ensuremath{\alpha}} forest absorption
  lines}.
\newblock {\em \mnras}, 450(2):1465--1476, June 2015.
\newblock \href {http://arxiv.org/abs/1502.05715} {\path{arXiv:1502.05715}},
  \href {http://dx.doi.org/10.1093/mnras/stv394} {\path{[DOI]}},
  {\small[\href{https://ui.adsabs.harvard.edu/abs/2015MNRAS.450.1465G}{ADS}]}.

\bibitem{2019A&A...621A..56P}
Lorenzo {Posti} and Amina {Helmi}.
\newblock {Mass and shape of the Milky Way's dark matter halo with globular
  clusters from Gaia and Hubble}.
\newblock {\em \aap}, 621:A56, January 2019.
\newblock \href {http://arxiv.org/abs/1805.01408} {\path{arXiv:1805.01408}},
  \href {http://dx.doi.org/10.1051/0004-6361/201833355} {\path{[DOI]}},
  {\small[\href{https://ui.adsabs.harvard.edu/abs/2019A&A...621A..56P}{ADS}]}.

\bibitem{2006SSRv..123..485G}
Jonathan~P. {Gardner et. al}.
\newblock {The James Webb Space Telescope}.
\newblock {\em \ssr}, 123(4):485--606, April 2006.
\newblock \href {http://arxiv.org/abs/astro-ph/0606175}
  {\path{arXiv:astro-ph/0606175}}, \href
  {http://dx.doi.org/10.1007/s11214-006-8315-7} {\path{[DOI]}},
  {\small[\href{https://ui.adsabs.harvard.edu/abs/2006SSRv..123..485G}{ADS}]}.

\bibitem{2002AJ....123..485S}
Chris {Stoughton et al}.
\newblock {Sloan Digital Sky Survey: Early Data Release}.
\newblock {\em \aj}, 123(1):485--548, January 2002.
\newblock \href {http://dx.doi.org/10.1086/324741} {\path{[DOI]}},
  {\small[\href{https://ui.adsabs.harvard.edu/abs/2002AJ....123..485S}{ADS}]}.

\bibitem{2011A&A...536A.105V}
J.~{Vernet et al.}
\newblock {X-shooter, the new wide band intermediate resolution spectrograph at
  the ESO Very Large Telescope}.
\newblock {\em \aap}, 536:A105, December 2011.
\newblock \href {http://arxiv.org/abs/1110.1944} {\path{arXiv:1110.1944}},
  \href {http://dx.doi.org/10.1051/0004-6361/201117752} {\path{[DOI]}},
  {\small[\href{https://ui.adsabs.harvard.edu/abs/2011A&A...536A.105V}{ADS}]}.

\bibitem{2000SPIE.4012....2W}
Martin~C. {Weisskopf}, Harvey~D. {Tananbaum}, Leon~P. {Van Speybroeck}, and
  Stephen~L. {O'Dell}.
\newblock {Chandra X-ray Observatory (CXO): overview}.
\newblock In Joachim~E. {Truemper} and Bernd {Aschenbach}, editors, {\em X-Ray
  Optics, Instruments, and Missions III}, volume 4012 of {\em Society of
  Photo-Optical Instrumentation Engineers (SPIE) Conference Series}, pages
  2--16, July 2000.
\newblock \href {http://arxiv.org/abs/astro-ph/0004127}
  {\path{arXiv:astro-ph/0004127}}, \href {http://dx.doi.org/10.1117/12.391545}
  {\path{[DOI]}},
  {\small[\href{https://ui.adsabs.harvard.edu/abs/2000SPIE.4012....2W}{ADS}]}.

\bibitem{2001A&A...365L...1J}
F.~{Jansen}, D.~{Lumb}, B.~{Altieri}, J.~{Clavel}, M.~{Ehle}, C.~{Erd},
  C.~{Gabriel}, M.~{Guainazzi}, P.~{Gondoin}, R.~{Much}, R.~{Munoz},
  M.~{Santos}, N.~{Schartel}, D.~{Texier}, and G.~{Vacanti}.
\newblock {XMM-Newton observatory. I. The spacecraft and operations}.
\newblock {\em \aap}, 365:L1--L6, January 2001.
\newblock \href {http://dx.doi.org/10.1051/0004-6361:20000036} {\path{[DOI]}},
  {\small[\href{https://ui.adsabs.harvard.edu/abs/2001A&A...365L...1J}{ADS}]}.

\bibitem{1957SvA.....1..678S}
V.~V. {Sobolev}.
\newblock {The Diffusion of L{\ensuremath{\alpha}} Radiation in Nebulae and
  Stellar Envelopes.}
\newblock {\em \sovast}, 1:678, October 1957.
\newblock
  {\small[\href{https://ui.adsabs.harvard.edu/abs/1957SvA.....1..678S}{ADS}]}.

\bibitem{2005ApJ...622.1356Z}
B.~{Zygelman}.
\newblock {Hyperfine Level-changing Collisions of Hydrogen Atoms and Tomography
  of the Dark Age Universe}.
\newblock {\em \apj}, 622(2):1356--1362, April 2005.
\newblock \href {http://dx.doi.org/10.1086/427682} {\path{[DOI]}},
  {\small[\href{https://ui.adsabs.harvard.edu/abs/2005ApJ...622.1356Z}{ADS}]}.

\bibitem{1968ApJ...153....1P}
P.~J.~E. {Peebles}.
\newblock {Recombination of the Primeval Plasma}.
\newblock {\em \apj}, 153:1, July 1968.
\newblock \href {http://dx.doi.org/10.1086/149628} {\path{[DOI]}},
  {\small[\href{https://ui.adsabs.harvard.edu/abs/1968ApJ...153....1P}{ADS}]}.

\bibitem{1969JETP...28..146Z}
Ya.~B. {Zel'dovich}, V.~G. {Kurt}, and R.~A. {Syunyaev}.
\newblock {Recombination of Hydrogen in the Hot Model of the Universe}.
\newblock {\em Soviet Journal of Experimental and Theoretical Physics}, 28:146,
  January 1969.
\newblock
  {\small[\href{https://ui.adsabs.harvard.edu/abs/1969JETP...28..146Z}{ADS}]}.

\bibitem{1986rpa..book.....R}
George~B. {Rybicki} and Alan~P. {Lightman}.
\newblock {\em {Radiative Processes in Astrophysics}}.
\newblock 1986.
\newblock
  {\small[\href{https://ui.adsabs.harvard.edu/abs/1986rpa..book.....R}{ADS}]}.

\bibitem{2011hea..book.....L}
Malcolm~S. {Longair}.
\newblock {\em {High Energy Astrophysics}}.
\newblock 2011.
\newblock
  {\small[\href{https://ui.adsabs.harvard.edu/abs/2011hea..book.....L}{ADS}]}.

\bibitem{2014MNRAS.444..420V}
P.~A.~M. {van Hoof}, R.~J.~R. {Williams}, K.~{Volk}, M.~{Chatzikos}, G.~J.
  {Ferland}, M.~{Lykins}, R.~L. {Porter}, and Y.~{Wang}.
\newblock {Accurate determination of the free-free Gaunt factor - I.
  Non-relativistic Gaunt factors}.
\newblock {\em \mnras}, 444(1):420--428, October 2014.
\newblock \href {http://arxiv.org/abs/1407.5048} {\path{arXiv:1407.5048}},
  \href {http://dx.doi.org/10.1093/mnras/stu1438} {\path{[DOI]}},
  {\small[\href{https://ui.adsabs.harvard.edu/abs/2014MNRAS.444..420V}{ADS}]}.

\bibitem{2015ApJ...810....3S}
Mayuri {Sathyanarayana Rao}, Ravi {Subrahmanyan}, N.~{Udaya Shankar}, and Jens
  {Chluba}.
\newblock {On the Detection of Spectral Ripples from the Recombination Epoch}.
\newblock {\em \apj}, 810(1):3, September 2015.
\newblock \href {http://arxiv.org/abs/1501.07191} {\path{arXiv:1501.07191}},
  \href {http://dx.doi.org/10.1088/0004-637X/810/1/3} {\path{[DOI]}},
  {\small[\href{https://ui.adsabs.harvard.edu/abs/2015ApJ...810....3S}{ADS}]}.

\bibitem{1957JETP....4..730K}
A.~S. {Kompaneets}.
\newblock {The Establishment of Thermal Equilibrium between Quanta and
  Electrons}.
\newblock {\em Soviet Journal of Experimental and Theoretical Physics},
  4(5):730--737, May 1957.
\newblock
  {\small[\href{https://ui.adsabs.harvard.edu/abs/1957JETP....4..730K}{ADS}]}.

\bibitem{1969Ap&SS...4..301Z}
Ya.~B. {Zeldovich} and R.~A. {Sunyaev}.
\newblock {The Interaction of Matter and Radiation in a Hot-Model Universe}.
\newblock {\em \apss}, 4(3):301--316, July 1969.
\newblock \href {http://dx.doi.org/10.1007/BF00661821} {\path{[DOI]}},
  {\small[\href{https://ui.adsabs.harvard.edu/abs/1969Ap&SS...4..301Z}{ADS}]}.

\bibitem{1981ApJ...244..392L}
A.~P. {Lightman}.
\newblock {Double Compton emission in radiation dominated thermal plasmas}.
\newblock {\em \apj}, 244:392--405, March 1981.
\newblock \href {http://dx.doi.org/10.1086/158716} {\path{[DOI]}},
  {\small[\href{https://ui.adsabs.harvard.edu/abs/1981ApJ...244..392L}{ADS}]}.

\bibitem{1981MNRAS.194..439T}
K.~S. {Thorne}.
\newblock {Relativistic radiative transfer - Moment formalisms}.
\newblock {\em \mnras}, 194:439--473, February 1981.
\newblock \href {http://dx.doi.org/10.1093/mnras/194.2.439} {\path{[DOI]}},
  {\small[\href{https://ui.adsabs.harvard.edu/abs/1981MNRAS.194..439T}{ADS}]}.

\bibitem{1982A&A...107...39D}
L.~{Danese} and G.~{de Zotti}.
\newblock {Double Compton process and the spectrum of the microwave
  background}.
\newblock {\em \aap}, 107(1):39--42, March 1982.
\newblock
  {\small[\href{https://ui.adsabs.harvard.edu/abs/1982A&A...107...39D}{ADS}]}.

\bibitem{1970Ap&SS...7...20S}
R.~A. {Sunyaev} and Ya.~B. {Zeldovich}.
\newblock {The interaction of matter and radiation in the hot model of the
  Universe, II}.
\newblock {\em \apss}, 7(1):20--30, April 1970.
\newblock \href {http://dx.doi.org/10.1007/BF00653472} {\path{[DOI]}},
  {\small[\href{https://ui.adsabs.harvard.edu/abs/1970Ap&SS...7...20S}{ADS}]}.

\bibitem{1991A&A...246...49B}
C.~{Burigana}, L.~{Danese}, and G.~{de Zotti}.
\newblock {Formation and evolution of early distortions of the microwave
  background spectrum - A numerical study}.
\newblock {\em \aap}, 246(1):49--58, June 1991.
\newblock
  {\small[\href{https://ui.adsabs.harvard.edu/abs/1991A&A...246...49B}{ADS}]}.

\bibitem{1993PhRvD..48..485H}
Wayne {Hu} and Joseph {Silk}.
\newblock {Thermalization and spectral distortions of the cosmic background
  radiation}.
\newblock {\em \prd}, 48(2):485--502, July 1993.
\newblock \href {http://dx.doi.org/10.1103/PhysRevD.48.485} {\path{[DOI]}},
  {\small[\href{https://ui.adsabs.harvard.edu/abs/1993PhRvD..48..485H}{ADS}]}.

\bibitem{2009A&A...507.1243P}
P.~{Procopio} and C.~{Burigana}.
\newblock {A numerical code for the solution of the Kompaneets equation in
  cosmological context}.
\newblock {\em \aap}, 507(3):1243--1256, December 2009.
\newblock \href {http://arxiv.org/abs/0905.2886} {\path{arXiv:0905.2886}},
  \href {http://dx.doi.org/10.1051/0004-6361/200912061} {\path{[DOI]}},
  {\small[\href{https://ui.adsabs.harvard.edu/abs/2009A&A...507.1243P}{ADS}]}.

\bibitem{2012MNRAS.419.1294C}
J.~{Chluba} and R.~A. {Sunyaev}.
\newblock {The evolution of CMB spectral distortions in the early Universe}.
\newblock {\em \mnras}, 419(2):1294--1314, January 2012.
\newblock \href {http://arxiv.org/abs/1109.6552} {\path{arXiv:1109.6552}},
  \href {http://dx.doi.org/10.1111/j.1365-2966.2011.19786.x} {\path{[DOI]}},
  {\small[\href{https://ui.adsabs.harvard.edu/abs/2012MNRAS.419.1294C}{ADS}]}.

\bibitem{2012JCAP...06..038K}
Rishi {Khatri} and Rashid~A. {Sunyaev}.
\newblock {Creation of the CMB spectrum: precise analytic solutions for the
  blackbody photosphere}.
\newblock {\em \jcap}, 2012(6):038, June 2012.
\newblock \href {http://arxiv.org/abs/1203.2601} {\path{arXiv:1203.2601}},
  \href {http://dx.doi.org/10.1088/1475-7516/2012/06/038} {\path{[DOI]}},
  {\small[\href{https://ui.adsabs.harvard.edu/abs/2012JCAP...06..038K}{ADS}]}.

\bibitem{1996ApJ...473..576F}
D.~J. {Fixsen}, E.~S. {Cheng}, J.~M. {Gales}, J.~C. {Mather}, R.~A. {Shafer},
  and E.~L. {Wright}.
\newblock {The Cosmic Microwave Background Spectrum from the Full COBE FIRAS
  Data Set}.
\newblock {\em \apj}, 473:576, December 1996.
\newblock \href {http://arxiv.org/abs/astro-ph/9605054}
  {\path{arXiv:astro-ph/9605054}}, \href {http://dx.doi.org/10.1086/178173}
  {\path{[DOI]}},
  {\small[\href{https://ui.adsabs.harvard.edu/abs/1996ApJ...473..576F}{ADS}]}.

\bibitem{2020A&A...641A...5P}
{Planck Collaboration} and N.. {Aghanim et. al}.
\newblock {Planck 2018 results. V. CMB power spectra and likelihoods}.
\newblock {\em \aap}, 641:A5, September 2020.
\newblock \href {http://arxiv.org/abs/1907.12875} {\path{arXiv:1907.12875}},
  \href {http://dx.doi.org/10.1051/0004-6361/201936386} {\path{[DOI]}},
  {\small[\href{https://ui.adsabs.harvard.edu/abs/2020A&A...641A...5P}{ADS}]}.

\bibitem{2018PhRvL.121h1101S}
Katelin {Schutz}, Tongyan {Lin}, Benjamin~R. {Safdi}, and Chih-Liang {Wu}.
\newblock {Constraining a Thin Dark Matter Disk with G a i a}.
\newblock {\em \prl}, 121(8):081101, August 2018.
\newblock \href {http://arxiv.org/abs/1711.03103} {\path{arXiv:1711.03103}},
  \href {http://dx.doi.org/10.1103/PhysRevLett.121.081101} {\path{[DOI]}},
  {\small[\href{https://ui.adsabs.harvard.edu/abs/2018PhRvL.121h1101S}{ADS}]}.

\bibitem{2022Natur.603..599X}
Maosheng {Xiang} and Hans-Walter {Rix}.
\newblock {A time-resolved picture of our Milky Way's early formation history}.
\newblock {\em \nat}, 603(7902):599--603, March 2022.
\newblock \href {http://arxiv.org/abs/2203.12110} {\path{arXiv:2203.12110}},
  \href {http://dx.doi.org/10.1038/s41586-022-04496-5} {\path{[DOI]}},
  {\small[\href{https://ui.adsabs.harvard.edu/abs/2022Natur.603..599X}{ADS}]}.

\bibitem{2017ApJ...837...30G}
S.~{Gillessen}, P.~M. {Plewa}, F.~{Eisenhauer}, R.~{Sari}, I.~{Waisberg},
  M.~{Habibi}, O.~{Pfuhl}, E.~{George}, J.~{Dexter}, S.~{von Fellenberg},
  T.~{Ott}, and R.~{Genzel}.
\newblock {An Update on Monitoring Stellar Orbits in the Galactic Center}.
\newblock {\em \apj}, 837(1):30, March 2017.
\newblock \href {http://arxiv.org/abs/1611.09144} {\path{arXiv:1611.09144}},
  \href {http://dx.doi.org/10.3847/1538-4357/aa5c41} {\path{[DOI]}},
  {\small[\href{https://ui.adsabs.harvard.edu/abs/2017ApJ...837...30G}{ADS}]}.

\bibitem{2013ApJS..208...20B}
C.~{Bennett et. al}.
\newblock {Nine-year Wilkinson Microwave Anisotropy Probe (WMAP) Observations:
  Final Maps and Results}.
\newblock {\em \apjs}, 208(2):20, October 2013.
\newblock \href {http://arxiv.org/abs/1212.5225} {\path{arXiv:1212.5225}},
  \href {http://dx.doi.org/10.1088/0067-0049/208/2/20} {\path{[DOI]}},
  {\small[\href{https://ui.adsabs.harvard.edu/abs/2013ApJS..208...20B}{ADS}]}.

\bibitem{2010AdAst2010E...5D}
W.~J.~G. {de Blok}.
\newblock {The Core-Cusp Problem}.
\newblock {\em Advances in Astronomy}, 2010:789293, January 2010.
\newblock \href {http://arxiv.org/abs/0910.3538} {\path{arXiv:0910.3538}},
  \href {http://dx.doi.org/10.1155/2010/789293} {\path{[DOI]}},
  {\small[\href{https://ui.adsabs.harvard.edu/abs/2010AdAst2010E...5D}{ADS}]}.

\bibitem{2000PhRvL..84.3760S}
David~N. {Spergel} and Paul~J. {Steinhardt}.
\newblock {Observational Evidence for Self-Interacting Cold Dark Matter}.
\newblock {\em \prl}, 84(17):3760--3763, April 2000.
\newblock \href {http://arxiv.org/abs/astro-ph/9909386}
  {\path{arXiv:astro-ph/9909386}}, \href
  {http://dx.doi.org/10.1103/PhysRevLett.84.3760} {\path{[DOI]}},
  {\small[\href{https://ui.adsabs.harvard.edu/abs/2000PhRvL..84.3760S}{ADS}]}.

\bibitem{2000JHEP...05..003D}
Sacha {Davidson}, Steen {Hannestad}, and Georg {Raffelt}.
\newblock {Updated bounds on milli-charged particles}.
\newblock {\em Journal of High Energy Physics}, 2000(5):003, May 2000.
\newblock \href {http://arxiv.org/abs/hep-ph/0001179}
  {\path{arXiv:hep-ph/0001179}}, \href
  {http://dx.doi.org/10.1088/1126-6708/2000/05/003} {\path{[DOI]}},
  {\small[\href{https://ui.adsabs.harvard.edu/abs/2000JHEP...05..003D}{ADS}]}.

\bibitem{2018JHEP...09..051C}
Jae~Hyeok {Chang}, Rouven {Essig}, and Samuel~D. {McDermott}.
\newblock {Supernova 1987A constraints on sub-GeV dark sectors, millicharged
  particles, the QCD axion, and an axion-like particle}.
\newblock {\em Journal of High Energy Physics}, 2018(9):51, September 2018.
\newblock \href {http://arxiv.org/abs/1803.00993} {\path{arXiv:1803.00993}},
  \href {http://dx.doi.org/10.1007/JHEP09(2018)051} {\path{[DOI]}},
  {\small[\href{https://ui.adsabs.harvard.edu/abs/2018JHEP...09..051C}{ADS}]}.

\bibitem{1987PhRvL..58.1490H}
K.~{Hirata}, T.~{Kajita}, M.~{Koshiba}, M.~{Nakahata}, Y.~{Oyama}, N.~{Sato},
  A.~{Suzuki}, M.~{Takita}, Y.~{Totsuka}, T.~{Kifune}, T.~{Suda},
  K.~{Takahashi}, T.~{Tanimori}, K.~{Miyano}, M.~{Yamada}, E.~W. {Beier}, L.~R.
  {Feldscher}, S.~B. {Kim}, A.~K. {Mann}, F.~M. {Newcomer}, R.~{van},
  W.~{Zhang}, and B.~G. {Cortez}.
\newblock {Observation of a neutrino burst from the supernova SN1987A}.
\newblock {\em \prl}, 58(14):1490--1493, April 1987.
\newblock \href {http://dx.doi.org/10.1103/PhysRevLett.58.1490} {\path{[DOI]}},
  {\small[\href{https://ui.adsabs.harvard.edu/abs/1987PhRvL..58.1490H}{ADS}]}.

\bibitem{1986PhRvD..34.2197K}
Edward~W. {Kolb}, Michael~S. {Turner}, and Terrence~P. {Walker}.
\newblock {Effect of interacting particles on primordial nucleosynthesis}.
\newblock {\em \prd}, 34(8):2197--2205, October 1986.
\newblock \href {http://dx.doi.org/10.1103/PhysRevD.34.2197} {\path{[DOI]}},
  {\small[\href{https://ui.adsabs.harvard.edu/abs/1986PhRvD..34.2197K}{ADS}]}.

\bibitem{2004PhRvD..70d3526S}
Pasquale~D. {Serpico} and Georg~G. {Raffelt}.
\newblock {MeV-mass dark matter and primordial nucleosynthesis}.
\newblock {\em \prd}, 70(4):043526, August 2004.
\newblock \href {http://arxiv.org/abs/astro-ph/0403417}
  {\path{arXiv:astro-ph/0403417}}, \href
  {http://dx.doi.org/10.1103/PhysRevD.70.043526} {\path{[DOI]}},
  {\small[\href{https://ui.adsabs.harvard.edu/abs/2004PhRvD..70d3526S}{ADS}]}.

\bibitem{2013JCAP...08..041B}
C{\'e}line {B{\oe}hm}, Matthew~J. {Dolan}, and Christopher {McCabe}.
\newblock {A lower bound on the mass of cold thermal dark matter from Planck}.
\newblock {\em \jcap}, 2013(8):041, August 2013.
\newblock \href {http://arxiv.org/abs/1303.6270} {\path{arXiv:1303.6270}},
  \href {http://dx.doi.org/10.1088/1475-7516/2013/08/041} {\path{[DOI]}},
  {\small[\href{https://ui.adsabs.harvard.edu/abs/2013JCAP...08..041B}{ADS}]}.

\bibitem{2014PhRvD..89h3508N}
Kenneth~M. {Nollett} and Gary {Steigman}.
\newblock {BBN and the CMB constrain light, electromagnetically coupled WIMPs}.
\newblock {\em \prd}, 89(8):083508, April 2014.
\newblock \href {http://arxiv.org/abs/1312.5725} {\path{arXiv:1312.5725}},
  \href {http://dx.doi.org/10.1103/PhysRevD.89.083508} {\path{[DOI]}},
  {\small[\href{https://ui.adsabs.harvard.edu/abs/2014PhRvD..89h3508N}{ADS}]}.

\bibitem{2015PhRvD..91h3505N}
Kenneth~M. {Nollett} and Gary {Steigman}.
\newblock {BBN and the CMB constrain neutrino coupled light WIMPs}.
\newblock {\em \prd}, 91(8):083505, April 2015.
\newblock \href {http://arxiv.org/abs/1411.6005} {\path{arXiv:1411.6005}},
  \href {http://dx.doi.org/10.1103/PhysRevD.91.083505} {\path{[DOI]}},
  {\small[\href{https://ui.adsabs.harvard.edu/abs/2015PhRvD..91h3505N}{ADS}]}.

\bibitem{2020JCAP...01..004S}
Nashwan {Sabti}, James {Alvey}, Miguel {Escudero}, Malcolm {Fairbairn}, and
  Diego {Blas}.
\newblock {Refined bounds on MeV-scale thermal dark sectors from BBN and the
  CMB}.
\newblock {\em \jcap}, 2020(1):004, January 2020.
\newblock \href {http://arxiv.org/abs/1910.01649} {\path{arXiv:1910.01649}},
  \href {http://dx.doi.org/10.1088/1475-7516/2020/01/004} {\path{[DOI]}},
  {\small[\href{https://ui.adsabs.harvard.edu/abs/2020JCAP...01..004S}{ADS}]}.

\bibitem{1992PhRvD..46.3372D}
Scott {Dodelson} and Michael~S. {Turner}.
\newblock {Nonequilibrium neutrino statistical mechanics in the expanding
  Universe}.
\newblock {\em \prd}, 46(8):3372--3387, October 1992.
\newblock \href {http://dx.doi.org/10.1103/PhysRevD.46.3372} {\path{[DOI]}},
  {\small[\href{https://ui.adsabs.harvard.edu/abs/1992PhRvD..46.3372D}{ADS}]}.

\bibitem{1992PZETF..56..129D}
A.~D. {Dolgov} and M.~{Fukugita}.
\newblock {Nonequilibrium neutrinos and primordial nucleosynthesis}.
\newblock {\em Pisma v Zhurnal Eksperimentalnoi i Teoreticheskoi Fiziki},
  56(3-4):129--132, August 1992.
\newblock
  {\small[\href{https://ui.adsabs.harvard.edu/abs/1992PZETF..56..129D}{ADS}]}.

\bibitem{1993PhRvD..47.4309F}
Brian~D. {Fields}, Scott {Dodelson}, and Michael~S. {Turner}.
\newblock {Effect of neutrino heating on primordial nucleosynthesis}.
\newblock {\em \prd}, 47(10):4309--4314, May 1993.
\newblock \href {http://arxiv.org/abs/astro-ph/9210007}
  {\path{arXiv:astro-ph/9210007}}, \href
  {http://dx.doi.org/10.1103/PhysRevD.47.4309} {\path{[DOI]}},
  {\small[\href{https://ui.adsabs.harvard.edu/abs/1993PhRvD..47.4309F}{ADS}]}.

\bibitem{1997NuPhB.503..426D}
A.~D. {Dolgov}, S.~H. {Hansen}, and D.~V. {Semikoz}.
\newblock {Non-equilibrium corrections to the spectra of massless neutrinos in
  the early universe}.
\newblock {\em Nuclear Physics B}, 503:426--444, February 1997.
\newblock \href {http://arxiv.org/abs/hep-ph/9703315}
  {\path{arXiv:hep-ph/9703315}}, \href
  {http://dx.doi.org/10.1016/S0550-3213(97)00479-3} {\path{[DOI]}},
  {\small[\href{https://ui.adsabs.harvard.edu/abs/1997NuPhB.503..426D}{ADS}]}.

\bibitem{2002PhR...370..333D}
A.~D. {Dolgov}.
\newblock {Neutrinos in cosmology}.
\newblock {\em \physrep}, 370(4-5):333--535, November 2002.
\newblock \href {http://arxiv.org/abs/hep-ph/0202122}
  {\path{arXiv:hep-ph/0202122}}, \href
  {http://dx.doi.org/10.1016/S0370-1573(02)00139-4} {\path{[DOI]}},
  {\small[\href{https://ui.adsabs.harvard.edu/abs/2002PhR...370..333D}{ADS}]}.

\bibitem{2019JCAP...09..070E}
Timon {Emken}, Rouven {Essig}, Chris {Kouvaris}, and Mukul {Sholapurkar}.
\newblock {Direct detection of strongly interacting sub-GeV dark matter via
  electron recoils}.
\newblock {\em \jcap}, 2019(9):070, September 2019.
\newblock \href {http://arxiv.org/abs/1905.06348} {\path{arXiv:1905.06348}},
  \href {http://dx.doi.org/10.1088/1475-7516/2019/09/070} {\path{[DOI]}},
  {\small[\href{https://ui.adsabs.harvard.edu/abs/2019JCAP...09..070E}{ADS}]}.

\bibitem{2000PhLB..480..181P}
M.~{Pospelov} and T.~{ter Veldhuis}.
\newblock {Direct and indirect limits on the electro-magnetic form factors of
  WIMPs}.
\newblock {\em Physics Letters B}, 480(1-2):181--186, May 2000.
\newblock \href {http://arxiv.org/abs/hep-ph/0003010}
  {\path{arXiv:hep-ph/0003010}}, \href
  {http://dx.doi.org/10.1016/S0370-2693(00)00358-0} {\path{[DOI]}},
  {\small[\href{https://ui.adsabs.harvard.edu/abs/2000PhLB..480..181P}{ADS}]}.

\bibitem{2012PhRvD..85g6007E}
Rouven {Essig}, Jeremy {Mardon}, and Tomer {Volansky}.
\newblock {Direct detection of sub-GeV dark matter}.
\newblock {\em \prd}, 85(7):076007, April 2012.
\newblock \href {http://arxiv.org/abs/1108.5383} {\path{arXiv:1108.5383}},
  \href {http://dx.doi.org/10.1103/PhysRevD.85.076007} {\path{[DOI]}},
  {\small[\href{https://ui.adsabs.harvard.edu/abs/2012PhRvD..85g6007E}{ADS}]}.

\bibitem{2011JCAP...07..025K}
A.~{Kogut et al.}
\newblock {The Primordial Inflation Explorer (PIXIE): a nulling polarimeter for
  cosmic microwave background observations}.
\newblock {\em \jcap}, 2011(7):025, July 2011.
\newblock \href {http://arxiv.org/abs/1105.2044} {\path{arXiv:1105.2044}},
  \href {http://dx.doi.org/10.1088/1475-7516/2011/07/025} {\path{[DOI]}},
  {\small[\href{https://ui.adsabs.harvard.edu/abs/2011JCAP...07..025K}{ADS}]}.

\bibitem{2001MNRAS.321..559B}
J.~S. {Bullock}, T.~S. {Kolatt}, Y.~{Sigad}, R.~S. {Somerville}, A.~V.
  {Kravtsov}, A.~A. {Klypin}, J.~R. {Primack}, and A.~{Dekel}.
\newblock {Profiles of dark haloes: evolution, scatter and environment}.
\newblock {\em \mnras}, 321(3):559--575, March 2001.
\newblock \href {http://arxiv.org/abs/astro-ph/9908159}
  {\path{arXiv:astro-ph/9908159}}, \href
  {http://dx.doi.org/10.1046/j.1365-8711.2001.04068.x} {\path{[DOI]}},
  {\small[\href{https://ui.adsabs.harvard.edu/abs/2001MNRAS.321..559B}{ADS}]}.

\bibitem{2008MNRAS.391.1940M}
Andrea~V. {Macci{\`o}}, Aaron~A. {Dutton}, and Frank~C. {van den Bosch}.
\newblock {Concentration, spin and shape of dark matter haloes as a function of
  the cosmological model: WMAP1, WMAP3 and WMAP5 results}.
\newblock {\em \mnras}, 391(4):1940--1954, December 2008.
\newblock \href {http://arxiv.org/abs/0805.1926} {\path{arXiv:0805.1926}},
  \href {http://dx.doi.org/10.1111/j.1365-2966.2008.14029.x} {\path{[DOI]}},
  {\small[\href{https://ui.adsabs.harvard.edu/abs/2008MNRAS.391.1940M}{ADS}]}.

\bibitem{1998ApJ...495...80B}
Greg~L. {Bryan} and Michael~L. {Norman}.
\newblock {Statistical Properties of X-Ray Clusters: Analytic and Numerical
  Comparisons}.
\newblock {\em \apj}, 495(1):80--99, March 1998.
\newblock \href {http://arxiv.org/abs/astro-ph/9710107}
  {\path{arXiv:astro-ph/9710107}}, \href {http://dx.doi.org/10.1086/305262}
  {\path{[DOI]}},
  {\small[\href{https://ui.adsabs.harvard.edu/abs/1998ApJ...495...80B}{ADS}]}.

\bibitem{2004MNRAS.352.1109A}
Y.~{Ascasibar}, G.~{Yepes}, S.~{Gottl{\"o}ber}, and V.~{M{\"u}ller}.
\newblock {On the physical origin of dark matter density profiles}.
\newblock {\em \mnras}, 352(4):1109--1120, August 2004.
\newblock \href {http://arxiv.org/abs/astro-ph/0312221}
  {\path{arXiv:astro-ph/0312221}}, \href
  {http://dx.doi.org/10.1111/j.1365-2966.2004.08005.x} {\path{[DOI]}},
  {\small[\href{https://ui.adsabs.harvard.edu/abs/2004MNRAS.352.1109A}{ADS}]}.

\bibitem{1997ApJ...486..581G}
Nickolay~Y. {Gnedin} and Jeremiah~P. {Ostriker}.
\newblock {Reionization of the Universe and the Early Production of Metals}.
\newblock {\em \apj}, 486(2):581--598, September 1997.
\newblock \href {http://arxiv.org/abs/astro-ph/9612127}
  {\path{arXiv:astro-ph/9612127}}, \href {http://dx.doi.org/10.1086/304548}
  {\path{[DOI]}},
  {\small[\href{https://ui.adsabs.harvard.edu/abs/1997ApJ...486..581G}{ADS}]}.

\bibitem{2010asph.book.....C}
Arnab~Rai {Choudhuri}.
\newblock {\em {Astrophysics for Physicists}}.
\newblock 2010.
\newblock
  {\small[\href{https://ui.adsabs.harvard.edu/abs/2010asph.book.....C}{ADS}]}.

\bibitem{cobedata}
{CMB Monopole Spectrum }.
\newblock URL:
  \url{https://lambda.gsfc.nasa.gov/product/cobe/firas_monopole_get.html}.

\end{thebibliography}
	\bibliographystyle{unsrtads}
\end{document}